\documentclass{aa}

\usepackage{natbib}
\bibpunct{(}{)}{;}{a}{}{,} 

\usepackage{graphicx}   
\usepackage{txfonts}
\usepackage[colorlinks=true,allcolors=cyan]{hyperref}

\newcommand{\MSUN}{{\rm M}_\odot}

\begin{document}

\titlerunning{X-ray AGN HOD}
\title{The cosmic web of X-ray active galactic nuclei seen through the eROSITA Final Equatorial Depth Survey (eFEDS)}

\authorrunning{Comparat et al.}
\author{
Johan Comparat\inst{1}\thanks{E-mail: comparat@mpe.mpg.de}, 
Wentao Luo\inst{2,3},
Andrea Merloni\inst{1},
Surhud More\inst{4,5},
Mara Salvato\inst{1},
Mirko Krumpe\inst{6},
Takamitsu Miyaji\inst{7},
William Brandt\inst{8,9,10},
Antonis Georgakakis\inst{11},
Masayuki Akiyama\inst{12},
Johannes Buchner\inst{1}, 
Tom Dwelly\inst{1}, 
Toshihiro Kawaguchi\inst{19},
Teng Liu\inst{1},
Tohru Nagao\inst{15},
Kirpal Nandra\inst{1},
John Silverman\inst{13,14},
Yoshiki Toba\inst{16, 17, 18, 15},
Scott F. Anderson\inst{20},
Juna Kollmeier\inst{21} 
}
\institute{
Max-Planck-Institut f\"{u}r extraterrestrische Physik (MPE), Giessenbachstrasse 1, D-85748 Garching bei M\"unchen, Germany
\and Key Laboratory for Research in Galaxies and Cosmology, School of Astronomy and Space Science, University of Science and Technology of China, Hefei, Anhui 230026, China
\and Department of Astronomy, School of Physical Sciences, University of Science and Technology of China, Hefei, Anhui 230026, China
\and The Inter-University Centre for Astronomy and Astrophysics (IUCAA), Post Bag 4, Ganeshkhind, Pune 411007, India 
\and Kavli Institute for the Physics and Mathematics of the Universe (IPMU), 5-1-5 Kashiwanoha, Kashiwa-shi, Chiba 277-8583, Japan
\and Leibniz-Institut für Astrophysik Potsdam, An der Sternwarte 16, 14482 Potsdam, Germany
\and Instituto de Astronomía sede Ensenada, Universidad Nacional Autónoma de México,
Km 107 Carretera Tijuana-Ensenada, 22860, Ensenada, Mexico
\and Department of Astronomy and Astrophysics, 525 Davey Lab, The Pennsylvania State University, University Park, PA 16802, USA
\and Institute for Gravitation and the Cosmos, The Pennsylvania State University, University Park, PA 16802, USA
\and Department of Physics, 104 Davey Laboratory, The Pennsylvania State University, University Park, PA 16802, USA
\and Institute for Astronomy and Astrophysics, National Observatory of Athens, V. Paulou \& I. Metaxa, 11532, Greece
\and Astronomical Institute, Tohoku University, 6-3 Aramaki, Aoba-ku, Sendai, Japan
\and Kavli Institute for the Physics and Mathematics of the Universe, The University of Tokyo, Kashiwa, Japan 277-8583 (Kavli IPMU, WPI)
\and Department of Astronomy, School of Science, The University of Tokyo, 7-3-1 Hongo, Bunkyo, Tokyo 113-0033, Japan
\and Research Center for Space and Cosmic Evolution, Ehime University, 2-5 Bunkyo-cho, Matsuyama, Ehime 790-8577, Japan
\and National Astronomical Observatory of Japan, 2-21-1 Osawa, Mitaka, Tokyo 181-8588, Japan
\and Department of Astronomy, Kyoto University, Kitashirakawa-Oiwake-cho, Sakyo-ku, Kyoto, Kyoto 606-8502, Japan
\and Academia Sinica Institute of Astronomy and Astrophysics, 11F Astronomy-Mathematics Building, AS/NTU, No.1, Section 4, Roosevelt Road, Taipei 10617, Taiwan
\and Department of Economics, Management and Information Science, Onomichi City University, Hisayamada 1600-2, Onomichi, Hiroshima 722-8506, Japan
\and Astronomy Department, University of Washington, Box 351580, Seattle, WA 98195, USA
\and The Carnegie Observatories, 813 Santa Barbara Street, Pasadena, CA 91101, USA
}
    
\date{\today}

\abstract{Which galaxies in the general population turn into active galactic nuclei (AGNs) is a keystone of galaxy formation and evolution.
Thanks to SRG/eROSITA's contiguous 140 square degree pilot survey field, we constructed a large, complete, and unbiased soft X-ray flux-limited ($F_X>6.5\times 10^{-15}$ erg s$^{-1}$ cm$^{-2}$) AGN sample at low redshift, $0.05<z<0.55$.
Two summary statistics, the clustering using spectra from SDSS-V and galaxy-galaxy lensing with imaging from HSC, are measured and interpreted with halo occupation distribution and abundance matching models.
Both models successfully account for the observations.
We obtain an exceptionally complete view of the AGN halo occupation distribution. 
The population of AGNs is broadly distributed among halos with a mean mass of $3.9 _{- 2.4 }^{+ 2.0 }\times10^{12}\MSUN$. This corresponds to a large-scale halo bias of $b(z=0.34)= 0.99 ^{+0.08}_{-0.10}$. 
The central occupation has a large transition parameter, $\sigma_{\log_{10}(M)}=1.28\pm0.2$. 
The satellite occupation distribution is characterized by a shallow slope, $\alpha_{{\rm sat}}=0.73\pm0.38$. 
We find that AGNs in satellites are rare, with $f_{{\rm sat}}<20\%$. 
Most soft X-ray-selected AGNs are hosted by central galaxies in their dark matter halo.
A weak correlation between soft X-ray luminosity and large-scale halo bias is confirmed (3.3$\sigma$). 
We discuss the implications of environmental-dependent AGN triggering. 
This study paves the way toward fully charting, in the coming decade, the coevolution of X-ray AGNs, their host galaxies, and dark matter halos by combining eROSITA with SDSS-V, 4MOST, DESI, LSST, and \textit{Euclid} data. }

\keywords{X-ray, active galactic nuclei}
\maketitle

\section{Introduction}
\label{sec:intro}

Active galactic nuclei (AGNs) are a keystone in galaxy evolution. 
How they are triggered and fueled are essential questions, and answering them will deepen our understanding of the coevolution between galaxies, the gas surrounding them, and their central supermassive black holes \citep[SMBHs; see reviews from][]{Padovani2017AARv..25....2P, Eckert2021Univ....7..142E}. 
This article focuses on the large-scale environment of X-ray-selected AGNs, namely the population of the dark matter halos that host them. 
X-ray selection provides AGN samples with higher completeness and purity than selections at different wavelengths \citep{Hickox2009ApJ...696..891H}. 
As devised in simulations, this population is diverse \citep{Georgakakis18, Comparat19}. 
To infer the population of dark matter halos that host a sample of galaxies, the best technique to date consists of interpreting the complementary signals from clustering and weak gravitational lensing \citep[see for example][]{Comparat2013MNRAS.433.1146C, More2015ApJ...806....2M, Coupon2015MNRAS.449.1352C, Favole16, Zhang_Luo2021AA...650A.155Z}.  

Previous studies of the clustering of X-ray-selected AGNs were limited by the total number of X-ray AGNs or a small survey area. 
They typically measured the large-scale halo bias of AGNs selected in different fashions \citep{Gilli09, Cappelluti10, Starikova11, Koutoulidis13, Koutoulidis18, Leauthaud15, Viitanen19, Allevato19}.    
The autocorrelation of X-ray-selected AGNs was studied locally ($z\sim 0.045$) with 199 AGNs in the \textit{Swift}-BAT (Burst Alert Telescope) all-sky survey by \cite{Cappelluti10}.
They found these bright low redshift AGNs to be hosted, on average, by dark matter halos of mass $1.6-2.5\times 10^{13}h^{-1} \MSUN$, corresponding to a large-scale halo bias of 1.2$\pm$0.1.  
At higher redshifts ($z\sim 1$) with deep pencil beam surveys (COSMOS observed with XMM and \textit{Chandra}, Bootes, and \textit{Chandra} compilations) and larger numbers of AGNs (ranging from 500 to 3,100), \citet{Gilli09}, \citet{Starikova11}, \citet{Koutoulidis13}, \citet{PlionisKoutoulidisKoulouridis_2018A&A...620A..17P}, \citet{Viitanen19}, and \citet{Allevato19} inferred a large-scale halo bias of $\sim 2 \pm 0.2,$ corresponding to halo masses of $4-10\times 10^{12}h^{-1}\MSUN$. 
In these studies, further splitting the samples as a function of AGN type, luminosity, or host-galaxy properties is not very conclusive due to small statistics. There are hints of a correlation with X-ray luminosity and an indication of a low satellite fraction. 
The study of the angular autocorrelation of photometrically selected AGNs, so with much larger samples, led to similar large-scale halo bias and typical dark matter halo masses \citep{Myers07, Donoso14, Koutoulidis18}.
Finally, \citet{Leauthaud15} studied the galaxy-galaxy lensing signal around 382 X-ray-selected AGNs in the COSMOS (Cosmic Evolution Survey) field \citep{Scoville2007ApJS..172....1S}. They find that the AGN host occupation is no different from that of galaxies. They explain the issue of quoting a mean for the halo mass when, instead, complete halo occupation distributions (HODs) should be discussed (see also \citealt{Georgakakis18} for an extended discussion).
Also, after controlling for stellar mass, \citet{Yang2018MNRAS.480.1022Y} found no clear dependence between the environment and the sample-averaged SMBH accretion rate or the AGN fraction, which indicates that environment-related physical mechanisms might not significantly affect SMBH growth. 

Cross-correlations with a controlled galaxy population, done to circumvent the low signal-to-noise ratio in the autocorrelation functions,  have recently been fruitful. Such studies relate AGN populations to their host dark matter halos \citep{Krumpe10, Krumpe12, Krumpe15, Krumpe18, Krumpe23, Mendez16, Mountrichas19, Zhang_Luo2021AA...650A.155Z}. They cross-correlated a similar number of X-ray-selected AGNs (between 300 and 1500) with spectroscopic galaxy surveys: 2MASS (Two Micron All Sky Survey), SDSS (Sloan Digital Sky Survey), VIPERS (VIMOS Public Extragalactic Redshift Survey), and COSMOS \citep{Skrutskie2006AJ....131.1163S, York2000, Guzzo2014AA...566A.108G, Scoville2007ApJS..172....1S}.
They obtain similar large-scale halo bias values as the autocorrelation studies and investigate the correlation with host-galaxy properties, hinting at possible correlations with stellar mass. 
This powerful technique works only with access to a well-studied galaxy sample \citep{Zehavi2011ApJ...736...59Z, Marulli2013AA...557A..17M}.
The limited signal-to-noise impedes establishing a clear definitive picture of how X-ray AGNs populate the cosmic web.

With the advent of eROSITA (extended ROentgen Survey with an Imaging Telescope Array) \citep{Predehl2021}, the number density of X-ray AGNs increased to more than a hundred per square degree in the eROSITA Final Equatorial Depth Survey \citep[eFEDS; 140 deg$^2$, $\sim$1,400 ks;][]{Brunner2021arXiv210614517B, Salvato2022AA...661A...3S}.
Accurate redshifts are required for precise clustering and lensing analysis. 
The dedicated spectroscopic observations of the X-ray sources detected in eFEDS \citep[SDSS-IV and SDSS-V;][Merloni et al. in preparation]{SDSS_DR17_2022ApJS..259...35A, Kollmeier17} enabled the accurate measurement of redshifts for about 11 thousand X-ray point sources in eFEDS (i.e., for $\sim 50\%$ of the sources).
This number of X-ray AGNs with spectra is already comparable to its predecessor follow-up of ROSAT (ROentgen SATellite) point sources \citep{Comparat20a}.

Outstanding weak-lensing data products are now available over wide areas thanks to the Hyper Suprime-Cam Subaru Strategic Program \citep[HSC-SSP;][]{Aihara2019PASJ...71..114A}. It measured accurate galaxy shapes for more than 20 source galaxies per square arcminute over vast areas (1,400 deg$^2$), which almost completely cover the eFEDS field \citep{Mandelbaum2018PASJ...70S..25M}. 

With these two outstanding observational advances, we measure the autocorrelation function and the galaxy-galaxy lensing signal of X-ray-selected AGNs to study their underlying dark matter halo distribution. 
We detail, in Sect. \ref{sec:data}, the construction of the X-ray AGN sample and the weak-lensing data products used. 
We describe the method for measuring the clustering and galaxy-galaxy lensing in Sect. \ref{sec:summary:statistics}. 
The HOD and sub-halo abundance matching (SHAM) models used are detailed in Sect. \ref{sec:models}.
Results are discussed in Sects. \ref{sec:results} and \ref{sec:discussion}.
Throughout, we assume a flat Lambda cold dark matter cosmology with $H_0=67.74$ km s$^{-1}$ Mpc$^{-1}$ and $\Omega_m(z=0)$=0.3089 \citep{PlanckCosmo2020AA...641A...6P}. 
The halo mass is defined as 200 times the background density. The halo mass function used is that of  \citet{TinkerRobertsonKravtsov_2010ApJ...724..878T}. 
The uncertainties are $1\sigma$ unless stated otherwise. 
Magnitudes are in the AB system \citep{Oke1983}. 
Throughout the article, we use the shorthand ``AGNs'' to designate ``X-ray-selected AGNs.'' 

\section{Data}
\label{sec:data}

In this section we describe the X-ray observations (Sect. \ref{subsec:data:efeds}) and the weak-lensing data products (Sect. \ref{sec:data:hsc}).

\subsection{eROSITA eFEDS}
\label{subsec:data:efeds}

\begin{figure*}
    \centering
\includegraphics[width=1.95\columnwidth]{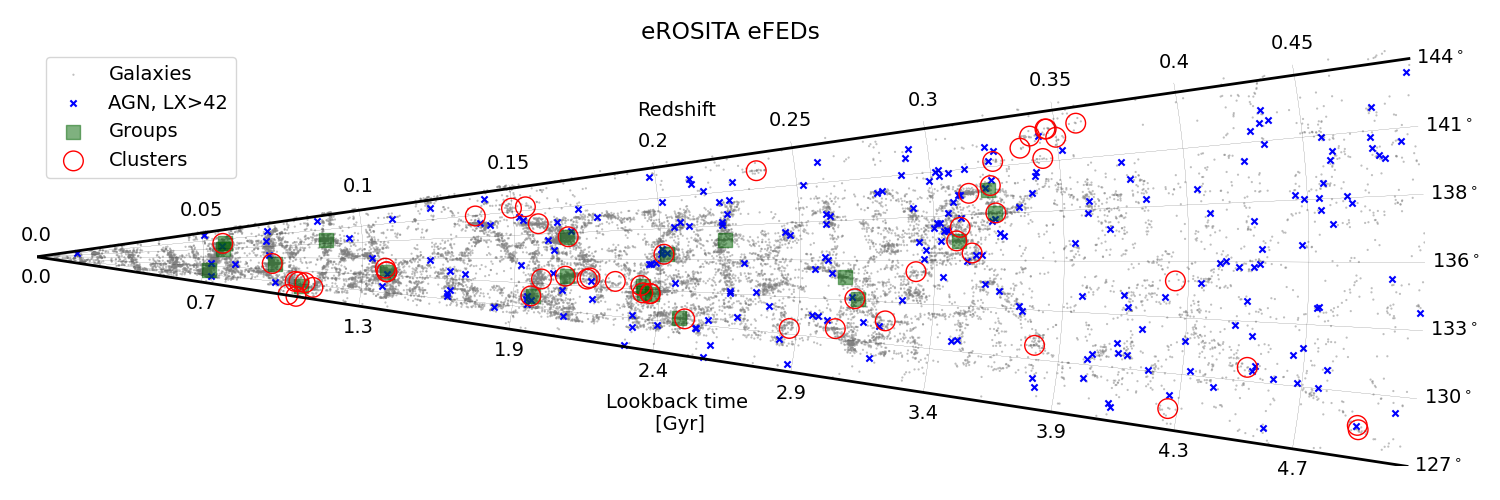}
    \caption{Slice of the light cone sampled by the X-ray-selected eFEDS AGN sample in the redshift range $0.05<z<0.55$ \citep[blue crosses;][]{Salvato2022AA...661A...3S}. The surrounding large-scale structure is sampled by GAMA galaxies \citep[gray points;][]{Driver2022MNRAS.513..439D} and GAMA galaxy groups \citep[green squares;][]{RobothamNorbergDriver_2011MNRAS.416.2640R} as well as by eROSITA eFEDS clusters \citep[red circles;][]{LiuAng2022AA...661A...2L}.}
    \label{fig:wedge}
\end{figure*}

We used the public Early Data Release eROSITA point source catalog of the eFEDS Performance Verification survey \citep{Brunner2021arXiv210614517B}. 
The catalog contains 20,191 primary sources, over 140 deg$^2$, detected with a likelihood greater than 8 ($\textrm{ERO}\_\textrm{DET}\_\textrm{LIKE} > 8$) and with a reliable counterpart ($\textrm{CTP}_\textrm{quality} \geq 2$) determined as described in  \citet{Salvato2022AA...661A...3S}. 
The catalog is flux-limited to F$_X>6.5 \times 10^{-15}$ erg s$^{-1}$ cm$^{-2}$ in the soft X-ray (0.5-2 keV). 
Simulations are only at X-ray wavelengths and not in the optical, so the impact of the determination of the counterpart is studied empirically in Sect. 4 of \citet{Salvato2022AA...661A...3S}. 
A study of the trade-off between purity and completeness  shows that counterparts with a threshold of $\textrm{CTP}_\textrm{quality} \geq 2$ ($\textrm{p}\_\textrm{any}>0.035$) have a purity and completeness both equal to 95\%. 
A total of 2,160 sources are classified as stars either via astrometry, spectroscopy, X-ray, and optical/IR colors or via a dedicated analysis as described in \citet{Schneider2022AA...661A...6S} and were removed from the rest of the study.
\noindent

As shown by simulations \citep{Liu_SIM_2022AA...661A..27L, Seppi2022AA...665A..78S}, faint clusters are contaminants of the point source catalog. In eFEDS, 129 clusters are present in the point source catalog \citep{Bulbul2021arXiv211009544B}. They are identified in \citet{Salvato2022AA...661A...3S} with the flag $\textrm{CLUSTER}\_\textrm{CLASS}\geq 3$ and are masked here.

After these cuts from the eFEDS point source catalog, we are left with 17,902 AGN candidates over 140 deg$^2$ (density of 127.9 per square degree).
Figure \ref{fig:wedge} illustrates the light cone considered in this analysis.

\subsubsection{Masks}

We had to propagate the masks applied to the source catalog to the random catalog to estimate clustering. 
The random catalog is a set of un-clustered data points that cover the same sky area as the observations (see the description in Sect. \ref{subsubsec:random:cat}). 
As the masking radius for each detected source, we used its radius of maximum signal-to-noise augmented by 40 percent. 
This radius was determined while extracting the X-ray spectrum of each source \citep{Liu_SPEC2022AA...661A...5L}; the details of this process are provided in Appendix \ref{appendix:mask:xray}.

The edges of the survey have a lower exposure time. 
We find that trimming the survey edges by requiring a minimum exposure time of 830 seconds minimizes the Kolmogorov-Smirnov test values (between random and data vectors) with a minimal area loss (see Sect. \ref{subsubsec:random:cat}).
After applying the minimum exposure time cut, we are left with 16,308 AGN candidates over 128 deg$^2$, resulting in a density of $\sim$127.4 deg$^{-2}$.

\subsubsection{Photometric redshifts}
Photometric redshift estimation for galaxies that host AGNs is complex \citep[e.g.,][]{Salvato18_photoZ_review}. 
In the eROSITA/eFEDS case, \citet{Salvato2022AA...661A...3S} measured photometric redshifts to have $\sigma_{NMAD}=1.48\times \textrm{median}\left(\frac{|z_{spec}-z_{phot}|}{1+z_{spec}}\right)\sim0.05$ and a fraction of outliers, with $\frac{|z_{spec}-z_{phot}|}{1+z_{spec}}>0.15$, on the order of 20\%. At the bright end (r<21.5), we find that $\sigma_{NMAD}$ decreases to $\sim0.03$, while the outlier fraction remains the same, 20\%. 

With the help of the simulation from \citet{Comparat19}, we find that the clustering measured using a photometric redshift with such a dispersion and fraction of outliers would result in losing between one-third and one-half of the amplitude of the clustering signal. 
So we did not use the photometric redshift to measure clustering statistics; instead, we focused on the subsample of 10,680 AGNs with spectroscopic observations (see Sect. \ref{subsubsec:specz:catalogue}).  

\subsubsection{Spectroscopic redshifts}
\label{subsubsec:specz:catalogue}

\begin{figure}
    \centering
\includegraphics[width=.95\columnwidth]{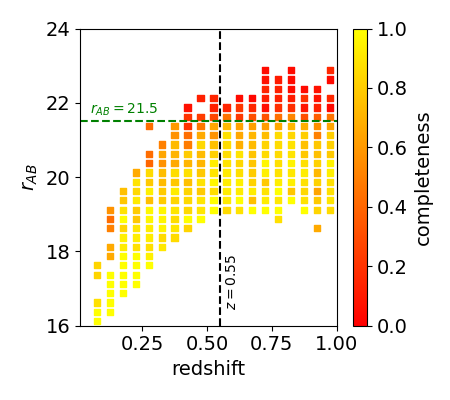} 

\includegraphics[width=.95\columnwidth]{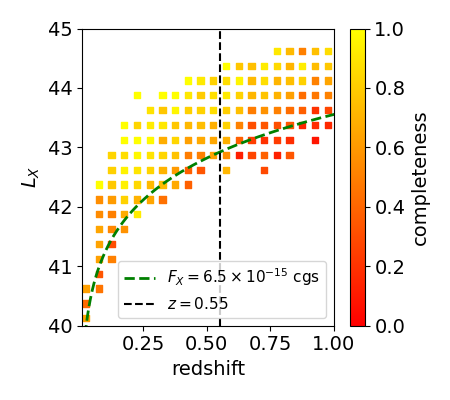}
    \caption{Spectroscopic completeness as a function of r-band magnitude vs. redshift (top) and soft X-ray luminosity (0.5-2 keV) vs. redshift (bottom). The completeness coverage is homogeneous below a redshift of 0.6. At higher redshifts, completeness at the faint end impacts the sample significantly.}
    \label{fig:mag:lx:selection}
\end{figure}

The eFEDS field was observed with the SDSS infrastructure \citep{Gunn2006,Smee13} in March-April 2020 with both BOSS (Baryon Oscillation Spectroscopic Survey) spectrographs (1000 fibers per plate; SDSS-IV; \citealt{blanton17}) and in March-April 2021 with a single BOSS spectrograph (500 fibers per plate; SDSS-V; \citealt{Kollmeier17}, Merloni et al. in preparation).
A total of 31 plates were observed (see the section entitled ``SPIDERS'' of  \citealt[][]{SDSS_DR17_2022ApJS..259...35A}), and the spectra are part of the SDSS Data Release (DR)18 \citep{SDSSDR18_2023arXiv230107688A}.
The total area covered by SDSS-IV and V spectroscopic observations is 133.77 deg$^2$ (95\% of the eFEDS area). 
The obtained spectroscopic redshift completeness depends on \textit{(i)} the position in the sky and \textit{(ii)} the optical magnitude of the source.
We considered the z-band AB magnitude measured as in the legacy survey DR8 \citep{DECALS_2018arXiv180408657D} and based on observations made with DECam \citep{Flaugher2015AJ....150..150F}.
Although photometric redshifts are not accurate enough for clustering studies, they are of sufficient quality to compare the distribution of magnitudes and fluxes in broad redshift bins. 
Overall, we find that at a z-band magnitude of 21.25 (19.0), the completeness is 50\% (90\%).  
We find that, up to redshift $\sim$0.55, the spectroscopic sample is a fair subsample (as a function of optical magnitude and X-ray flux) of the entire population. 
Since SDSS-V observations are limited to z-band magnitudes brighter than 21.5, beyond a redshift of 0.55 we are missing a significant fraction of the AGNs that are optically faint X-ray-selected AGNs (see Fig. \ref{fig:mag:lx:selection}). 

We estimated the spectroscopic completeness in $\sim$3.5 deg$^{2}$ equal area pixels (half the size of an SDSS plate, $\sim$7 deg$^{2}$). 
The minimum (maximum) completeness measured in a pixel is 13\% (69\%).
The relative variations in the spectroscopic redshift distribution as a function of completeness are within the expected fluctuations for pixels with completeness levels above 40\%.
So, we discarded areas with completeness lower than 40\%.
This removed about 20 deg$^2$ of the area located at the edge of the eFEDS field (most of it overlaps with the low-exposure regions).

\subsubsection{AGN sample}
In summary, to measure the clustering, we created an AGN sample that covered redshifts between 0.05 and 0.55, where the spectroscopic sample is not biased compared to the parent photo-z sample.
We obtained a sample of 1,992 AGNs with spectroscopic redshift covering 122.3 deg$^2$. 
Figure \ref{fig:wedge} illustrates how the AGNs considered in this analysis sample the large-scale structure observed with galaxies and groups from the GAMA (Galaxy and Mass Assembly) survey and eROSITA eFEDS clusters.
Table \ref{tab:properties:sample} summarizes the main properties of the considered sample. 
The mean redshift of the sample is 0.34, with a standard deviation of 0.13.
The sample's mean X-ray luminosity in the soft 0.5--2 keV band is 42.91. The distribution around the mean is broad and has a standard deviation of 0.65. 

Among the 1992 AGNs studied here, 1648 (82.7\%) have their spectroscopic redshifts measurements from SDSS observations, and 270 (13.5\%) from GAMA observations \citep{Liske2015MNRAS.452.2087L}. 
The other spectroscopic redshifts originate from:  
 WiggleZ Dark Energy Survey \citep[46;][]{Drinkwater2018MNRAS.474.4151D},
 LAMOST (Large sky Area Multi-Object Fiber Spectroscopic Telescope) DR5 v3 \citep[8;][]{Luo2015RAA....15.1095L},
2SLAQ (2dF-SDSS LRG and QSO Survey) \citep[7;][]{Cannon2006MNRAS.372..425C}, 2MASS \citep[5;][]{Skrutskie2006AJ....131.1163S},
HSC SSP (Hyper Suprime-Cam Subaru Strategic Program) \citep[3;][]{Oguri2018PASJ...70S..20O},
 6dFGS (Six-degree Field Galaxy Survey) \citep[2;][]{Jones2009MNRAS.399..683J},
HYPERLEDA (HYPER Lyon-Meudon Extragalactic Database) \citep[1;][]{Paturel2003AA...412...45P},
 \citet[1]{VeronVeron2010}, 
and RCSEDv2 (Reference Catalog of Spectral Energy Distributions) \citep[1;][]{Chilingarian2021arXiv211204866C}.

We verified that the edges of the selection do not impact the clustering and lensing summary statistics: by moving the redshift cut from 0.05 to 0.1 and from 0.55 to 0.5 and by adding a minimum luminosity threshold of 41.5. 
Given the negligible impact of the verification on the summary statistics, we did not cut the sample further.

Additional splitting the sample in soft X-ray luminosity-limited samples (or following a visual inspection of the optical spectra) significantly decreases the signal-to-noise in the measurements, and HOD model parameters become unconstrained. Larger numbers of AGNs with spectroscopic redshifts are required to investigate the trends with parameters that define the sample.

\begin{table}
    \centering
    \caption{Properties of the sample. The minimum, mean, maximum, and standard deviation of the redshift, the soft-band X-ray luminosity, and the g, r, and z magnitudes are from the legacy survey \citep{DECALS_2018arXiv180408657D}.}
    \begin{tabular}{c|c c c c}
      \hline \hline
      property   & min   & mean  & max   & std  \\ 
      \hline
redshift & 0.05 & 0.34 & 0.55 & 0.13 \\
soft L$_X$ & 40.49 & 42.91 & 45.12 & 0.65 \\
g$_{AB}$ & 15.16 & 20.1 & 23.13 & 1.41 \\
r$_{AB}$ & 14.26 & 19.23 & 22.39 & 1.35 \\
z$_{AB}$ & 13.62 & 18.57 & 21.82 & 1.32 \\

      \hline
    \end{tabular}
    \label{tab:properties:sample}
\end{table}

\subsubsection{Random catalog}
\label{subsubsec:random:cat} 

 To measure clustering, one compares the set of observed points to a group of points with no clustering but all other aspects equal (window function, redshift distribution). In this section, we explain how the set of random points is constructed. 

We drew a set of random points with a large uniform density of $\sim$81,000 deg$^{-2}$ on the sky (about 11.5 million points on eFEDS). 
We first trimmed it to precisely follow the edges of the survey. 
We then followed the methodology of \citet{Georgakakis08} to down-sample the uniform random catalog with the sensitivity map (see details in Appendix \ref{appendix:random:cat}). 
Heuristically, this step applies the X-ray flux limit and its variations across the field to the set of random points. 
The total number of random points remaining after down-sampling, masking (see the previous section), and trimming (low exposure time region) is 3,713,726. The density of random points, $\sim30,000$ deg$^{-2}$, is more than 200 times larger than that of the data points (127.4 deg$^{-2}$), which is largely sufficient.

We down-sampled the random catalog to follow the spectroscopic redshift completeness map and its dependency on right ascension and declination. We cut areas where spectroscopic completeness is lower than 40 \%. 
As the relative variation in the redshift distribution is independent of the completeness (see the previous section), we shuffled the set of observed redshifts and assigned them to the random points, regardless of the completeness level. 

\subsection{HSC-SSP weak-lensing data}
\label{sec:data:hsc}

We used the HSC S19A weak-lensing products based on the HSC-SSP accumulated $i$-band imaging data from 2014 to 2019. The original HSC-SSP S19A wide-layer data cover about 512 deg$^2$, and it reduces to 433.48deg$^2$ after the full-color full-depth selection. With the $i$-band magnitude cut of 24.5, the observed number density reached up to 22.9 arcmin$^{-2}$. 
The deep imaging data enable a comprehensive redshift coverage ranging from 0 to 3, and the calibrated bias residual shows no dependence on the redshift. We refer the readers to the year 3 shape catalog paper by \cite{Li2022PASJ...74..421L} for more details.

We briefly describe the photometric redshift accuracy, the year 3 shape catalog, and the measurements of the shear estimator in the following subsections.

\subsubsection{Photometry and photometric redshifts}
\label{sec:hsc:photometry}

Good photometry is a precondition for the measurement of photometric redshifts. 
The performance of the HSC photometry was tested in two ways. The first was an internal test that compared the point spread function (PSF) magnitude to the Kron magnitude for a bright star sample ($i$<21.5) in the Wide XMM-LSS (X-ray Multi-Mirror satellite Large-Scale Structure Survey) field.
The standard deviation of the difference (PsfMag-KronMag) achieved is better than 1\%, independently of the filters and fields.
In addition, the difference between the CModel magnitude and PSF magnitude is below 0.2\%. 
For the external test, Pan-STARRS-1 (PS1) stars brighter than $r$-band 20 mag are used (\citealt{Chambers2016arXiv161205560C}).
The scatter level is also at about the 1\% level, indicating good photometric performance. 

The photometric redshifts used in this work are \texttt{dNNz} (A. J. Nishizawa et al. in preparation). 
The accuracy of the photometric redshifts is quantified by a set of metrics, namely the bias defined as $\Delta z=(z_{phot}-z_{ref})/(1+z_{ref})$, the dispersion $\sigma_{zphot}=1.48\times MAD(\delta z)$ (MAD is the median absolute deviation), the outlier rate $f_{outlier}=N(\Delta z)>0.15/N_{total}$, and the loss function $L(\Delta z)=1-1/(1+(\Delta z/\gamma)^2)$ with $\gamma=0.15$. 
The photometric and public spectroscopic surveys that overlap with HSC-SSP provide a wealth of data for photometric redshift calibration \citep[e.g.,][]{Lilly2009ApJS..184..218L, Bradshaw2013MNRAS.433..194B, Mclure2013ASSP...37..323M, Skelton2014ApJS..214...24S, Silverman2015ApJS..220...12S, Momcheva2016ApJS..225...27M, Driver2022MNRAS.513..439D}. 
There are about 170 thousand spectroscopic redshifts (spec-$z$) and 37 thousand high-quality g/prism-$z$ that we used as $z_{ref}$.
The \texttt{dNNz} method achieves an accuracy of $\Delta z=10^{-4}$ bias, a dispersion $\sigma_{zphot}=3\%$, and an outlier rate smaller than $f_{outlier}\leq 10\%$.

\subsubsection{Shape catalog}
\label{sec:hsc:shape}
The HSC galaxy sample was selected following a series of basic flag cuts, such as \texttt{i\_detect\_isprimary}, \texttt{i\_extendedness\_value}, and \texttt{i\_sdsscentroid\_flag}. The detailed descriptions are listed in Table 2 of \cite{Li2022PASJ...74..421L}. 
The shapes of galaxies were measured via a re-Gaussianization method \citep{Hirata2003MNRAS.343..459H} (\texttt{reGauss}, which has been merged with \texttt{GalSim}, \citealt{Rowe2015A&C....10..121R}), and the PSF effects were corrected during the measurement process. 
HSC covers six discrete fields, named after overlapping regions from previous surveys: XMM, HECTOMAP, WIDE12H, GAMA09H, GAMA15H, and VVDS. The eFEDS region overlaps with GAMA09H. We note that the HSC region GAMA09H covers a larger area than the original GAMA09H field and completely encompasses the eFEDS field.

The final shape catalog contains the two components of the ellipticity:
\begin{equation}
    \rm (e_1,e_2) =\frac{1-(b/a)^2}{1+(b/a)^2}(cos2\phi,sin2\phi),
\end{equation}
where $b/a$ is the ratio between the minor axis and major axis, and $\phi$ is the position angle of the major axis with respect to the sky coordinates. The shear distortion, $\gamma_i$, is then related to the $e_i$ (i=1,2) such that
\begin{equation}
    \rm \gamma_i=\frac{1}{2\mathcal{R}}\langle e_i \rangle (i=1,2),
\end{equation}
where $\mathcal{R}$ is the response of the galaxy ellipticity to a small distortion defined in \cite{Kaiser1995ApJ...449..460K} and \cite{Bernstein2002AJ....123..583B}. 
The response was calculated from the calibrated parameters $e_{rms}$ and $\sigma_e$ based on simulations \citep{Mandelbaum2018MNRAS.481.3170M, Li2022PASJ...74..421L} as follows
\begin{equation}
\label{eq:resp}
    \rm \mathcal{R}=1-\frac{\sum_i w_i e_{rms,i}^2}{\sum_i w_i}.
\end{equation}
The weighting term in Eq. \ref{eq:resp},  $\rm w_i$, is composed of the per-component error from the simulation due to the photon noise, $\sigma_{e;i}$, and the rms of galaxy shape distribution, $e_{rms;i}$, $\rm w_i=1/(\sigma^2_{e;i}+e_{rms;i}^2)$.

The \texttt{reGauss} algorithm suffers from several estimation biases, for example model bias, noise bias, and selection bias, which can be classified into multiplicative bias, $\rm m_i$, and additive bias, $\rm c_i$ (i=1,2), such that
\begin{equation}
    \rm \gamma_i = (1+m_i)\gamma_i^{true} + c_i.
\end{equation}
The final shear estimator was obtained with Eq. \ref{eq:estimatorShear}. 
It does not incorporate the geometry factor $\Sigma_{crit}$ described in Sect. \ref{sec:method:gglensing}:
\begin{equation}
\label{eq:estimatorShear}
    \rm \langle \gamma_i \rangle = \frac{\sum_jw_ie_{i;j}}{2\mathcal{R}(1+\langle m_i\rangle)\sum_jw_j}-\frac{\langle c_i\rangle}{1+\langle m\rangle}.
\end{equation}
Both multiplicative and additive biases are calibrated based on the simulations mentioned above.
The two biases were then assigned to each galaxy as a function of S/N and resolution $R_2$. Additionally, there is selection bias as well as weight bias. The overall bias was quantified as the residuals for both multiplicative bias, $\delta m$, and additive bias, $\delta a$, both of which can reach below the 1\% level for the HSC-SSP Y3 shape catalog \citep{Li2022PASJ...74..421L}. 

\section{Summary statistics}
\label{sec:summary:statistics}
Galaxy clustering and gravitational lensing probe the galaxy and matter over-density field's auto- and cross-correlations as a function of scale via a biasing function \citep{Tegmark1998ApJ...500L..79T, Tegmark1999ApJ...518L..69T, Dekel1999ApJ...520...24D}. 
These measurements are well suited to constrain the biasing function, also more generically named the galaxy-halo connection \citep{Sheth1999MNRAS.308..119S, Wechsler2018ARAA..56..435W}. 
With the data described above, we computed two summary statistics: the AGN-AGN autocorrelation (clustering; Sect. \ref{sec:method:clustering}) and galaxy-galaxy lensing with the AGN population as the lenses (Sect. \ref{sec:method:gglensing}).

\subsection{Clustering measurement}
\label{sec:method:clustering}

We used the \citet{Landy1993ApJ...412...64L} estimator to measure the projected two-point correlation function, labeled $w_p(r_p)$ \citep[for a detailed definition, see, e.g.,][]{Davis1983ApJ...267..465D}. 
To count pairs and integrate along the line of sight, we used the \textsc{corrfunc} software \citep{Sinha2020MNRAS.491.3022S}.
For the integration, we used $\pi_{max}=40\; h^{-1}$Mpc. 
We carried out measurements with shorter and longer $\pi_{max}$ and found that with 40, we would obtain the largest signal-to-noise in the clustering measurement.

\begin{figure}
    \centering
\includegraphics[width=0.95\columnwidth]{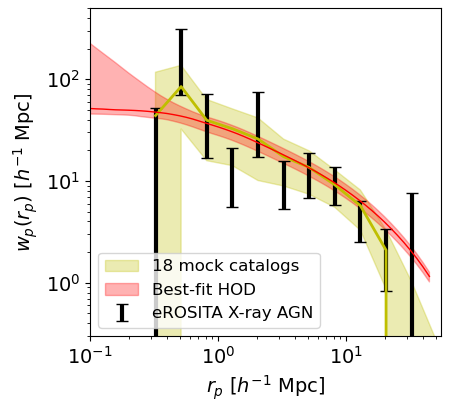}
    \caption{Projected clustering measurement of the X-ray flux-limited eFEDS AGN sample in the redshift range $0.05<z<0.55$ (black). The prediction from the 18 eFEDS simulations appears in yellow. The best-fit model (jointly with weak-lensing observations) is in red.}
    \label{fig:wp}
\end{figure}

We randomly down-sampled the catalog of random points for the clustering measurement to have 20 times the number of AGNs. 
To have consistent 3D positions between the optical spectra and the X-ray sources, we computed the clustering using the position on the sky of the optical counterparts \citep{Salvato2022AA...661A...3S}; we did not use the positions of the X-ray sources. 
The projected correlation function obtained is shown in Fig. \ref{fig:wp} (black error bars). 
The clustering measurement's uncertainty was estimated using the diagonal component of the covariance matrix obtained with 18 eFEDS simulated catalogs \citep{Liu_SIM_2022AA...661A..27L}. 
These simulated eFEDS observations are based on the empirical models of the X-ray cosmic web from \citet{Comparat19,Comparat2020OJAp....3E..13C}. The yellow shaded area in Fig. \ref{fig:wp} shows the prediction from the 18 mocks. We find that the forecast is faithful to the observations. 

Following \citet{Driver2010MNRAS.407.2131D}, we estimated the cosmic variance in this field to be 1\%, which we added as a constant systematic uncertainty at all scales to the clustering measurement. 
We note that it is small compared to the statistical uncertainties. 
Using the eFEDS simulations, we find that clustering summary statistics are significantly biased low for separations $r_p>$40 $h^{-1}$Mpc. 
This is due to the finite volume observed.
So, we excluded from the fitting procedure clustering measurements with a separation larger than 40 $ h^{-1}$Mpc.
The total signal-to-noise in the clustering measurement is 17.7, split into 11 radial bins. We sampled the separation range with five bins per decade evenly log-spaced (0.2 dex steps) between 0.25 ($10^{-0.6}$) and 39.8 ($10^{0.6}$) $h^{-1}$Mpc. 

The fiber collision radius in SDSS is 62 arcseconds ($\sim$0.25$h^{-1}$Mpc at the mean redshift of the sample). 
The eROSITA PSF is 30 arcseconds, so X-ray-selected AGN pairs with a separation smaller than one arcminute are hardly detected. 
Moreover, since the AGN sample considered here is sparse (120 deg$^{-2}\sim$0.03 arc minutes$^{-2}$) and spread over a long line of sight, AGN close pairs are small in number. 
Using the mock catalogs limited to redshifts $0.05<z<0.55$, we estimate the number of expected pairs with an angular separation smaller than 62 arcseconds to typically be a handful (i.e., fewer than ten). 
Only half have physical separations smaller than 40 $h^{-1}$Mpc. 
So, the number of missed pairs due to fiber collisions is negligible.
In our case, we thus consider that fiber collisions are not an issue, and we defined our lowest separation bin at 0.25 $h^{-1}$Mpc. 

\subsection{Galaxy-galaxy lensing measurement}
\label{sec:method:gglensing}

The galaxy-galaxy lensing measurement is a cross-correlation between positions of foreground lenses (AGNs in our case) and shapes of background galaxies acting as sources (HSC galaxies; see the reviews \citealt{Bartelmann2001PhR...340..291B} and \citealt{Refregier2003ARAA..41..645R}). This measurement directly traces the galaxy halo connection \citep[e.g.,][]{Mandelbaum2005MNRAS.362.1451M, Seljak2005PhRvD..71d3511S}. Numerous studies have used galaxy-galaxy lensing (sometimes combined with galaxy clustering) to trace the galaxy-halo connection in general \citep{Leauthaud2011ApJ...738...45L, Coupon2015MNRAS.449.1352C, Zu2015MNRAS.454.1161Z, Dvornik2018MNRAS.479.1240D, Zacharegkas2022MNRAS.509.3119Z}.

We combined the X-ray point sources from the eFEDS region and the HSC shape catalog to compute the galaxy-galaxy lensing using each source galaxy and its probability distribution function as a function of redshift ($p(z)$). 
The physical interpretation of the galaxy-galaxy lensing signal is the difference between the average density inside a certain projected radius, $R$, and the average density at that same radius, so an excess surface density (ESD; $\Delta\Sigma$), that is
\begin{equation}
    \rm \Delta\Sigma (R)= \bar{\Sigma}(\leq R)-\Sigma(R).
\end{equation}
We followed the measurement procedure described in \cite{Miyatake2019ApJ...875...63M} and \cite{Luo2022arXiv220403817L}, that
\begin{equation}
    \rm \Delta \Sigma (R) = \frac{1}{2\mathcal{R} (R)}\frac{\sum_l^{N_l}w_l\sum_s^{N_s}w_{ls}e_{t,ls}[\langle \Sigma_{cr}^{-1}\rangle]^{-1}}{[1+K(R)]\sum_l^{N_l}w_l\sum_s^{N_s}w_{ls}}.
\end{equation}

In the above ESD estimator, $\mathcal{R}(R)$ is the response of the shape estimator, which, for this work, takes a value of 0.84. The
$\rm w_{ls}$ is the weight for each lens-source galaxy pair, and 
$\rm w_l$ is a weight assigned to each lens galaxy. We used $\rm w_l=1$, meaning there are no particular requirements on redshift, stellar mass, or other properties for the lens catalog. The 
$e_{t,ls}$ is the tangential component of the source galaxy shape with respect to the lens. 
The factor $\rm K(R)$ accounts for the multiplicative bias calibrated based on a suite of simulations developed in \cite{Mandelbaum2018MNRAS.481.3170M} and \cite{Li2022PASJ...74..421L}.

Two blinding schemes are provided for systematic sanity tests, the low- and the high-accuracy blinding scheme. We used the former, in which a value of $\delta m$ is added to the original calibrated additive bias, $m$, where $e_i=(1+m_i)e_i+c_i$ (i=1,2). In the low-accuracy scheme, only $\delta m_1$ is added and encrypted for each user of the shape catalog. It is then decrypted and removed by subtracting the $\delta m_1$ term. 

We applied an extra selection function for each lens-source pair following \cite{Medezinski2018PASJ...70...30M} so that the accumulated probability of the P(z) satisfies
\begin{equation}
    \rm P(z_s\geq z_l+0.2)=\int_{z_l+0.2}^{\infty} p(z)dz \geq 0.98.
\end{equation}

\begin{figure}
    \centering
\includegraphics[width=0.95\columnwidth]{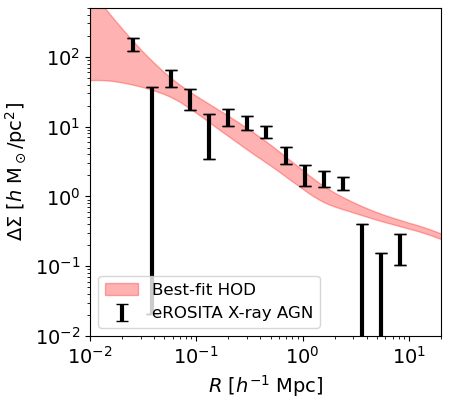}
    \caption{Galaxy-galaxy lensing measurement. The excess surface mass density measurement with HSC S19A data uses as lenses the X-ray flux-limited eFEDS AGN sample in the redshift range $0.05<z<0.55$ (black). The best-fit model (jointly with clustering measurements) is shown in red.}
    \label{fig:esd}
\end{figure}

Figure \ref{fig:esd} shows the obtained galaxy-galaxy lensing measurements (black) as well as the best-fit HOD model (red). 
Measurements with a separation smaller than 20 $h^{-1}$Mpc were included in the fitting procedure.
The total signal-to-noise in the lensing measurement is 46, split into 15 radial bins. 
We sampled the separation range with 5.5 bins per decade in evenly log-spaced (0.18 dex) steps between 0.025 ($10^{-1.6}$) and 8.3 ($10^{0.9}$) $h^{-1}$Mpc. 
The measurement extends to separations that are one-tenth of the size of the clustering measurement.
We measured the same using the public HSC one-year S16A (KIDS DR4) lensing products. 
They are consistent but have a lower total signal-to-noise of 21.4 (6.5) compared to 46.

%
%
%
%
\section{Models}
\label{sec:models}

We interpreted the clustering and lensing summary statistics using models of halo occupation statistics \citep{Cooray2002PhR...372....1C, Guo2010}.
There are three flavors of models: HOD \cite[][]{Berlind2002, Kravtsov2004, Zheng2005, Zheng2007}, SHAM \citep{Conroy2006, Trujillo2011, Klypin2013}, and emulators of the large-scale structure of the Universe \citep[e.g., \textsc{DarkEmulator;}][]{Nishimichi_2019, Nishizawa2020arXiv200301511N}.
The HOD and SHAM models reproduce measurements of the galaxy clustering as a function of luminosity and color \citep{Zehavi2011ApJ...736...59Z, Marulli2013AA...557A..17M}. 
The exploration and parametrization of the assembly bias in such models is still a matter of debate \citep{Contreras2021MNRAS.504.5205C, Xu2021MNRAS.502.3242X}. 
Emulators enable a precise prediction of the cross-correlation between halos and the dark matter.\ However, due to the finite resolution of the simulations they build upon, they are currently limited to predicting statistics at the high mass end only \citep{Nishimichi_2019}. 
So, we first adjusted the HOD model parameters to the measurements obtained in the previous section (see Sect. \ref{sec:models:hodM15}). 
Then, we compared measurements to the prediction of the SHAM model from \citet[see our Sect. \ref{sec:models:SHAM}]{Comparat19}.  

\subsection{Halo occupation distribution model}
\label{sec:models:hodM15}

\begin{table}
        \centering
        \caption{\label{tab:parameters:HOD:FIT} HOD parameters obtained (median of the posterior) by jointly fitting the autocorrelation function and the galaxy-galaxy lensing of the X-ray flux-limited AGN sample. The uncertainties quoted are $1\sigma$ (15.9--84.1 percentiles). Priors are flat in linear space.}
        \begin{tabular}{c c c c c c c c }
                \hline \hline
                parameters           & min & max & $0.05<z<0.55$  \\
                \hline
$M_{{\rm min}}$              & 8.0  & 15.0 & 13.06  $\pm$ 0.44 \\
$\sigma_{\log_{10}M}$  & 0.05 & 1.5  & 1.28   $\pm$ 0.2  \\
$\alpha_{{\rm sat}}$         & 0.1  & 1.5  & 0.73   $\pm$ 0.38 \\
$M_{{\rm sat}}-M_{{\rm min}}$      & -3.0 & 2.45 & 1.46   $\pm$ 0.52 \\
$M^*_{12}$             & -4.0 & 0.1  & -0.96  $\pm$ 0.45 \\
evidence (logZ)        &      &      & -39.88 $\pm$ 0.14 \\

\multicolumn{4}{r}{  deduced parameters  }  \\
\multicolumn{4}{r}{  $b(z=\bar{z})= 0.991 ^{+ 0.078 }_{- 0.096 }$  }  \\
\multicolumn{4}{r}{  $b(z=0.1)= 0.915 ^{+ 0.065 }_{- 0.08 }$       }  \\
\multicolumn{4}{r}{  $f_{{\rm sat}}<20.6\%$       }  \\

\hline
\multicolumn{4}{c}{4-parameter fit, $\sigma_{\log_{10}M}=1.3$} \\
\hline
$M_{{\rm min}}$              & & &  13.09 $\pm$ 0.19 \\
$\alpha_{{\rm sat}}$         & & &  0.75 $\pm$ 0.39 \\
$M_{{\rm sat}}-M_{{\rm min}}$      & & &  1.56 $\pm$ 0.46 \\
$M^*_{12}$             & & &  -0.97 $\pm$ 0.46 \\
evidence (logZ)        & & & -38.6 $\pm$ 0.12 \\

\multicolumn{4}{r}{  deduced parameters  }  \\
\multicolumn{4}{r}{  $b(z=\bar{z})= 1.001 ^{+ 0.075 }_{- 0.094 }$  }  \\
\multicolumn{4}{r}{  $b(z=0.1)= 0.918 ^{+ 0.065 }_{- 0.076 }$      }  \\
\multicolumn{4}{r}{  $f_{{\rm sat}}<16.8\%$       }  \\
\hline
\multicolumn{4}{c}{4-parameter fit, $\sigma_{\log_{10}M}=1.0$} \\
\hline
$M_{{\rm min}}$              & & &  12.47 $\pm$ 0.26  \\
$\alpha_{{\rm sat}}$         & & &  0.75 $\pm$ 0.39   \\
$M_{{\rm sat}}-M_{{\rm min}}$      & & &  1.84 $\pm$ 0.53   \\
$M^*_{12}$             & & &  -1.0 $\pm$ 0.5    \\
evidence (logZ)        & & & -40.34 $\pm$ 0.11 \\

\multicolumn{4}{r}{  deduced parameters  }  \\
\multicolumn{4}{r}{  $b(z=\bar{z})= 0.996 ^{+ 0.067 }_{- 0.097 }$  }  \\
\multicolumn{4}{r}{  $b(z=0.1)= 0.919 ^{+ 0.059 }_{- 0.081 }$    }  \\
\multicolumn{4}{r}{  $f_{{\rm sat}}<66.4\%$       }  \\

\hline
        \end{tabular}
\end{table}

As a baseline, we used the HOD model formulated by \citet{More2015ApJ...806....2M}.
It is described by the two equations below, in which
$\langle N_C \rangle$ ($\langle N_S \rangle$) gives the average occupation of a dark matter halo of mass $M$ by a central (satellite) galaxy:\begin{align}
        \langle N_C \rangle & ( M, \vec{\theta} ) = \frac{f_A}{2} \left( 1 + \operatorname{erf}\left( \frac{(log_{10}(M) - M_{{\rm min}})}{\sigma_{\log_{10}M}} \right) \right)
\end{align}
\begin{equation}
        \langle N_S \rangle  ( M, \vec{\theta} ) = \langle N_C \rangle (M, \vec{\theta})  \left( \frac{M - 10^{M_{{\rm sat}}-1}}{ 10^{M_{{\rm sat}}} } \right) ^ {\alpha_{{\rm sat}}}
.\end{equation}
The model has five parameters $\vec{\theta}=( M_{{\rm min}}, \sigma_{\log_{10}M}, \alpha_{{\rm sat}}, M_{{\rm sat}}, M^*_{12} $).
Only a fraction of distinct halos host a central galaxy with an AGN. The $f_A$ parameter, as introduced by \citet[][Eq. 24]{Miyaji2011ApJ...726...83M}, can be interpreted as the duty cycle of halo centers being an AGN. 
In this study, since the correlation function measurements do not depend on the normalization of the occupation distribution, we arbitrarily set $f_A$ to 1.

To avoid sampling un-physical values of $M_{{\rm sat}}$, the parameter passed to the fitting routine is $M_{{\rm sat}}-M_{{\rm min}}$ with the boundaries specified in Table \ref{tab:parameters:HOD:FIT}.
To fit for the $\Delta\Sigma$ measurement at small separations and benefit from the signal present, we needed to add a prediction for a point-like mass term that represents the baryonic lensing mass of the AGN host galaxies. We added the parameter $M^*_{12}$ as follows:
\begin{equation}
    \Delta\Sigma^*(r) = \frac{10^{M^*_{12}+12}}{\pi r^2}.
\end{equation}

The posterior of this parameter represents the mean baryonic lensing mass of the galaxies that host AGNs. 
This mass is related to stellar mass (inferred with stellar population synthesis models) but will also encompass gas in and around the galaxy. 
This baryonic lensing mass can be considered the upper limit of the mean stellar mass of galaxies that host AGNs. 

In total, we fit for five parameters on the two measurements, $\Delta\Sigma$ ($w_p(r_p)$), which have S/N=46 (17.7) in 15 (11) radial bins.
The parameters were sampled with a flat prior (in linear space) within broad boundaries, as specified in Table \ref{tab:parameters:HOD:FIT}.

\subsection{Sub-halo abundance matching models}
\label{sec:models:SHAM}

The \citet{Comparat19, Comparat20a} empirical AGN model statistically links the dark matter halos to the probability of hosting an AGN and its spectral energy distribution. 
By construction, it follows the X-ray luminosity function from \citet{Aird15}. 
Importantly for interpretation, the assignment is done regardless of the environment in which the halos exist. 
The model has two parameters: the fraction of AGNs in satellites sub-halos and the scatter in the abundance matching relation between stellar mass and hard X-ray luminosity.

We show the direct $w_r(r_p)$ prediction from the mock catalogs of \citet{Liu_SIM_2022AA...661A..27L} in Fig. \ref{fig:wp}. 
It is consistent with observations. 
It was obtained with $f_{{\rm sat}}=10\%$ (fraction of AGNs that are satellites) and $\sigma=1$ (scatter in the abundance matching procedure between hard X-ray luminosity and stellar mass). 
These parameters were chosen by hand by \citet{Comparat19, Comparat2020OJAp....3E..13C}. 
At that time (before this study), such parameters resulted in reasonable predictions. 

Creating a complete SHAM-base mock catalog is time-consuming (order of a few CPU hours) and thus impractical for fitting purposes.
Furthermore, with current light cones constructed with replications, predicting the galaxy-galaxy lensing signal is tedious as the dark matter particles are not kept. 
So, instead of predicting summary statistics as measured as a function of the SHAM parameters, we directly predicted the HOD curves as a function of $f_{{\rm sat}}$ and $\sigma$.
Thus, we sampled a small and finite number of ($f_{{\rm sat}}$, $\sigma$) combinations and created individual mock catalogs to predict the HOD curves.

\begin{figure*}
    \centering
\includegraphics[width=1.95\columnwidth]{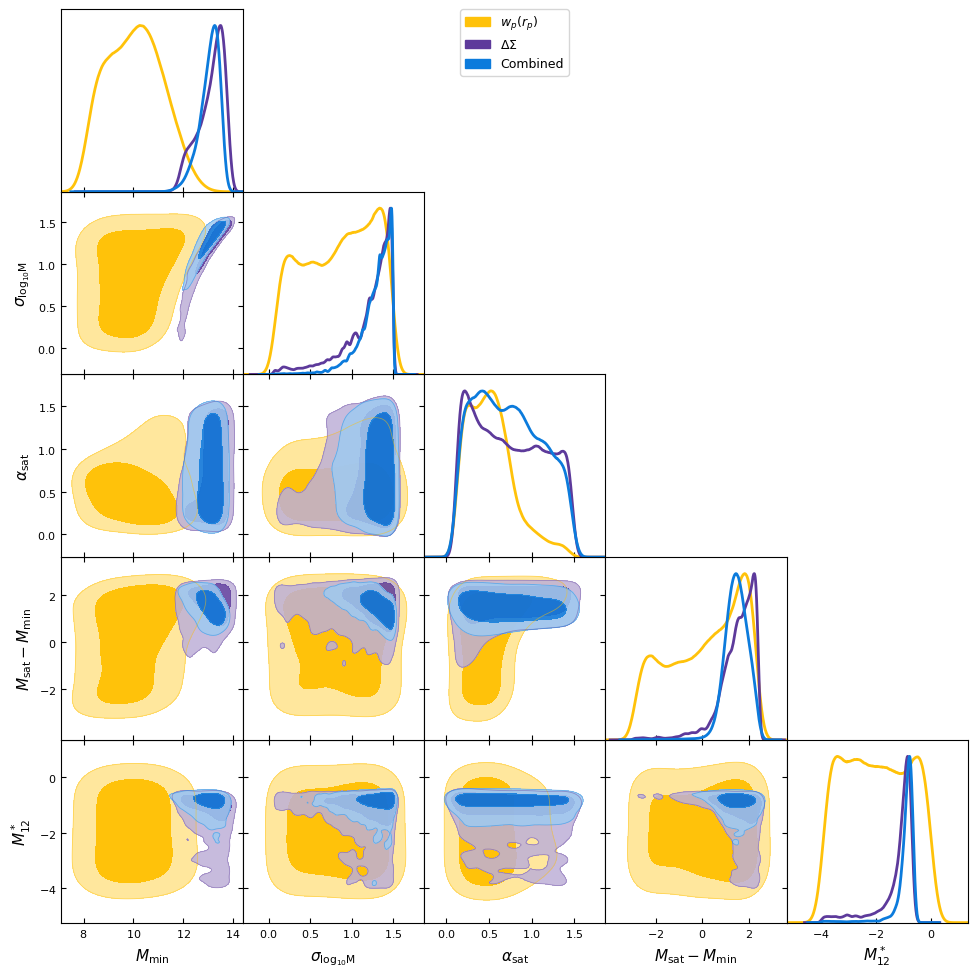}
    \caption{Constraints obtained on the HOD parameters when fitting only the clustering measurement (yellow), only the lensing measurement (purple), or both jointly (blue). Contours show 1 and 2 $\sigma$ constraints. Most of the constraining power comes from galaxy-galaxy lensing.}
    \label{fig:wp:esd:CORNER}
\end{figure*}

%
%
%
%
%
%

\section{Results}
\label{sec:results}

We discuss here the results of the fitting procedure and the comparison between models.
In Sects. \ref{sec:results:HOD} and \ref{sec:results:SHAM} we discuss the results obtained with the HOD and SHAM models, respectively.
For the first time, we measure with relatively small uncertainties the complete HOD of a low redshift flux-limited X-ray-selected sample of AGNs.
We obtain a global view of the distribution of halos that host X-ray AGNs (see Fig. \ref{fig:HOD}).

\subsection{HOD results}
\label{sec:results:HOD}

\begin{figure*}
    \centering
\includegraphics[width=.97\columnwidth]{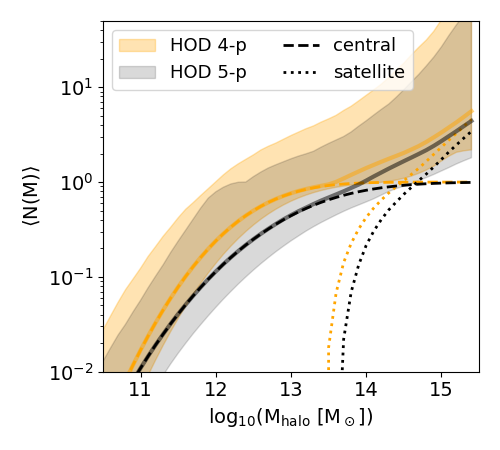}
\includegraphics[width=.97\columnwidth]{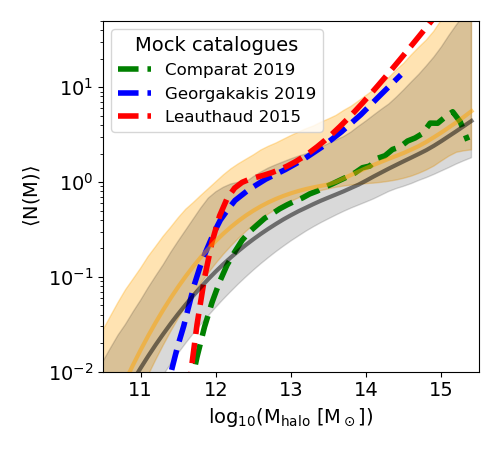}

\caption{Inferred HOD (solid) split into central (dashes) and satellite (dots) for the four- and five-parameter HOD fits (orange and black).
The four-parameter best-fit model is within the $1\sigma$ uncertainty of the five-parameter best-fit model.
Due to the degeneracy between $M_{{\rm min}}$ and $\sigma_{\log_{10}M}$, the four-parameter fit HOD with $\sigma_{\log_{10}M}=1$ is skewed toward lower masses compared to the five-parameter fit.
The direct predictions  from the \citet{Leauthaud15}, \citet{Georgakakis18}, and \citet{Comparat19} mock catalogs are shown in the right panel. They are within the fitted contours obtained.
The mocks from \citet{Leauthaud15} and \citet{Georgakakis18} have a lower $\sigma_{\log_{10}M}$ value (sharper transition) and are thus more in line with the four-parameter fit.
The mock from \citet{Comparat19} has a higher $\sigma_{\log_{10}M}$ value and is comparable to the five-parameter fit. }
    \label{fig:HOD}
\end{figure*}

We fit the parameters of the HOD model with a nested sampling method, \textsc{ultranest} \citep{Buchner2021JOSS....6.3001B}.
The resulting parameters are given in Table \ref{tab:parameters:HOD:FIT}. 
The constraints on the HOD parameters (when fitting each summary statistic individually or both jointly) are shown in Fig. \ref{fig:wp:esd:CORNER}. 
It illustrates the complementary nature of the two measurements.
The comparison between the joint best-fit model and the clustering measurements and lensing measurements are shown in Figs. \ref{fig:wp} and \ref{fig:esd}. 
The models are meaningful, they account for the observations.

The five parameters are meaningful, although not precisely constrained by the joint fit of both summary statistics.
For the central halos, $M_{{\rm min}}$ takes a median posterior value of 13.06$\pm$0.44, and the width of the error function is found at $\sigma_{\log_{10}M}=$ 1.3 $\pm$ 0.2. 
There is a low 1$\sigma$ level tension between the constraints on these parameters obtained by each summary statistic (see Fig. \ref{fig:wp:esd:CORNER}).
Due to the higher signal-to-noise on the lensing statistic, the combined best-fit values are closer to the individual best-fit value on the lensing statistic.
For the satellites, the slope is fit best at $\alpha_{{\rm sat}}=$0.73$\pm$0.38, and the transition occurs in halos 10 to 100 times more massive than the typical halo : $M_{{\rm sat}}-M_{{\rm min}}=$ 1.46 $\pm$ 0.52. Both summary statistics point to these parameter values (Fig. \ref{fig:wp:esd:CORNER}).
The typical baryonic lensing mass of galaxies that host these AGNs is $M^*_{12}=$ -0.96 $\pm$ 0.45. This sets an upper limit to the mean stellar mass of galaxies that host AGNs of $\sim 10^{11}\MSUN$.
The 1$\sigma$ boundaries encompass $3.9\times10^{10}$ and $3.1\times10^{11}\MSUN$, which is in fair agreement with expectations from the AGN host stellar mass function \citep{Bongiorno16, Yang2018MNRAS.475.1887Y}.
This parameter is not degenerate with others and is constrained only by the lensing measurements at small separations.

We noted a degeneracy between $M_{{\rm min}}$ and $\sigma_{\log_{10}M}$, which we investigated.
We decreased the number of parameters by fixing $\sigma_{\log_{10}M}$ to 1.3 (similar to the best-fit value) and 1 (to force a less broad halo distribution and a sharper transition).
We obtained a set of best-fit parameters (see Table \ref{tab:parameters:HOD:FIT}) that are compatible with the five-parameter fit.
When fixing $\sigma_{\log_{10}M}$ to 1.3, we obtained $M_{{\rm min}}13.09\pm0.19$, a value similar to the five-parameter fit but with half the uncertainty. Results for other parameters remain unchanged.
When fixing $\sigma_{\log_{10}M}$ to 1, $M_{{\rm min}}$ is logically forced to lower values, to $\sim 12.5,$ to be able to fit the overall signal.
The HOD posterior is close to that of the five-parameter fit (see the black and orange contours in Fig. \ref{fig:HOD}
 ).

We obtain a global view of the distribution of halos that X-ray AGNs (see Fig. \ref{fig:HOD}).
The four-parameter best-fit model is within the $1\sigma$ uncertainty of the five-parameter best-fit model.
Due to the degeneracy between $M_{{\rm min}}$ and $\sigma_{\log_{10}M}$, the four-parameter fit HOD with $\sigma_{\log_{10}M}=1$ is skewed toward lower masses compared to the five-parameter fit.
With the HOD model, we derive an average halo mass hosting a central (central or satellite) AGN of $3.93 _{- 2.44 }^{+ 2.03 }\times10^{12}\MSUN$ ($4.95 _{- 1.99 }^{+ 2.63 }\times10^{12}\MSUN$). These values are comparable to the findings of \citet{Rodrigo-Torres17}. 
We find the distribution of halo masses to be broad. We thus confirm that quoting a typical halo mass will be extremely sensitive to the definition of what ``typical'' means (see the discussion in \citealt{Leauthaud15}).

The direct HOD predictions from mock catalogs from \citet{Leauthaud15}, \citet{Georgakakis18}, and \citet{Comparat19} are shown in the right panel of Fig. \ref{fig:HOD}. 
They are within the fitted contours obtained.
The mocks from \citet{Leauthaud15} and \citet{Georgakakis18} have a lower $\sigma_{\log_{10}M}$ value (sharper transition) and are thus more in line with the four-parameter fit.
The mock from \citet{Comparat19} has a higher $\sigma_{\log_{10}M}$ value and is comparable to the five-parameter fit (see more discussion in the SHAM section below).
The normalization of $\langle N(M) \rangle$ can be added as a parameter and possibly constrained by jointly fitting the clustering and lensing summary statistics with the stellar mass (or luminosity) function of galaxies that host X-ray AGNs.
However, measuring reliable host-galaxy stellar masses in the case of type 1 AGNs is complex \citep{ciesla15, Zou2022ApJS..262...15Z} and is left for future studies. 

\begin{figure}
    \centering
\includegraphics[width=.95\columnwidth]{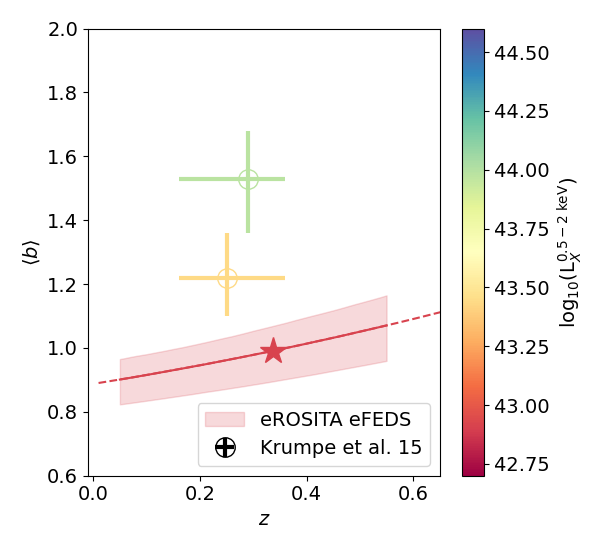}
\includegraphics[width=.95\columnwidth]{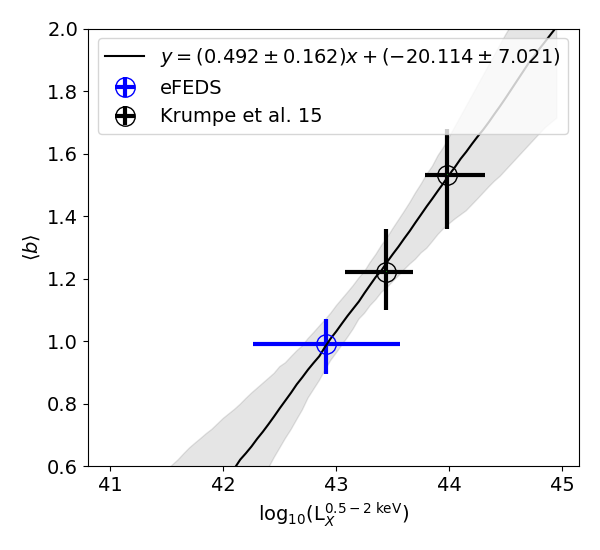}
    \caption{Inferred large-scale halo bias as a function of redshift (top panel) and luminosity (bottom panel) compared to \citet[redshift range 0.16-0.36]{Krumpe15}. 
    We confirm the trend with soft X-ray luminosity and obtain a best fit of $y=(0.482\pm0.143) x + (-19.684\pm6.173)$ between the soft X-ray luminosity and the large-scale halo bias.
    }
    \label{fig:biasComparison}
\end{figure}

The large-scale halo bias inferred is given in Table \ref{tab:parameters:HOD:FIT} and shown in Fig. \ref{fig:biasComparison}. 
At the mean redshift (0.34), it takes a value of $b(\bar{z}=0.34)= 0.99 ^{+ 0.08 }_{- 0.10 }$, which extrapolated to redshift $z=0.1$ becomes $b(z=0.1)= 0.92 ^{+ 0.07 }_{- 0.08 }$.
The deduced large-scale halo bias is the same if we fit the HOD model with four or five parameters (see Table \ref{tab:parameters:HOD:FIT}).
\citet{Krumpe15} measured the bias of X-ray-selected AGNs in a similar redshift range but for intrinsically more luminous AGNs.
With our analysis, we add a new bias measurement at lower soft X-ray luminosity: $8.1\times 10^{42}$ erg s$^{-1}$.
We confirm the weak positive correlation between bias and soft X-ray luminosity found by \citet[see the bottom panel of our Fig. \ref{fig:biasComparison}  ]{Krumpe15}. 
We fit a linear relationship between the quantities and obtain $b=(0.48\pm0.14) L_X + (-19.68\pm6.17)$. The slope value obtained is $3.3\sigma$ (0.48/0.14=3.3) away from 0. 

Other X-ray-selected AGN clustering studies were either at lower redshift \citep{Cappelluti10} or higher redshift \citep{Gilli09,Starikova11,Koutoulidis13,Viitanen19,Allevato19} and always covered higher luminosities. This new study is complementary to them.

\subsection{Halo abundance matching results}
\label{sec:results:SHAM}

\begin{figure*}
\centering
\includegraphics[width=0.95\columnwidth]{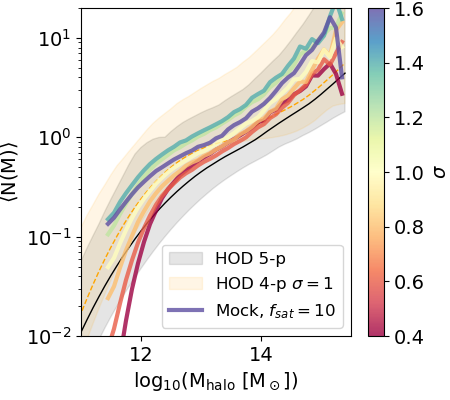}
\includegraphics[width=0.95\columnwidth]{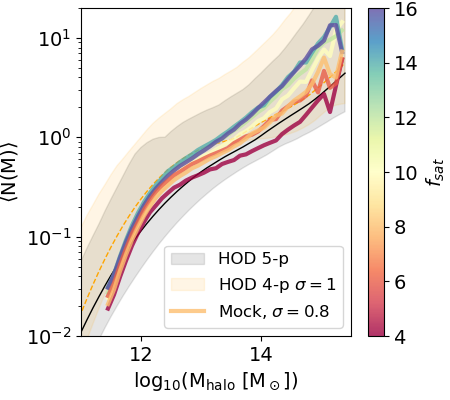}
\includegraphics[width=0.95\columnwidth]{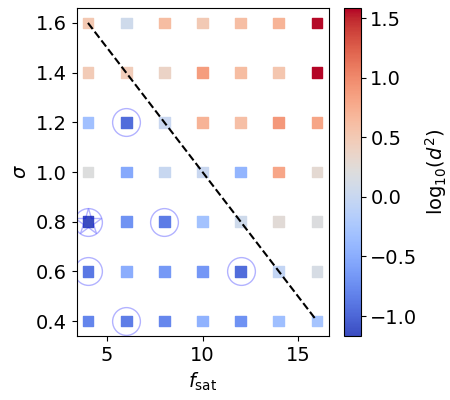}
\includegraphics[width=0.95\columnwidth]{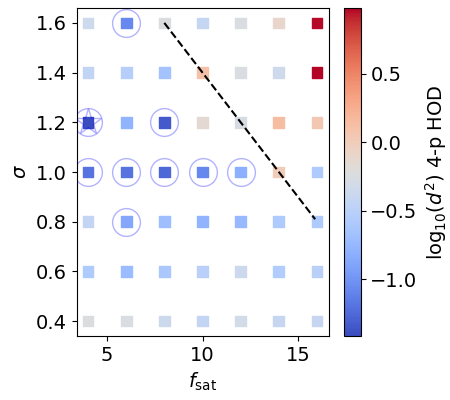}
\caption{SHAM predictions plotted with the fitted HOD model 1 $\sigma$ contour of the four- and five-parameter HOD fit results.
In the top-left panel, the satellite fraction is fixed to 10\%, and the $\sigma$ parameter is varied.
In the top-right panel, $\sigma$ is fixed to 0.8, and the $f_{{\rm sat}}$ is varied.
In the bottom-left (bottom-right) panel, shown as a function of $\sigma$ and $f_{{\rm sat}}$, is the distance between the HOD predicted by SHAM models and the five-parameter (four-parameter) HOD inferred from the observations.
The dashed black line in the bottom-left panel corresponds to $\sigma=2-f_{{\rm sat}}/10$. 
When compared to the five-parameter HOD fit, the bottom-left half, below the $\sigma<2-f_{{\rm sat}}/10$ line of the parameter space, is preferred.
Compared to the four-parameter HOD fit (bottom-right panel), the bottom-left half is also preferred. The dashed line represents $\sigma=2.4-f_{{\rm sat}}/10$.
The star identifies the lowest distance model. Empty circles identify the six lowest distance models.}
    \label{fig:HOD:SHAM}
\end{figure*}

The model has two parameters: the fraction of satellite AGNs and the scatter in the relation between stellar mass and X-ray luminosity. 
The predicted curves extend to halo masses of $10^{11.5}\MSUN$ (and not lower) due to the resolution of the simulation used.
Both parameters impact the shape and amplitude of the clustering signal and the HOD.
Figure \ref{fig:HOD:SHAM} shows the predicted HOD curves for a subset of the parameter space explored.
In the top-left panel, the satellite fraction is fixed to 10\%, and the $\sigma$ parameter varies.
The lower the $\sigma$, the sharper the transition is. 
The $\sigma$ parameter from SHAM is related to the $\sigma_{\log_{10}M}$ parameter from the HOD model.
Mock catalogs with higher $\sigma$ have a distribution of dark matter halos more extended toward lower masses.
In the top-right panel, $\sigma$ is fixed to 0.8, and the $f_{{\rm sat}}$ varies.
The higher the $f_{{\rm sat}}$, the steeper the slope of the satellite occupation curve (the larger the $\alpha$ parameter).
The $f_{{\rm sat}}$ parameter is related to both the $\alpha_{{\rm sat}}$ and the $M_{{\rm sat}}$ HOD parameters.

We computed a distance, denoted $d$, between each predicted HOD curve ($N^{SHAM}(M)$) and the 50th percentile of the inferred HOD model as follows,
\begin{equation}
d = \Sigma^{M=15.5}_{M=11.5} \frac{\left[N^{SHAM}(M, f_{{\rm sat}}, \sigma)-N^{50\%}_{HOD}(M)\right]^2}{\left[N^{84.1\%}_{HOD}(M)-N^{15.9\%}_{HOD}(M)\right]^2}.
\end{equation}
Figure \ref{fig:HOD:SHAM} (bottom panels) shows the distances as a function of $\sigma$ and $f_{{\rm sat}}$.
We find that mock catalogs constructed with parameters satisfying $\sigma<2-f_{{\rm sat}}/10$ predict HODs well within the contours of the five-parameter best-fit HOD inferred from the observations (Fig. \ref{fig:HOD:SHAM} bottom-left panel).
Parameter combinations such as $\sigma>2-f_{{\rm sat}}/10$ are less preferred (top-right corner of the bottom-left panel). 
The $d$ is minimized for $\sigma=0.8$ and $f_{{\rm sat}}=4\%$ (see the star in the figure).
The six smallest distances are indicated with empty circles.
When comparing to the four-parameter best-fit HOD contours (Fig. \ref{fig:HOD:SHAM}, bottom-right panel), parameters within $\sigma<2.4-f_{{\rm sat}}/10$ and $\sigma>0.5$ are acceptable.
Here we find that solutions with low $\sigma$ are less preferred.
In that case, $d$ is minimized for $\sigma=1.2$, and $f_{{\rm sat}}=4$ (see the star in the figure).
The six smallest distances are indicated with empty circles.
In both cases (comparing either to the four-parameter of the five-parameter HOD fit), the best solutions point toward low $f_{{\rm sat}}$ values.

Mocks with both high $\sigma$ and high $f_{{\rm sat}}$ differ significantly from the observations and are ruled out.

\section{Summary and discussion}
\label{sec:discussion}

This article provides a complete picture of how soft X-ray AGNs populate the cosmic web (Fig. \ref{fig:wedge}).
This achievement is possible thanks to two factors: \textit{(i)} the combination of the eROSITA eFEDS X-ray survey with its dedicated SDSS spectroscopic follow-up and with the HSC S19A lensing products, and \textit{(ii)} the complementary nature of the two fitted summary statistics (Figs. \ref{fig:wp} and \ref{fig:esd}). 
We obtain meaningful HOD constraints for an X-ray-selected AGN sample (Fig. \ref{fig:wp:esd:CORNER}, \ref{fig:HOD}).
We interpreted the summary statistics with state-of-the-art HOD and SHAM models (Sect. \ref{sec:results}).
We provide evidence supporting the idea that the mass distribution of halos that host X-ray-selected AGN is broad, as hinted at by previous studies.
Both models point to a shallower satellite slope than for galaxy surveys, meaning that the satellite fraction for X-ray-selected AGN is low, similar to the findings of \citet{Miyaji2011ApJ...726...83M}. 
Interestingly, we find a relatively large $\sigma_{\log_{10}M}$ that is likely related to the width of the specific accretion rate distribution.
Contrasting our results with those of \citet{Krumpe15}, the large-scale halo bias of X-ray-selected AGNs appears to correlate (3.3$\sigma$ significance) with soft-band X-ray luminosity (Fig. \ref{fig:biasComparison}). 
We compared the results with predictions from SHAM models and can rule out a portion of the parameter space (Fig. \ref{fig:HOD:SHAM}).

\subsection{On the $\sigma$ and $\sigma_{\log_{10}M}$ parameters}

The $\sigma$ parameter in the SHAM model is the scatter in the abundance matching relation between the stellar mass of the galaxy hosting the AGN and the AGN hard X-ray luminosity (2--10 keV). 
The probability distribution function of the specific accretion rate resulting from a broad range of stellar masses of galaxies  (that host AGNs) is close to a power law (with slope -1) in the range $31.5<\log_{10}(\lambda_{SAR})<33.5$ (see Fig. 5 of \citealt{Comparat19}). 
The distribution obtained deviates from the power law at high accretion rates. Indeed, the scatter induces an exponential cutoff. 
The distribution at the faint end of the function would require higher-resolution simulations to be populated.  
The HOD results obtained here are compatible with those of SHAM models if $\sigma\in[0.8,1.2]$ and incompatible for low values of $\sigma<0.5$ (for the four-parameter HOD fit) or high values (for the five-parameter HOD fit).  
Indeed, low values of $\sigma$ induce a steeper probability distribution function of the specific accretion rate and are excluded by observations \citep{Georgakakis17}. 
In the opposite regime, large values of $\sigma$ induce a shallow (tending to flat) probability distribution function of the specific accretion rate when considering the entire population. 
In a sense, the $\sigma$ SHAM parameter is related to how broad the distribution of the specific accretion rate and its slope is. 

The $\sigma_{\log_{10}M}$ parameter characterizes how broad the host halo distribution is. It is related to the diversity of host galaxies and their stellar mass via the stellar-to-halo mass relation \citep{Moster13, Behroozi13}. 
The relatively high $\sigma_{\log_{10}M}\sim1.3$ parameter obtained indicates that the host halo mass distribution and, thus, the host stellar mass distribution are both broad. 
This is consistent with studies of the AGN host-galaxy stellar mass function \citep{Bongiorno16}. 
So, it seems that both models point to the same general interpretation: the distribution of host-galaxy stellar mass and that of the specific accretion rate are ``broad,'' which strengthens the validity of direct observations of these distributions \citep{Bongiorno16, Georgakakis17}, even if they might be subject to systematic effects in the measurement of the stellar mass of type 1 AGNs \citep{ciesla15}.

With the innermost lensing measurements, we measure the baryonic lensing mass for this sample ($M^*_{12}$) to be within $4\times10^{10}$ and $3\times10^{11}\MSUN$. From the mock catalog, we predict a broad stellar mass distribution of AGN host galaxies with a median of $4\times10^{10}\MSUN$ and a large standard deviation of 0.6 dex. Therefore, the SHAM method   predicted a stellar mass smaller than the HOD-inferred baryonic lensing mass; we find that interpretations from the two models are consistent.

\subsection{On the satellite occupation}

As suggested in \citet{Leauthaud15}, the combination of clustering and lensing best constrains satellite occupation statistics. 
Compared to previous studies, here we take a significant step forward. 
Satellite fractions inferred from clustering studies are limited by the precision of redshift in the presence of broad-line AGNs, and for example \citet{Shen13} or \citet{Rodrigo-Torres17} could not constrain it. 
Lensing studies were limited by small numbers of X-ray-selected AGNs \citep{Leauthaud15} and showed large uncertainties in the satellite occupation statistics. 
By combining eFEDS with HSC, we find a preference for low satellite fractions (HOD upper limit is $f_{{\rm sat}}<20\%$ and SHAM best fits are with $f_{{\rm sat}}<12\%$). The HOD result shows a preference for a shallow satellite slope ($\sim0.75$) that is smaller than that measured for galaxy samples ($\alpha\sim1-1.1$ and $f_{{\rm sat}}$ of 40\% for galaxies with a stellar mass of 3$\times10^{10}\MSUN$; \citealt{Zehavi2011ApJ...736...59Z, Zu2015MNRAS.454.1161Z}). 

The low satellite fraction could, in part, be due to the soft X-ray selection of the AGN. 
Indeed, satellite AGNs could be obscured and only detectable in hard X-ray or the infrared \citep{Kocevski2015ApJ...814..104K, Krumpe18}. 
\citet{Krumpe18} compared the cross-correlation functions of {\it Swift} BAT AGNs with 2MASS redshift survey galaxies and their HODs. 
Since the {\sl Swift} BAT AGN sample is hard (14-195 keV) X-ray-selected, it contains a larger fraction of type 2 obscured AGNs than {eROSITA}-based samples. 
They find clear suppression of the one-halo term in type 1 AGN cross-correlation functions compared to type 2. The HOD analysis shows $\alpha\sim 1$ for the type 2 AGN HOD, while that of the type 1 AGN was $\alpha \la 0.6$. 
\citet{powell18} obtained similar results. 
A possible scenario causing the low $\alpha$ is the suppression of sub-halo mergers in high velocity encounters in high mass halos \citep{altamirano16,oogi20}. 

At low redshift, previous dynamical analyses of clusters ($\textrm{M}_{\rm halo}>10^{14}\textrm{M}_\odot$) have shown that luminous X-ray-selected AGNs are preferentially in-falling onto clusters and that the cluster environment suppresses their X-ray activity \citep[e.g.,][]{HainesPereiraSanderson_2012ApJ...754...97H, EhlertAllenBrandt_2013MNRAS.428.3509E, EhlertvonderLindenAllen_2014MNRAS.437.1942E}. 
More recent studies of AGN abundance as a function of cluster-centric radius involving larger numbers of clusters confirmed this picture up to a redshift of 1 \citep{KoulouridisRicciGiles_2018A&A...620A..20K, KoulouridisBartalucci_2019A&A...623L..10K, MishraDai_2020AJ....159...69M}.
This activity may in part be due to ram pressure feeding and stripping \citep{PoggiantiJaffeMoretti_2017Natur.548..304P, PoggiantiMorettiGullieuszik_2017ApJ...844...48P,PelusoVulcaniPoggianti_2022ApJ...927..130P} or to close pair interaction of galaxy mergers \citep{KoulouridisPlionisChavushyan_2013A&A...552A.135K,EhlertAllenBrandt_2015MNRAS.446.2709E}.
\citet{PimbbletShabalaHaines_2013MNRAS.429.1827P} reached similar conclusions using optically selected AGNs.
An alternative interpretation of the apparent shallow slope is therefore that the satellite HOD slope is not shallow but rather the satellite distribution profile within the dark matter halo does not follow the mass density (Navarro-Frenk-White) profile assumed in the HOD modeling. 
If the satellite distribution is suppressed toward the outer (inner) part of the halo, the ordinary HOD modeling would result in low (high) fitted $\alpha$. 
Indeed, it would appear as if the satellites were suppressed in high-mass halos with large virial radii.
However, one should be cautious as such interpretations are still a matter of debate. 

\citet{PetterHickoxAlexander_2023arXiv230200690P} also find a low satellite fraction (between 5 and 20\%) at redshift 1.4. But the luminous AGN fraction in clusters was found to evolve (increase) rapidly with redshift \citep{MartiniMillerBrodwin_2013ApJ...768....1M, MishraDai_2020AJ....159...69M}. We still lack a clear picture of how AGNs are satellites of galaxy clusters. Upcoming large samples of X-ray-selected clusters and AGNs with spectroscopic redshifts are key to settling this issue \citep{Merloni2019, Finoguenov2019Msngr.175...39F}.

\subsection{Triggering mechanism for soft X-ray AGNs in the cosmic web}
The general SHAM scheme applied to populate mock catalogs with AGNs satisfactorily accounts for the observations. 
One important assumption made in the SHAM model is that the assignment of an AGN to a galaxy is independent of the environment: it ignores the properties of the neighboring halos. 
It implies that, to the first order, the larger-scale environment, beyond the galaxy host halo, is not the primary driver to turn on the AGN. 
Instead, the local environment (within the virial radius) -- that is, the circumgalactic medium, the interstellar medium, and the stellar populations -- are likely more decisive parameters. This is in agreement with the findings of \citet{Yang2018MNRAS.480.1022Y}, \citet{Allevato19}, and \citet{Siudek2023MNRAS.518..724S}. 
It emphasizes internal processes and their role as AGN triggers, for example, disc instabilities \citep[e.g.,][]{Bournaud2011ApJ...741L..33B} or the presence of bars \citep[e.g.,][]{Ohta2007ApJS..169....1O}. 

The fact that the satellite slope is shallower than that of galaxies with equivalent stellar masses means that the in-fall of a satellite on a larger structure makes it less likely to host an AGN; it is increasingly unlikely the larger the structure is.
This likely illustrates that the gas stripping from satellite galaxies in deep potential wells suppresses AGNs. 
It is compatible with the environment quenching mechanism described by \citet{Peng2010ApJ...721..193P, Peng2012ApJ...757....4P}.

\subsection{Outlook}
The eROSITA eFEDS observations constitute about 1\% of the full eROSITA All-Sky Survey (eRASS). 
This study paves the way toward charting the coevolution of X-ray AGNs
and their host galaxies and dark matter halos. 

In the coming decade, by combining eROSITA with SDSS-V, 4MOST, and DESI (Dark Energy Spectroscopic Instrument) spectroscopic redshifts \citep{Kollmeier17, Merloni2019, DESI16} and with LSST (Vera C. Rubin Observatory 10-year Legacy Survey of Space and Time) and \textit{Euclid} lensing products, one will be able to carry out a similar analysis over a larger area and on an extended redshift range, up to z=1.
Between eFEDS (120 deg$^2$, $z<0.55$) and future analyses (13,000 deg$^2$, $z<1$), the comoving volume will increase by a factor of 450, and HOD parameters should be inferred to the percent level. 
We will accurately measure the HODs as a function of host-galaxy properties and AGN properties with the aim of characterizing possible correlations between HOD parameters and host-galaxy, AGN, and environmental properties. 
With that, one should be able to unravel the role of AGNs in shaping the galaxy population and its hot circumgalactic medium \citep{Hopkins06, Comparat2022arXiv220105169C}. 

Complementary to HOD analysis are the direct or partial correlations with host galaxy properties (see reviews from \citealt{Brandt2021arXiv211101156B, Brandt15}). 
In recent years, spectral energy distribution fitting has become dramatically better at retrieving unbiased galaxy stellar parameters of galaxies that host AGNs (e.g., \citealt{Mountrichas2021AA...646A..29M, Yang2022ApJ...927..192Y,Zou2022ApJS..262...15Z}, Buchner et al. in prep). 
The upcoming Rubin Observatory LSST survey\footnote{\url{https://www.lsst.org}} \citep{Ivezic_LSST_2019ApJ...873..111I} will provide deep multi-band imaging to be used to determine host galaxy properties. 
In addition, the future \textit{Euclid}\footnote{\url{https://www.cosmos.esa.int/web/euclid}} imaging space mission \citep{Laureijs_EUCLID_2011arXiv1110.3193L} will enable accurate morphological measurements of AGN hosts on a significant fraction of the extragalactic sky.
Together they will allow the physics of the connection between AGNs, host galaxy morphology, and stellar properties to be charted \citep[e.g.,][]{Yang2019MNRAS.485.3721Y, Ni2019MNRAS.490.1135N} and provide further insight into the ecology of the cosmic web of X-ray AGNs.

\begin{acknowledgements}
 WL acknowledge the support from the National Key R\&D Program of China (2021YFC2203100), the 111 Project for "Observational and Theoretical Research on Dark Matter and Dark Energy" (B23042), NSFC(NO. 11833005, 12192224) as well as the Fundamental Research Funds for the Central Universities (WK3440000006). MK is supported by the DFG grant KR 3338/4-1. 
T. M. is supported by UNAM-DGAPA PAPIIT 111319, 114423,  and CONACyT Ciencias B\'asica 252531. \\

This work is based on data from eROSITA, the soft X-ray instrument aboard SRG, a joint Russian-German science mission supported by the Russian Space Agency (Roskosmos), in the interests of the Russian Academy of Sciences represented by its Space Research Institute (IKI), and the Deutsches Zentrum für Luft- und Raumfahrt (DLR). The SRG spacecraft was built by Lavochkin Association (NPOL) and its subcontractors, and is operated by NPOL with support from the Max Planck Institute for Extraterrestrial Physics (MPE).\\

The development and construction of the eROSITA X-ray instrument was led by MPE, with contributions from the Dr. Karl Remeis Observatory Bamberg \& ECAP (FAU Erlangen-Nuernberg), the University of Hamburg Observatory, the Leibniz Institute for Astrophysics Potsdam (AIP), and the Institute for Astronomy and Astrophysics of the University of Tübingen, with the support of DLR and the Max Planck Society. The Argelander Institute for Astronomy of the University of Bonn and the Ludwig Maximilians Universität Munich also participated in the science preparation for eROSITA.\\

The eROSITA data shown here were processed using the eSASS/NRTA software system developed by the German eROSITA consortium. \\

The Hyper Suprime-Cam (HSC) collaboration includes the astronomical communities of Japan and Taiwan, and Princeton University. The HSC instrumentation and software were developed by the National Astronomical Observatory of Japan (NAOJ), the Kavli Institute for the Physics and Mathematics of the Universe (Kavli IPMU), the University of Tokyo, the High Energy Accelerator Research Organization (KEK), the Academia Sinica Institute for Astronomy and Astrophysics in Taiwan (ASIAA), and Princeton University. Funding was contributed by the FIRST program from the Japanese Cabinet Office, the Ministry of Education, Culture, Sports, Science and Technology (MEXT), the Japan Society for the Promotion of Science (JSPS), Japan Science and Technology Agency (JST), the Toray Science Foundation, NAOJ, Kavli IPMU, KEK, ASIAA, and Princeton University. \\

This paper makes use of software developed for Vera C. Rubin Observatory. We thank the Rubin Observatory for making their code available as free software at http://pipelines.lsst.io/.\\

This paper is based on data collected at the Subaru Telescope and retrieved from the HSC data archive system, which is operated by the Subaru Telescope and Astronomy Data Center (ADC) at NAOJ. Data analysis was in part carried out with the cooperation of Center for Computational Astrophysics (CfCA), NAOJ. We are honored and grateful for the opportunity of observing the Universe from Maunakea, which has the cultural, historical and natural significance in Hawaii. \\

Funding for the Sloan Digital Sky Survey V has been provided by the Alfred P. Sloan Foundation, the Heising-Simons Foundation, the National Science Foundation, and the Participating Institutions. SDSS acknowledges support and resources from the Center for High-Performance Computing at the University of Utah. The SDSS web site is \url{www.sdss5.org}.\\

SDSS is managed by the Astrophysical Research Consortium for the Participating Institutions of the SDSS Collaboration, including the Carnegie Institution for Science, Chilean National Time Allocation Committee (CNTAC) ratified researchers, the Gotham Participation Group, Harvard University, Heidelberg University, The Johns Hopkins University, L'Ecole polytechnique f\'{e}d\'{e}rale de Lausanne (EPFL), Leibniz-Institut f\"{u}r Astrophysik Potsdam (AIP), Max-Planck-Institut f\"{u}r Astronomie (MPIA Heidelberg), Max-Planck-Institut f\"{u}r Extraterrestrische Physik (MPE), Nanjing University, National Astronomical Observatories of China (NAOC), New Mexico State University, The Ohio State University, Pennsylvania State University, Smithsonian Astrophysical Observatory, Space Telescope Science Institute (STScI), the Stellar Astrophysics Participation Group, Universidad Nacional Aut\'{o}noma de M\'{e}xico, University of Arizona, University of Colorado Boulder, University of Illinois at Urbana-Champaign, University of Toronto, University of Utah, University of Virginia, Yale University, and Yunnan University.\\

Funding for the Sloan Digital Sky Survey IV has been provided by the Alfred P. Sloan Foundation, the U.S. Department of Energy Office of Science, and the Participating Institutions. SDSS acknowledges support and resources from the Center for High-Performance Computing at the University of Utah. The SDSS web site is www.sdss.org.\\

SDSS is managed by the Astrophysical Research Consortium for the Participating Institutions of the SDSS Collaboration including the Brazilian Participation Group, the Carnegie Institution for Science, Carnegie Mellon University, Center for Astrophysics | Harvard \& Smithsonian (CfA), the Chilean Participation Group, the French Participation Group, Instituto de Astrofísica de Canarias, The Johns Hopkins University, Kavli Institute for the Physics and Mathematics of the Universe (IPMU) / University of Tokyo, the Korean Participation Group, Lawrence Berkeley National Laboratory, Leibniz Institut für Astrophysik Potsdam (AIP), Max-Planck-Institut für Astronomie (MPIA Heidelberg), Max-Planck-Institut für Astrophysik (MPA Garching), Max-Planck-Institut für Extraterrestrische Physik (MPE), National Astronomical Observatories of China, New Mexico State University, New York University, University of Notre Dame, Observatório Nacional / MCTI, The Ohio State University, Pennsylvania State University, Shanghai Astronomical Observatory, United Kingdom Participation Group, Universidad Nacional Autónoma de México, University of Arizona, University of Colorado Boulder, University of Oxford, University of Portsmouth, University of Utah, University of Virginia, University of Washington, University of Wisconsin, Vanderbilt University, and Yale University.

\end{acknowledgements}

\bibliographystyle{aa}
\bibliography{references}

\begin{thebibliography}{166}
\expandafter\ifx\csname natexlab\endcsname\relax\def\natexlab#1{#1}\fi

\bibitem[{{Abdurro'uf} {et~al.}(2022){Abdurro'uf}, {Accetta}, {Aerts}, {Silva
  Aguirre}, {Ahumada}, {Ajgaonkar}, {Filiz Ak}, {Alam}, {Allende Prieto},
  {Almeida}, {Anders}, {Anderson}, {Andrews}, {Anguiano}, {Aquino-Ort{\'\i}z},
  {Arag{\'o}n-Salamanca}, {Argudo-Fern{\'a}ndez}, {Ata}, {Aubert},
  {Avila-Reese}, {Badenes}, {Barb{\'a}}, {Barger}, {Barrera-Ballesteros},
  {Beaton}, {Beers}, {Belfiore}, {Bender}, {Bernardi}, {Bershady}, {Beutler},
  {Bidin}, {Bird}, {Bizyaev}, {Blanc}, {Blanton}, {Boardman}, {Bolton},
  {Boquien}, {Borissova}, {Bovy}, {Brandt}, {Brown}, {Brownstein}, {Brusa},
  {Buchner}, {Bundy}, {Burchett}, {Bureau}, {Burgasser}, {Cabang}, {Campbell},
  {Cappellari}, {Carlberg}, {Wanderley}, {Carrera}, {Cash}, {Chen}, {Chen},
  {Cherinka}, {Chiappini}, {Choi}, {Chojnowski}, {Chung}, {Clerc}, {Cohen},
  {Comerford}, {Comparat}, {da Costa}, {Covey}, {Crane}, {Cruz-Gonzalez},
  {Culhane}, {Cunha}, {Dai}, {Damke}, {Darling}, {Davidson}, {Davies},
  {Dawson}, {De Lee}, {Diamond-Stanic}, {Cano-D{\'\i}az}, {S{\'a}nchez},
  {Donor}, {Duckworth}, {Dwelly}, {Eisenstein}, {Elsworth}, {Emsellem},
  {Eracleous}, {Escoffier}, {Fan}, {Farr}, {Feng}, {Fern{\'a}ndez-Trincado},
  {Feuillet}, {Filipp}, {Fillingham}, {Frinchaboy}, {Fromenteau}, {Galbany},
  {Garc{\'\i}a}, {Garc{\'\i}a-Hern{\'a}ndez}, {Ge}, {Geisler}, {Gelfand},
  {G{\'e}ron}, {Gibson}, {Goddy}, {Godoy-Rivera}, {Grabowski}, {Green},
  {Greener}, {Grier}, {Griffith}, {Guo}, {Guy}, {Hadjara}, {Harding},
  {Hasselquist}, {Hayes}, {Hearty}, {Hern{\'a}ndez}, {Hill}, {Hogg},
  {Holtzman}, {Horta}, {Hsieh}, {Hsu}, {Hsu}, {Huber}, {Huertas-Company},
  {Hutchinson}, {Hwang}, {Ibarra-Medel}, {Chitham}, {Ilha}, {Imig}, {Jaekle},
  {Jayasinghe}, {Ji}, {Johnson}, {Jones}, {J{\"o}nsson}, {Katkov}, {Khalatyan},
  {Kinemuchi}, {Kisku}, {Knapen}, {Kneib}, {Kollmeier}, {Kong}, {Kounkel},
  {Kreckel}, {Krishnarao}, {Lacerna}, {Lane}, {Langgin}, {Lavender}, {Law},
  {Lazarz}, {Leung}, {Leung}, {Lewis}, {Li}, {Li}, {Lian}, {Liang}, {Lin},
  {Lin}, {Lin}, {Lintott}, {Long}, {Longa-Pe{\~n}a}, {L{\'o}pez-Cob{\'a}},
  {Lu}, {Lundgren}, {Luo}, {Mackereth}, {de la Macorra}, {Mahadevan},
  {Majewski}, {Manchado}, {Mandeville}, {Maraston}, {Margalef-Bentabol},
  {Masseron}, {Masters}, {Mathur}, {McDermid}, {Mckay}, {Merloni},
  {Merrifield}, {Meszaros}, {Miglio}, {Di Mille}, {Minniti}, {Minsley},
  {Monachesi}, {Moon}, {Mosser}, {Mulchaey}, {Muna}, {Mu{\~n}oz}, {Myers},
  {Myers}, {Nadathur}, {Nair}, {Nandra}, {Neumann}, {Newman}, {Nidever},
  {Nikakhtar}, {Nitschelm}, {O'Connell}, {Garma-Oehmichen}, {Luan Souza de
  Oliveira}, {Olney}, {Oravetz}, {Ortigoza-Urdaneta}, {Osorio}, {Otter},
  {Pace}, {Padilla}, {Pan}, {Pan}, {Parikh}, {Parker}, {Peirani}, {Pe{\~n}a
  Ram{\'\i}rez}, {Penny}, {Percival}, {Perez-Fournon}, {Pinsonneault},
  {Poidevin}, {Poovelil}, {Price-Whelan}, {B{\'a}rbara de Andrade Queiroz},
  {Raddick}, {Ray}, {Rembold}, {Riddle}, {Riffel}, {Riffel}, {Rix}, {Robin},
  {Rodr{\'\i}guez-Puebla}, {Roman-Lopes}, {Rom{\'a}n-Z{\'u}{\~n}iga}, {Rose},
  {Ross}, {Rossi}, {Rubin}, {Salvato}, {S{\'a}nchez}, {S{\'a}nchez-Gallego},
  {Sanderson}, {Santana Rojas}, {Sarceno}, {Sarmiento}, {Sayres}, {Sazonova},
  {Schaefer}, {Schiavon}, {Schlegel}, {Schneider}, {Schultheis}, {Schwope},
  {Serenelli}, {Serna}, {Shao}, {Shapiro}, {Sharma}, {Shen}, {Shetrone}, {Shu},
  {Simon}, {Skrutskie}, {Smethurst}, {Smith}, {Sobeck}, {Spoo}, {Sprague},
  {Stark}, {Stassun}, {Steinmetz}, {Stello}, {Stone-Martinez},
  {Storchi-Bergmann}, {Stringfellow}, {Stutz}, {Su}, {Taghizadeh-Popp},
  {Talbot}, {Tayar}, {Telles}, {Teske}, {Thakar}, {Theissen}, {Tkachenko},
  {Thomas}, {Tojeiro}, {Hernandez Toledo}, {Troup}, {Trump}, {Trussler},
  {Turner}, {Tuttle}, {Unda-Sanzana}, {V{\'a}zquez-Mata}, {Valentini},
  {Valenzuela}, {Vargas-Gonz{\'a}lez}, {Vargas-Maga{\~n}a}, {Alfaro},
  {Villanova}, {Vincenzo}, {Wake}, {Warfield}, {Washington}, {Weaver},
  {Weijmans}, {Weinberg}, {Weiss}, {Westfall}, {Wild}, {Wilde}, {Wilson},
  {Wilson}, {Wilson}, {Wolf}, {Wood-Vasey}, {Yan}, {Zamora}, {Zasowski},
  {Zhang}, {Zhao}, {Zheng}, {Zheng}, \& {Zhu}}]{SDSS_DR17_2022ApJS..259...35A}
{Abdurro'uf}, {Accetta}, K., {Aerts}, C., {et~al.} 2022, \apjs, 259, 35

\bibitem[{{Aihara} {et~al.}(2019){Aihara}, {AlSayyad}, {Ando}, {Armstrong},
  {Bosch}, {Egami}, {Furusawa}, {Furusawa}, {Goulding}, {Harikane}, {Hikage},
  {Ho}, {Hsieh}, {Huang}, {Ikeda}, {Imanishi}, {Ito}, {Iwata}, {Jaelani},
  {Kakuma}, {Kawana}, {Kikuta}, {Kobayashi}, {Koike}, {Komiyama}, {Li},
  {Liang}, {Lin}, {Luo}, {Lupton}, {Lust}, {MacArthur}, {Matsuoka}, {Mineo},
  {Miyatake}, {Miyazaki}, {More}, {Murata}, {Namiki}, {Nishizawa}, {Oguri},
  {Okabe}, {Okamoto}, {Okura}, {Ono}, {Onodera}, {Onoue}, {Osato}, {Ouchi},
  {Shibuya}, {Strauss}, {Sugiyama}, {Suto}, {Takada}, {Takagi}, {Takata},
  {Takita}, {Tanaka}, {Terai}, {Toba}, {Uchiyama}, {Utsumi}, {Wang}, {Wang}, \&
  {Yamada}}]{Aihara2019PASJ...71..114A}
{Aihara}, H., {AlSayyad}, Y., {Ando}, M., {et~al.} 2019, \pasj, 71, 114

\bibitem[{{Aird} {et~al.}(2015){Aird}, {Coil}, {Georgakakis}, {Nandra},
  {Barro}, \& {P{\'e}rez-Gonz{\'a}lez}}]{Aird15}
{Aird}, J., {Coil}, A.~L., {Georgakakis}, A., {et~al.} 2015, \mnras, 451, 1892

\bibitem[{{Allevato} {et~al.}(2019){Allevato}, {Viitanen}, {Finoguenov},
  {Civano}, {Suh}, {Shankar}, {Bongiorno}, {Ferrara}, {Gilli}, {Miyaji},
  {Marchesi}, {Cappelluti}, \& {Salvato}}]{Allevato19}
{Allevato}, V., {Viitanen}, A., {Finoguenov}, A., {et~al.} 2019, \aap, 632, A88

\bibitem[{{Almeida} {et~al.}(2023){Almeida}, {Anderson},
  {Argudo-Fern{\'a}ndez}, {Badenes}, {Barger}, {Barrera-Ballesteros}, {Bender},
  {Benitez}, {Besser}, {Bizyaev}, {Blanton}, {Bochanski}, {Bovy}, {Brandt},
  {Brownstein}, {Buchner}, {Bulbul}, {Burchett}, {Cano D{\'\i}az}, {Carlberg},
  {Casey}, {Chandra}, {Cherinka}, {Chiappini}, {Coker}, {Comparat}, {Conroy},
  {Contardo}, {Cortes}, {Covey}, {Crane}, {Cunha}, {Dabbieri}, {Davidson},
  {Davis}, {De Lee}, {M{\'e}ndez Delgado}, {Demasi}, {Di Mille}, {Donor},
  {Dow}, {Dwelly}, {Eracleous}, {Eriksen}, {Fan}, {Farr}, {Frederick}, {Fries},
  {Frinchaboy}, {Gaensicke}, {Ge}, {Gonz{\'a}lez {\'A}vila}, {Grabowski},
  {Grier}, {Guiglion}, {Gupta}, {Hall}, {Hawkins}, {Hayes}, {Hermes},
  {Hern{\'a}ndez-Garc{\'\i}a}, {Hogg}, {Holtzman}, {Ibarra-Medel}, {Ji},
  {Jofre}, {Johnson}, {Jones}, {Kinemuchi}, {Kluge}, {Koekemoer}, {Kollmeier},
  {Kounkel}, {Krishnarao}, {Krumpe}, {Lacerna}, {Jakson Assuncao Lago},
  {Laporte}, {Liu}, {Liu}, {Liu}, {Lopes}, {Macktoobian}, {Malanushenko},
  {Maoz}, {Masseron}, {Masters}, {Matijevic}, {McBride}, {Medan}, {Merloni},
  {Morrison}, {Myers}, {M{\'e}sz{\'a}ros}, {Negrete}, {Nidever}, {Nitschelm},
  {Oravetz}, {Oravetz}, {Pan}, {Peng}, {Pinsonneault}, {Pogge}, {Qiu},
  {Queiroz}, {Ramirez}, {Rix}, {Fern{\'a}ndez Rosso}, {Runnoe}, {Salvato},
  {Sanchez}, {Santana}, {Saydjari}, {Sayres}, {Schlaufman}, {Schneider},
  {Schwope}, {Serna}, {Shen}, {Sobeck}, {Song}, {Souto}, {Spoo}, {Stassun},
  {Steinmetz}, {Straumit}, {Stringfellow}, {S{\'a}nchez-Gallego},
  {Taghizadeh-Popp}, {Tayar}, {Thakar}, {Tissera}, {Tkachenko}, {Hernandez
  Toledo}, {Trakhtenbrot}, {Fernandez Trincado}, {Troup}, {Trump}, {Tuttle},
  {Ulloa}, {Vazquez-Mata}, {Alfaro}, {Villanova}, {Wachter}, {Weijmans},
  {Wheeler}, {Wilson}, {Wojno}, {Wolf}, {Xue}, {Ybarra}, {Zari}, \&
  {Zasowski}}]{SDSSDR18_2023arXiv230107688A}
{Almeida}, A., {Anderson}, S.~F., {Argudo-Fern{\'a}ndez}, M., {et~al.} 2023,
  arXiv e-prints, arXiv:2301.07688

\bibitem[{{Altamirano-D{\'e}vora} {et~al.}(2016){Altamirano-D{\'e}vora},
  {Miyaji}, {Aceves}, {Castro}, {Ca{\~n}as}, \& {Tamayo}}]{altamirano16}
{Altamirano-D{\'e}vora}, L., {Miyaji}, T., {Aceves}, H., {et~al.} 2016, \rmxaa,
  52, 11

\bibitem[{{Bartelmann} \& {Schneider}(2001)}]{Bartelmann2001PhR...340..291B}
{Bartelmann}, M. \& {Schneider}, P. 2001, \physrep, 340, 291

\bibitem[{{Behroozi} {et~al.}(2013){Behroozi}, {Wechsler}, \&
  {Wu}}]{Behroozi13}
{Behroozi}, P., {Wechsler}, R., \& {Wu}, H.-Y. 2013, \apj, 762, 109

\bibitem[{{Berlind} \& {Weinberg}(2002)}]{Berlind2002}
{Berlind}, A.~A. \& {Weinberg}, D.~H. 2002, \apj, 575, 587

\bibitem[{{Bernstein} \& {Jarvis}(2002)}]{Bernstein2002AJ....123..583B}
{Bernstein}, G.~M. \& {Jarvis}, M. 2002, \aj, 123, 583

\bibitem[{{Blanton} {et~al.}(2017){Blanton}, {Bershady}, {Abolfathi},
  {Albareti}, {Allende Prieto}, {Almeida}, {Alonso-Garc{\'\i}a}, {Anders},
  {Anderson}, {Andrews}, \& et~al.}]{blanton17}
{Blanton}, M.~R., {Bershady}, M.~A., {Abolfathi}, B., {et~al.} 2017, \aj, 154,
  28

\bibitem[{{Bongiorno} {et~al.}(2016){Bongiorno}, {Schulze}, {Merloni},
  {Zamorani}, {Ilbert}, {La Franca}, {Peng}, {Piconcelli}, {Mainieri},
  {Silverman}, {Brusa}, {Fiore}, {Salvato}, \& {Scoville}}]{Bongiorno16}
{Bongiorno}, A., {Schulze}, A., {Merloni}, A., {et~al.} 2016, \aap, 588, A78

\bibitem[{{Bournaud} {et~al.}(2011){Bournaud}, {Dekel}, {Teyssier}, {Cacciato},
  {Daddi}, {Juneau}, \& {Shankar}}]{Bournaud2011ApJ...741L..33B}
{Bournaud}, F., {Dekel}, A., {Teyssier}, R., {et~al.} 2011, \apjl, 741, L33

\bibitem[{{Bradshaw} {et~al.}(2013){Bradshaw}, {Almaini}, {Hartley}, {Smith},
  {Conselice}, {Dunlop}, {Simpson}, {Chuter}, {Cirasuolo}, {Foucaud}, {McLure},
  {Mortlock}, \& {Pearce}}]{Bradshaw2013MNRAS.433..194B}
{Bradshaw}, E.~J., {Almaini}, O., {Hartley}, W.~G., {et~al.} 2013, \mnras, 433,
  194

\bibitem[{{Brandt} \& {Alexander}(2015)}]{Brandt15}
{Brandt}, W.~N. \& {Alexander}, D.~M. 2015, Astronomy and Astrophysics Review,
  23, 1

\bibitem[{{Brandt} \& {Yang}(2021)}]{Brandt2021arXiv211101156B}
{Brandt}, W.~N. \& {Yang}, G. 2021, arXiv e-prints, arXiv:2111.01156

\bibitem[{{Brunner} {et~al.}(2022){Brunner}, {Liu}, {Lamer}, {Georgakakis},
  {Merloni}, {Brusa}, {Bulbul}, {Dennerl}, {Friedrich}, {Liu}, {Maitra},
  {Nandra}, {Ramos-Ceja}, {Sanders}, {Stewart}, {Boller}, {Buchner}, {Clerc},
  {Comparat}, {Dwelly}, {Eckert}, {Finoguenov}, {Freyberg}, {Ghirardini},
  {Gueguen}, {Haberl}, {Kreykenbohm}, {Krumpe}, {Osterhage}, {Pacaud},
  {Predehl}, {Reiprich}, {Robrade}, {Salvato}, {Santangelo}, {Schrabback},
  {Schwope}, \& {Wilms}}]{Brunner2021arXiv210614517B}
{Brunner}, H., {Liu}, T., {Lamer}, G., {et~al.} 2022, \aap, 661, A1

\bibitem[{{Buchner}(2021)}]{Buchner2021JOSS....6.3001B}
{Buchner}, J. 2021, The Journal of Open Source Software, 6, 3001

\bibitem[{{Bulbul} {et~al.}(2022){Bulbul}, {Liu}, {Pasini}, {Comparat},
  {Hoang}, {Klein}, {Ghirardini}, {Salvato}, {Merloni}, {Seppi}, {Wolf},
  {Anderson}, {Bahar}, {Brusa}, {Br{\"u}ggen}, {Buchner}, {Dwelly},
  {Ibarra-Medel}, {Ider Chitham}, {Liu}, {Nandra}, {Ramos-Ceja}, {Sanders}, \&
  {Shen}}]{Bulbul2021arXiv211009544B}
{Bulbul}, E., {Liu}, A., {Pasini}, T., {et~al.} 2022, \aap, 661, A10

\bibitem[{{Cannon} {et~al.}(2006){Cannon}, {Drinkwater}, {Edge}, {Eisenstein},
  {Nichol}, {Outram}, {Pimbblet}, {de Propris}, {Roseboom}, {Wake}, {Allen},
  {Bland-Hawthorn}, {Bridges}, {Carson}, {Chiu}, {Colless}, {Couch}, {Croom},
  {Driver}, {Fine}, {Hewett}, {Loveday}, {Ross}, {Sadler}, {Shanks}, {Sharp},
  {Smith}, {Stoughton}, {Weilbacher}, {Brunner}, {Meiksin}, \&
  {Schneider}}]{Cannon2006MNRAS.372..425C}
{Cannon}, R., {Drinkwater}, M., {Edge}, A., {et~al.} 2006, \mnras, 372, 425

\bibitem[{{Cappelluti} {et~al.}(2010){Cappelluti}, {Ajello}, {Burlon},
  {Krumpe}, {Miyaji}, {Bonoli}, \& {Greiner}}]{Cappelluti10}
{Cappelluti}, N., {Ajello}, M., {Burlon}, D., {et~al.} 2010, \apjl, 716, L209

\bibitem[{{Chambers} {et~al.}(2016){Chambers}, {Magnier}, {Metcalfe},
  {Flewelling}, {Huber}, {Waters}, {Denneau}, {Draper}, {Farrow}, {Finkbeiner},
  {Holmberg}, {Koppenhoefer}, {Price}, {Rest}, {Saglia}, {Schlafly}, {Smartt},
  {Sweeney}, {Wainscoat}, {Burgett}, {Chastel}, {Grav}, {Heasley}, {Hodapp},
  {Jedicke}, {Kaiser}, {Kudritzki}, {Luppino}, {Lupton}, {Monet}, {Morgan},
  {Onaka}, {Shiao}, {Stubbs}, {Tonry}, {White}, {Ba{\~n}ados}, {Bell},
  {Bender}, {Bernard}, {Boegner}, {Boffi}, {Botticella}, {Calamida},
  {Casertano}, {Chen}, {Chen}, {Cole}, {Deacon}, {Frenk}, {Fitzsimmons},
  {Gezari}, {Gibbs}, {Goessl}, {Goggia}, {Gourgue}, {Goldman}, {Grant},
  {Grebel}, {Hambly}, {Hasinger}, {Heavens}, {Heckman}, {Henderson}, {Henning},
  {Holman}, {Hopp}, {Ip}, {Isani}, {Jackson}, {Keyes}, {Koekemoer}, {Kotak},
  {Le}, {Liska}, {Long}, {Lucey}, {Liu}, {Martin}, {Masci}, {McLean}, {Mindel},
  {Misra}, {Morganson}, {Murphy}, {Obaika}, {Narayan}, {Nieto-Santisteban},
  {Norberg}, {Peacock}, {Pier}, {Postman}, {Primak}, {Rae}, {Rai}, {Riess},
  {Riffeser}, {Rix}, {R{\"o}ser}, {Russel}, {Rutz}, {Schilbach}, {Schultz},
  {Scolnic}, {Strolger}, {Szalay}, {Seitz}, {Small}, {Smith}, {Soderblom},
  {Taylor}, {Thomson}, {Taylor}, {Thakar}, {Thiel}, {Thilker}, {Unger},
  {Urata}, {Valenti}, {Wagner}, {Walder}, {Walter}, {Watters}, {Werner},
  {Wood-Vasey}, \& {Wyse}}]{Chambers2016arXiv161205560C}
{Chambers}, K.~C., {Magnier}, E.~A., {Metcalfe}, N., {et~al.} 2016, arXiv
  e-prints, arXiv:1612.05560

\bibitem[{{Chilingarian} {et~al.}(2021){Chilingarian}, {Borisov},
  {Goradzhanov}, {Grishin}, {Kasparova}, {Katkov}, {Klochkov}, {Rubtsov}, \&
  {Toptun}}]{Chilingarian2021arXiv211204866C}
{Chilingarian}, I., {Borisov}, S., {Goradzhanov}, V., {et~al.} 2021, arXiv
  e-prints, arXiv:2112.04866

\bibitem[{{Ciesla} {et~al.}(2015){Ciesla}, {Charmandaris}, {Georgakakis},
  {Bernhard}, {Mitchell}, {Buat}, {Elbaz}, {LeFloc'h}, {Lacey}, {Magdis}, \&
  {Xilouris}}]{ciesla15}
{Ciesla}, L., {Charmandaris}, V., {Georgakakis}, A., {et~al.} 2015, \aap, 576,
  A10

\bibitem[{{Comparat} {et~al.}(2020{\natexlab{a}}){Comparat}, {Eckert},
  {Finoguenov}, {Schmidt}, {Sanders}, {Nagai}, {Lau}, {Kafer}, {Pacaud},
  {Clerc}, {Reiprich}, {Bulbul}, {Chitham}, {Chiang}, {Ghirardini},
  {Gonzalez-Perez}, {Gozaliasl}, {Fitzpatrick}, {Klypin}, {Merloni}, {Nandra},
  {Liu}, {Prada}, {Ramos-Ceja}, {Salvato}, {Seppi}, {Tempel}, \&
  {Yepes}}]{Comparat2020OJAp....3E..13C}
{Comparat}, J., {Eckert}, D., {Finoguenov}, A., {et~al.} 2020{\natexlab{a}},
  The Open Journal of Astrophysics, 3, 13

\bibitem[{{Comparat} {et~al.}(2013){Comparat}, {Jullo}, {Kneib}, {Schimd},
  {Shan}, {Erben}, {Ilbert}, {Brownstein}, {Ealet}, {Escoffier}, {Moraes},
  {Mostek}, {Newman}, {Pereira}, {Prada}, {Schlegel}, {Schneider}, \&
  {Brandt}}]{Comparat2013MNRAS.433.1146C}
{Comparat}, J., {Jullo}, E., {Kneib}, J.-P., {et~al.} 2013, \mnras, 433, 1146

\bibitem[{{Comparat} {et~al.}(2020{\natexlab{b}}){Comparat}, {Merloni},
  {Dwelly}, {Salvato}, {Schwope}, {Coffey}, {Wolf}, {Arcodia}, {Liu},
  {Buchner}, {Nandra}, {Georgakakis}, {Clerc}, {Brusa}, {Brownstein},
  {Schneider}, {Pan}, \& {Bizyaev}}]{Comparat20a}
{Comparat}, J., {Merloni}, A., {Dwelly}, T., {et~al.} 2020{\natexlab{b}}, \aap,
  636, A97

\bibitem[{{Comparat} {et~al.}(2019){Comparat}, {Merloni}, {Salvato}, {Nandra},
  {Boller}, {Georgakakis}, {Finoguenov}, {Dwelly}, {Buchner}, {Del Moro},
  {Clerc}, {Wang}, {Zhao}, {Prada}, {Yepes}, {Brusa}, {Krumpe}, \&
  {Liu}}]{Comparat19}
{Comparat}, J., {Merloni}, A., {Salvato}, M., {et~al.} 2019, \mnras, 487, 2005

\bibitem[{{Comparat} {et~al.}(2022){Comparat}, {Truong}, {Merloni},
  {Pillepich}, {Ponti}, {Driver}, {Bellstedt}, {Liske}, {Aird}, {Br{\"u}ggen},
  {Bulbul}, {Davies}, {Villalba}, {Georgakakis}, {Haberl}, {Liu}, {Maitra},
  {Nandra}, {Popesso}, {Predehl}, {Robotham}, {Salvato}, {Thorne}, \&
  {Zhang}}]{Comparat2022arXiv220105169C}
{Comparat}, J., {Truong}, N., {Merloni}, A., {et~al.} 2022, \aap, 666, A156

\bibitem[{{Conroy} {et~al.}(2006){Conroy}, {Wechsler}, \&
  {Kravtsov}}]{Conroy2006}
{Conroy}, C., {Wechsler}, R.~H., \& {Kravtsov}, A.~V. 2006, \apj, 647, 201

\bibitem[{{Contreras} {et~al.}(2021){Contreras}, {Angulo}, \&
  {Zennaro}}]{Contreras2021MNRAS.504.5205C}
{Contreras}, S., {Angulo}, R.~E., \& {Zennaro}, M. 2021, \mnras, 504, 5205

\bibitem[{{Cooray} \& {Sheth}(2002)}]{Cooray2002PhR...372....1C}
{Cooray}, A. \& {Sheth}, R. 2002, \physrep, 372, 1

\bibitem[{{Coupon} {et~al.}(2015){Coupon}, {Arnouts}, {van Waerbeke},
  {Moutard}, {Ilbert}, {van Uitert}, {Erben}, {Garilli}, {Guzzo}, {Heymans},
  {Hildebrandt}, {Hoekstra}, {Kilbinger}, {Kitching}, {Mellier}, {Miller},
  {Scodeggio}, {Bonnett}, {Branchini}, {Davidzon}, {De Lucia}, {Fritz}, {Fu},
  {Hudelot}, {Hudson}, {Kuijken}, {Leauthaud}, {Le F{\`e}vre}, {McCracken},
  {Moscardini}, {Rowe}, {Schrabback}, {Semboloni}, \&
  {Velander}}]{Coupon2015MNRAS.449.1352C}
{Coupon}, J., {Arnouts}, S., {van Waerbeke}, L., {et~al.} 2015, \mnras, 449,
  1352

\bibitem[{{Davis} \& {Peebles}(1983)}]{Davis1983ApJ...267..465D}
{Davis}, M. \& {Peebles}, P.~J.~E. 1983, \apj, 267, 465

\bibitem[{{Dekel} \& {Lahav}(1999)}]{Dekel1999ApJ...520...24D}
{Dekel}, A. \& {Lahav}, O. 1999, \apj, 520, 24

\bibitem[{{DESI Collaboration} {et~al.}(2016){DESI Collaboration}, {Aghamousa},
  {Aguilar}, {Ahlen}, {Alam}, {Allen}, {Allende Prieto}, {Annis}, {Bailey},
  {Balland}, \& et~al.}]{DESI16}
{DESI Collaboration}, {Aghamousa}, A., {Aguilar}, J., {et~al.} 2016, ArXiv
  e-prints [\eprint[arXiv]{1611.00036}]

\bibitem[{{Dey} {et~al.}(2019){Dey}, {Schlegel}, {Lang}, {Blum}, {Burleigh},
  {Fan}, {Findlay}, {Finkbeiner}, {Herrera}, {Juneau}, {Landriau}, {Levi},
  {McGreer}, {Meisner}, {Myers}, {Moustakas}, {Nugent}, {Patej}, {Schlafly},
  {Walker}, {Valdes}, {Weaver}, {Y{\`e}che}, {Zou}, {Zhou}, {Abareshi},
  {Abbott}, {Abolfathi}, {Aguilera}, {Alam}, {Allen}, {Alvarez}, {Annis},
  {Ansarinejad}, {Aubert}, {Beechert}, {Bell}, {BenZvi}, {Beutler}, {Bielby},
  {Bolton}, {Brice{\~n}o}, {Buckley-Geer}, {Butler}, {Calamida}, {Carlberg},
  {Carter}, {Casas}, {Castander}, {Choi}, {Comparat}, {Cukanovaite}, {Delubac},
  {DeVries}, {Dey}, {Dhungana}, {Dickinson}, {Ding}, {Donaldson}, {Duan},
  {Duckworth}, {Eftekharzadeh}, {Eisenstein}, {Etourneau}, {Fagrelius},
  {Farihi}, {Fitzpatrick}, {Font-Ribera}, {Fulmer}, {G{\"a}nsicke},
  {Gaztanaga}, {George}, {Gerdes}, {Gontcho}, {Gorgoni}, {Green}, {Guy},
  {Harmer}, {Hernandez}, {Honscheid}, {Huang}, {James}, {Jannuzi}, {Jiang},
  {Joyce}, {Karcher}, {Karkar}, {Kehoe}, {Kneib}, {Kueter-Young}, {Lan},
  {Lauer}, {Le Guillou}, {Le Van Suu}, {Lee}, {Lesser}, {Perreault Levasseur},
  {Li}, {Mann}, {Marshall}, {Mart{\'\i}nez-V{\'a}zquez}, {Martini}, {du Mas des
  Bourboux}, {McManus}, {Meier}, {M{\'e}nard}, {Metcalfe},
  {Mu{\~n}oz-Guti{\'e}rrez}, {Najita}, {Napier}, {Narayan}, {Newman}, {Nie},
  {Nord}, {Norman}, {Olsen}, {Paat}, {Palanque-Delabrouille}, {Peng},
  {Poppett}, {Poremba}, {Prakash}, {Rabinowitz}, {Raichoor}, {Rezaie},
  {Robertson}, {Roe}, {Ross}, {Ross}, {Rudnick}, {Safonova}, {Saha},
  {S{\'a}nchez}, {Savary}, {Schweiker}, {Scott}, {Seo}, {Shan}, {Silva},
  {Slepian}, {Soto}, {Sprayberry}, {Staten}, {Stillman}, {Stupak}, {Summers},
  {Sien Tie}, {Tirado}, {Vargas-Maga{\~n}a}, {Vivas}, {Wechsler}, {Williams},
  {Yang}, {Yang}, {Yapici}, {Zaritsky}, {Zenteno}, {Zhang}, {Zhang}, {Zhou}, \&
  {Zhou}}]{DECALS_2018arXiv180408657D}
{Dey}, A., {Schlegel}, D.~J., {Lang}, D., {et~al.} 2019, \aj, 157, 168

\bibitem[{{Donoso} {et~al.}(2014){Donoso}, {Yan}, {Stern}, \&
  {Assef}}]{Donoso14}
{Donoso}, E., {Yan}, L., {Stern}, D., \& {Assef}, R.~J. 2014, \apj, 789, 44

\bibitem[{{Drinkwater} {et~al.}(2018){Drinkwater}, {Byrne}, {Blake},
  {Glazebrook}, {Brough}, {Colless}, {Couch}, {Croton}, {Croom}, {Davis},
  {Forster}, {Gilbank}, {Hinton}, {Jelliffe}, {Jurek}, {Li}, {Martin},
  {Pimbblet}, {Poole}, {Pracy}, {Sharp}, {Smillie}, {Spolaor}, {Wisnioski},
  {Woods}, {Wyder}, \& {Yee}}]{Drinkwater2018MNRAS.474.4151D}
{Drinkwater}, M.~J., {Byrne}, Z.~J., {Blake}, C., {et~al.} 2018, \mnras, 474,
  4151

\bibitem[{{Driver} {et~al.}(2022){Driver}, {Bellstedt}, {Robotham}, {Baldry},
  {Davies}, {Liske}, {Obreschkow}, {Taylor}, {Wright}, {Alpaslan}, {Bamford},
  {Bauer}, {Bland-Hawthorn}, {Bilicki}, {Bravo}, {Brough}, {Casura}, {Cluver},
  {Colless}, {Conselice}, {Croom}, {de Jong}, {D'Eugenio}, {De Propris},
  {Dogruel}, {Drinkwater}, {Dvornik}, {Farrow}, {Frenk}, {Giblin}, {Graham},
  {Grootes}, {Gunawardhana}, {Hashemizadeh}, {H{\"a}u{\ss}ler}, {Heymans},
  {Hildebrandt}, {Holwerda}, {Hopkins}, {Jarrett}, {Heath Jones}, {Kelvin},
  {Koushan}, {Kuijken}, {Lara-L{\'o}pez}, {Lange}, {L{\'o}pez-S{\'a}nchez},
  {Loveday}, {Mahajan}, {Meyer}, {Moffett}, {Napolitano}, {Norberg}, {Owers},
  {Radovich}, {Raouf}, {Peacock}, {Phillipps}, {Pimbblet}, {Popescu}, {Said},
  {Sansom}, {Seibert}, {Sutherland}, {Thorne}, {Tuffs}, {Turner}, {van der
  Wel}, {van Kampen}, \& {Wilkins}}]{Driver2022MNRAS.513..439D}
{Driver}, S.~P., {Bellstedt}, S., {Robotham}, A. S.~G., {et~al.} 2022, \mnras,
  513, 439

\bibitem[{{Driver} \& {Robotham}(2010)}]{Driver2010MNRAS.407.2131D}
{Driver}, S.~P. \& {Robotham}, A. S.~G. 2010, \mnras, 407, 2131

\bibitem[{{Dvornik} {et~al.}(2018){Dvornik}, {Hoekstra}, {Kuijken},
  {Schneider}, {Amon}, {Nakajima}, {Viola}, {Choi}, {Erben}, {Farrow},
  {Heymans}, {Hildebrandt}, {Sif{\'o}n}, \&
  {Wang}}]{Dvornik2018MNRAS.479.1240D}
{Dvornik}, A., {Hoekstra}, H., {Kuijken}, K., {et~al.} 2018, \mnras, 479, 1240

\bibitem[{{Eckert} {et~al.}(2021){Eckert}, {Gaspari}, {Gastaldello}, {Le Brun},
  \& {O'Sullivan}}]{Eckert2021Univ....7..142E}
{Eckert}, D., {Gaspari}, M., {Gastaldello}, F., {Le Brun}, A. M.~C., \&
  {O'Sullivan}, E. 2021, Universe, 7, 142

\bibitem[{{Ehlert} {et~al.}(2015){Ehlert}, {Allen}, {Brandt}, {Canning}, {Luo},
  {Mantz}, {Morris}, {von der Linden}, \&
  {Xue}}]{EhlertAllenBrandt_2015MNRAS.446.2709E}
{Ehlert}, S., {Allen}, S.~W., {Brandt}, W.~N., {et~al.} 2015, \mnras, 446, 2709

\bibitem[{{Ehlert} {et~al.}(2013){Ehlert}, {Allen}, {Brandt}, {Xue}, {Luo},
  {von der Linden}, {Mantz}, \&
  {Morris}}]{EhlertAllenBrandt_2013MNRAS.428.3509E}
{Ehlert}, S., {Allen}, S.~W., {Brandt}, W.~N., {et~al.} 2013, \mnras, 428, 3509

\bibitem[{{Ehlert} {et~al.}(2014){Ehlert}, {von der Linden}, {Allen}, {Brandt},
  {Xue}, {Luo}, {Mantz}, {Morris}, {Applegate}, \&
  {Kelly}}]{EhlertvonderLindenAllen_2014MNRAS.437.1942E}
{Ehlert}, S., {von der Linden}, A., {Allen}, S.~W., {et~al.} 2014, \mnras, 437,
  1942

\bibitem[{{Favole} {et~al.}(2016){Favole}, {Comparat}, {Prada}, {Yepes},
  {Jullo}, {Niemiec}, {Kneib}, {Rodr{\'{\i}}guez-Torres}, {Klypin}, {Skibba},
  {McBride}, {Eisenstein}, {Schlegel}, {Nuza}, {Chuang}, {Delubac},
  {Y{\`e}che}, \& {Schneider}}]{Favole16}
{Favole}, G., {Comparat}, J., {Prada}, F., {et~al.} 2016, \mnras, 461, 3421

\bibitem[{{Finoguenov} {et~al.}(2019){Finoguenov}, {Merloni}, {Comparat},
  {Nandra}, {Salvato}, {Tempel}, {Raichoor}, {Richard}, {Kneib}, {Pillepich},
  {Sahl{\'e}n}, {Popesso}, {Norberg}, {McMahon}, \& {4MOST
  Collaboration}}]{Finoguenov2019Msngr.175...39F}
{Finoguenov}, A., {Merloni}, A., {Comparat}, J., {et~al.} 2019, The Messenger,
  175, 39

\bibitem[{{Flaugher} {et~al.}(2015){Flaugher}, {Diehl}, {Honscheid}, {Abbott},
  {Alvarez}, {Angstadt}, {Annis}, {Antonik}, {Ballester}, {Beaufore},
  {Bernstein}, {Bernstein}, {Bigelow}, {Bonati}, {Boprie}, {Brooks},
  {Buckley-Geer}, {Campa}, {Cardiel-Sas}, {Castander}, {Castilla}, {Cease},
  {Cela-Ruiz}, {Chappa}, {Chi}, {Cooper}, {da Costa}, {Dede}, {Derylo},
  {DePoy}, {de Vicente}, {Doel}, {Drlica-Wagner}, {Eiting}, {Elliott}, {Emes},
  {Estrada}, {Fausti Neto}, {Finley}, {Flores}, {Frieman}, {Gerdes},
  {Gladders}, {Gregory}, {Gutierrez}, {Hao}, {Holland}, {Holm}, {Huffman},
  {Jackson}, {James}, {Jonas}, {Karcher}, {Karliner}, {Kent}, {Kessler},
  {Kozlovsky}, {Kron}, {Kubik}, {Kuehn}, {Kuhlmann}, {Kuk}, {Lahav}, {Lathrop},
  {Lee}, {Levi}, {Lewis}, {Li}, {Mandrichenko}, {Marshall}, {Martinez},
  {Merritt}, {Miquel}, {Mu{\~n}oz}, {Neilsen}, {Nichol}, {Nord}, {Ogando},
  {Olsen}, {Palaio}, {Patton}, {Peoples}, {Plazas}, {Rauch}, {Reil}, {Rheault},
  {Roe}, {Rogers}, {Roodman}, {Sanchez}, {Scarpine}, {Schindler}, {Schmidt},
  {Schmitt}, {Schubnell}, {Schultz}, {Schurter}, {Scott}, {Serrano}, {Shaw},
  {Smith}, {Soares-Santos}, {Stefanik}, {Stuermer}, {Suchyta}, {Sypniewski},
  {Tarle}, {Thaler}, {Tighe}, {Tran}, {Tucker}, {Walker}, {Wang}, {Watson},
  {Weaverdyck}, {Wester}, {Woods}, {Yanny}, \& {DES
  Collaboration}}]{Flaugher2015AJ....150..150F}
{Flaugher}, B., {Diehl}, H.~T., {Honscheid}, K., {et~al.} 2015, \aj, 150, 150

\bibitem[{{Georgakakis} {et~al.}(2017){Georgakakis}, {Aird}, {Schulze},
  {Dwelly}, {Salvato}, {Nandra}, {Merloni}, \& {Schneider}}]{Georgakakis17}
{Georgakakis}, A., {Aird}, J., {Schulze}, A., {et~al.} 2017, \mnras, 471, 1976

\bibitem[{{Georgakakis} {et~al.}(2018){Georgakakis}, {Comparat}, {Merloni},
  {Ciesla}, {Aird}, \& {Finoguenov}}]{Georgakakis18}
{Georgakakis}, A., {Comparat}, J., {Merloni}, A., {et~al.} 2018, \mnras, 3272

\bibitem[{{Georgakakis} {et~al.}(2008){Georgakakis}, {Nandra}, {Laird}, {Aird},
  \& {Trichas}}]{Georgakakis08}
{Georgakakis}, A., {Nandra}, K., {Laird}, E.~S., {Aird}, J., \& {Trichas}, M.
  2008, \mnras, 388, 1205

\bibitem[{{Gilli} {et~al.}(2009){Gilli}, {Zamorani}, {Miyaji}, {Silverman},
  {Brusa}, {Mainieri}, {Cappelluti}, {Daddi}, {Porciani}, {Pozzetti}, {Civano},
  {Comastri}, {Finoguenov}, {Fiore}, {Salvato}, {Vignali}, {Hasinger}, {Lilly},
  {Impey}, {Trump}, {Capak}, {McCracken}, {Scoville}, {Taniguchi}, {Carollo},
  {Contini}, {Kneib}, {Le Fevre}, {Renzini}, {Scodeggio}, {Bardelli},
  {Bolzonella}, {Bongiorno}, {Caputi}, {Cimatti}, {Coppa}, {Cucciati}, {de La
  Torre}, {de Ravel}, {Franzetti}, {Garilli}, {Iovino}, {Kampczyk}, {Knobel},
  {Kova{\v{c}}}, {Lamareille}, {Le Borgne}, {Le Brun}, {Maier}, {Mignoli},
  {Pell{\`o}}, {Peng}, {Perez Montero}, {Ricciardelli}, {Tanaka}, {Tasca},
  {Tresse}, {Vergani}, {Zucca}, {Abbas}, {Bottini}, {Cappi}, {Cassata},
  {Fumana}, {Guzzo}, {Leauthaud}, {Maccagni}, {Marinoni}, {Memeo}, {Meneux},
  {Oesch}, {Scaramella}, \& {Walcher}}]{Gilli09}
{Gilli}, R., {Zamorani}, G., {Miyaji}, T., {et~al.} 2009, \aap, 494, 33

\bibitem[{{Gunn} {et~al.}(2006){Gunn}, {Siegmund}, {Mannery}, {Owen}, {Hull},
  {Leger}, {Carey}, {Knapp}, {York}, {Boroski}, {Kent}, {Lupton}, {Rockosi},
  {Evans}, {Waddell}, {Anderson}, {Annis}, {Barentine}, {Bartoszek}, {Bastian},
  {Bracker}, {Brewington}, {Briegel}, {Brinkmann}, {Brown}, {Carr},
  {Czarapata}, {Drennan}, {Dombeck}, {Federwitz}, {Gillespie}, {Gonzales},
  {Hansen}, {Harvanek}, {Hayes}, {Jordan}, {Kinney}, {Klaene}, {Kleinman},
  {Kron}, {Kresinski}, {Lee}, {Limmongkol}, {Lindenmeyer}, {Long}, {Loomis},
  {McGehee}, {Mantsch}, {Neilsen}, {Neswold}, {Newman}, {Nitta}, {Peoples},
  {Pier}, {Prieto}, {Prosapio}, {Rivetta}, {Schneider}, {Snedden}, \&
  {Wang}}]{Gunn2006}
{Gunn}, J.~E., {Siegmund}, W.~A., {Mannery}, E.~J., {et~al.} 2006, \aj, 131,
  2332

\bibitem[{{Guo} {et~al.}(2010){Guo}, {White}, {Li}, \&
  {Boylan-Kolchin}}]{Guo2010}
{Guo}, Q., {White}, S., {Li}, C., \& {Boylan-Kolchin}, M. 2010, \mnras, 404,
  1111

\bibitem[{{Guzzo} {et~al.}(2014){Guzzo}, {Scodeggio}, {Garilli}, {Granett},
  {Fritz}, {Abbas}, {Adami}, {Arnouts}, {Bel}, {Bolzonella}, {Bottini},
  {Branchini}, {Cappi}, {Coupon}, {Cucciati}, {Davidzon}, {De Lucia}, {de la
  Torre}, {Franzetti}, {Fumana}, {Hudelot}, {Ilbert}, {Iovino}, {Krywult}, {Le
  Brun}, {Le F{\`e}vre}, {Maccagni}, {Ma{\l}ek}, {Marulli}, {McCracken},
  {Paioro}, {Peacock}, {Polletta}, {Pollo}, {Schlagenhaufer}, {Tasca},
  {Tojeiro}, {Vergani}, {Zamorani}, {Zanichelli}, {Burden}, {Di Porto},
  {Marchetti}, {Marinoni}, {Mellier}, {Moscardini}, {Nichol}, {Percival},
  {Phleps}, \& {Wolk}}]{Guzzo2014AA...566A.108G}
{Guzzo}, L., {Scodeggio}, M., {Garilli}, B., {et~al.} 2014, \aap, 566, A108

\bibitem[{{Haines} {et~al.}(2012){Haines}, {Pereira}, {Sanderson}, {Smith},
  {Egami}, {Babul}, {Edge}, {Finoguenov}, {Moran}, \&
  {Okabe}}]{HainesPereiraSanderson_2012ApJ...754...97H}
{Haines}, C.~P., {Pereira}, M.~J., {Sanderson}, A.~J.~R., {et~al.} 2012, \apj,
  754, 97

\bibitem[{{Hickox} {et~al.}(2009){Hickox}, {Jones}, {Forman}, {Murray},
  {Kochanek}, {Eisenstein}, {Jannuzi}, {Dey}, {Brown}, {Stern}, {Eisenhardt},
  {Gorjian}, {Brodwin}, {Narayan}, {Cool}, {Kenter}, {Caldwell}, \&
  {Anderson}}]{Hickox2009ApJ...696..891H}
{Hickox}, R.~C., {Jones}, C., {Forman}, W.~R., {et~al.} 2009, \apj, 696, 891

\bibitem[{{Hirata} \& {Seljak}(2003)}]{Hirata2003MNRAS.343..459H}
{Hirata}, C. \& {Seljak}, U. 2003, \mnras, 343, 459

\bibitem[{{Hopkins} {et~al.}(2006){Hopkins}, {Hernquist}, {Cox}, {Di Matteo},
  {Robertson}, \& {Springel}}]{Hopkins06}
{Hopkins}, P.~F., {Hernquist}, L., {Cox}, T.~J., {et~al.} 2006, \apjs, 163, 1

\bibitem[{{Ivezi{\'c}} {et~al.}(2019){Ivezi{\'c}}, {Kahn}, {Tyson}, {Abel},
  {Acosta}, {Allsman}, {Alonso}, {AlSayyad}, {Anderson}, {Andrew}, {Angel},
  {Angeli}, {Ansari}, {Antilogus}, {Araujo}, {Armstrong}, {Arndt}, {Astier},
  {Aubourg}, {Auza}, {Axelrod}, {Bard}, {Barr}, {Barrau}, {Bartlett}, {Bauer},
  {Bauman}, {Baumont}, {Bechtol}, {Bechtol}, {Becker}, {Becla}, {Beldica},
  {Bellavia}, {Bianco}, {Biswas}, {Blanc}, {Blazek}, {Blandford}, {Bloom},
  {Bogart}, {Bond}, {Booth}, {Borgland}, {Borne}, {Bosch}, {Boutigny},
  {Brackett}, {Bradshaw}, {Brandt}, {Brown}, {Bullock}, {Burchat}, {Burke},
  {Cagnoli}, {Calabrese}, {Callahan}, {Callen}, {Carlin}, {Carlson},
  {Chandrasekharan}, {Charles-Emerson}, {Chesley}, {Cheu}, {Chiang}, {Chiang},
  {Chirino}, {Chow}, {Ciardi}, {Claver}, {Cohen-Tanugi}, {Cockrum}, {Coles},
  {Connolly}, {Cook}, {Cooray}, {Covey}, {Cribbs}, {Cui}, {Cutri}, {Daly},
  {Daniel}, {Daruich}, {Daubard}, {Daues}, {Dawson}, {Delgado}, {Dellapenna},
  {de Peyster}, {de Val-Borro}, {Digel}, {Doherty}, {Dubois},
  {Dubois-Felsmann}, {Durech}, {Economou}, {Eifler}, {Eracleous}, {Emmons},
  {Fausti Neto}, {Ferguson}, {Figueroa}, {Fisher-Levine}, {Focke}, {Foss},
  {Frank}, {Freemon}, {Gangler}, {Gawiser}, {Geary}, {Gee}, {Geha}, {Gessner},
  {Gibson}, {Gilmore}, {Glanzman}, {Glick}, {Goldina}, {Goldstein}, {Goodenow},
  {Graham}, {Gressler}, {Gris}, {Guy}, {Guyonnet}, {Haller}, {Harris},
  {Hascall}, {Haupt}, {Hernandez}, {Herrmann}, {Hileman}, {Hoblitt}, {Hodgson},
  {Hogan}, {Howard}, {Huang}, {Huffer}, {Ingraham}, {Innes}, {Jacoby}, {Jain},
  {Jammes}, {Jee}, {Jenness}, {Jernigan}, {Jevremovi{\'c}}, {Johns}, {Johnson},
  {Johnson}, {Jones}, {Juramy-Gilles}, {Juri{\'c}}, {Kalirai}, {Kallivayalil},
  {Kalmbach}, {Kantor}, {Karst}, {Kasliwal}, {Kelly}, {Kessler}, {Kinnison},
  {Kirkby}, {Knox}, {Kotov}, {Krabbendam}, {Krughoff}, {Kub{\'a}nek},
  {Kuczewski}, {Kulkarni}, {Ku}, {Kurita}, {Lage}, {Lambert}, {Lange},
  {Langton}, {Le Guillou}, {Levine}, {Liang}, {Lim}, {Lintott}, {Long},
  {Lopez}, {Lotz}, {Lupton}, {Lust}, {MacArthur}, {Mahabal}, {Mandelbaum},
  {Markiewicz}, {Marsh}, {Marshall}, {Marshall}, {May}, {McKercher}, {McQueen},
  {Meyers}, {Migliore}, {Miller}, {Mills}, {Miraval}, {Moeyens}, {Moolekamp},
  {Monet}, {Moniez}, {Monkewitz}, {Montgomery}, {Morrison}, {Mueller},
  {Muller}, {Mu{\~n}oz Arancibia}, {Neill}, {Newbry}, {Nief}, {Nomerotski},
  {Nordby}, {O'Connor}, {Oliver}, {Olivier}, {Olsen}, {O'Mullane}, {Ortiz},
  {Osier}, {Owen}, {Pain}, {Palecek}, {Parejko}, {Parsons}, {Pease},
  {Peterson}, {Peterson}, {Petravick}, {Libby Petrick}, {Petry},
  {Pierfederici}, {Pietrowicz}, {Pike}, {Pinto}, {Plante}, {Plate}, {Plutchak},
  {Price}, {Prouza}, {Radeka}, {Rajagopal}, {Rasmussen}, {Regnault}, {Reil},
  {Reiss}, {Reuter}, {Ridgway}, {Riot}, {Ritz}, {Robinson}, {Roby}, {Roodman},
  {Rosing}, {Roucelle}, {Rumore}, {Russo}, {Saha}, {Sassolas}, {Schalk},
  {Schellart}, {Schindler}, {Schmidt}, {Schneider}, {Schneider}, {Schoening},
  {Schumacher}, {Schwamb}, {Sebag}, {Selvy}, {Sembroski}, {Seppala}, {Serio},
  {Serrano}, {Shaw}, {Shipsey}, {Sick}, {Silvestri}, {Slater}, {Smith},
  {Smith}, {Sobhani}, {Soldahl}, {Storrie-Lombardi}, {Stover}, {Strauss},
  {Street}, {Stubbs}, {Sullivan}, {Sweeney}, {Swinbank}, {Szalay}, {Takacs},
  {Tether}, {Thaler}, {Thayer}, {Thomas}, {Thornton}, {Thukral}, {Tice},
  {Trilling}, {Turri}, {Van Berg}, {Vanden Berk}, {Vetter}, {Virieux},
  {Vucina}, {Wahl}, {Walkowicz}, {Walsh}, {Walter}, {Wang}, {Wang}, {Warner},
  {Wiecha}, {Willman}, {Winters}, {Wittman}, {Wolff}, {Wood-Vasey}, {Wu},
  {Xin}, {Yoachim}, \& {Zhan}}]{Ivezic_LSST_2019ApJ...873..111I}
{Ivezi{\'c}}, {\v{Z}}., {Kahn}, S.~M., {Tyson}, J.~A., {et~al.} 2019, \apj,
  873, 111

\bibitem[{{Jones} {et~al.}(2009){Jones}, {Read}, {Saunders}, {Colless},
  {Jarrett}, {Parker}, {Fairall}, {Mauch}, {Sadler}, {Watson}, {Burton},
  {Campbell}, {Cass}, {Croom}, {Dawe}, {Fiegert}, {Frankcombe}, {Hartley},
  {Huchra}, {James}, {Kirby}, {Lahav}, {Lucey}, {Mamon}, {Moore}, {Peterson},
  {Prior}, {Proust}, {Russell}, {Safouris}, {Wakamatsu}, {Westra}, \&
  {Williams}}]{Jones2009MNRAS.399..683J}
{Jones}, D.~H., {Read}, M.~A., {Saunders}, W., {et~al.} 2009, \mnras, 399, 683

\bibitem[{{Kaiser} {et~al.}(1995){Kaiser}, {Squires}, \&
  {Broadhurst}}]{Kaiser1995ApJ...449..460K}
{Kaiser}, N., {Squires}, G., \& {Broadhurst}, T. 1995, \apj, 449, 460

\bibitem[{{Klypin} {et~al.}(2013){Klypin}, {Prada}, {Yepes}, {Hess}, \&
  {Gottlober}}]{Klypin2013}
{Klypin}, A., {Prada}, F., {Yepes}, G., {Hess}, S., \& {Gottlober}, S. 2013,
  ArXiv e-prints [\eprint[arXiv]{1310.3740}]

\bibitem[{{Kocevski} {et~al.}(2015){Kocevski}, {Brightman}, {Nandra},
  {Koekemoer}, {Salvato}, {Aird}, {Bell}, {Hsu}, {Kartaltepe}, {Koo}, {Lotz},
  {McIntosh}, {Mozena}, {Rosario}, \& {Trump}}]{Kocevski2015ApJ...814..104K}
{Kocevski}, D.~D., {Brightman}, M., {Nandra}, K., {et~al.} 2015, \apj, 814, 104

\bibitem[{{Kollmeier} {et~al.}(2017){Kollmeier}, {Zasowski}, {Rix}, {Johns},
  {Anderson}, {Drory}, {Johnson}, {Pogge}, {Bird}, {Blanc}, {Brownstein},
  {Crane}, {De Lee}, {Klaene}, {Kreckel}, {MacDonald}, {Merloni}, {Ness},
  {O'Brien}, {Sanchez-Gallego}, {Sayres}, {Shen}, {Thakar}, {Tkachenko},
  {Aerts}, {Blanton}, {Eisenstein}, {Holtzman}, {Maoz}, {Nandra}, {Rockosi},
  {Weinberg}, {Bovy}, {Casey}, {Chaname}, {Clerc}, {Conroy}, {Eracleous},
  {G{\"a}nsicke}, {Hekker}, {Horne}, {Kauffmann}, {McQuinn}, {Pellegrini},
  {Schinnerer}, {Schlafly}, {Schwope}, {Seibert}, {Teske}, \& {van
  Saders}}]{Kollmeier17}
{Kollmeier}, J.~A., {Zasowski}, G., {Rix}, H.-W., {et~al.} 2017, arXiv e-prints
  [\eprint[arXiv]{1711.03234}]

\bibitem[{{Koulouridis} \&
  {Bartalucci}(2019)}]{KoulouridisBartalucci_2019A&A...623L..10K}
{Koulouridis}, E. \& {Bartalucci}, I. 2019, \aap, 623, L10

\bibitem[{{Koulouridis} {et~al.}(2013){Koulouridis}, {Plionis}, {Chavushyan},
  {Dultzin}, {Krongold}, {Georgantopoulos}, \&
  {Le{\'o}n-Tavares}}]{KoulouridisPlionisChavushyan_2013A&A...552A.135K}
{Koulouridis}, E., {Plionis}, M., {Chavushyan}, V., {et~al.} 2013, \aap, 552,
  A135

\bibitem[{{Koulouridis} {et~al.}(2018){Koulouridis}, {Ricci}, {Giles}, {Adami},
  {Ramos-Ceja}, {Pierre}, {Plionis}, {Lidman}, {Georgantopoulos}, {Chiappetti},
  \& et~al.}]{KoulouridisRicciGiles_2018A&A...620A..20K}
{Koulouridis}, E., {Ricci}, M., {Giles}, P., {et~al.} 2018, \aap, 620, A20

\bibitem[{{Koutoulidis} {et~al.}(2018){Koutoulidis}, {Georgantopoulos},
  {Mountrichas}, {Plionis}, {Georgakakis}, {Akylas}, \&
  {Rovilos}}]{Koutoulidis18}
{Koutoulidis}, L., {Georgantopoulos}, I., {Mountrichas}, G., {et~al.} 2018,
  \mnras, 481, 3063

\bibitem[{{Koutoulidis} {et~al.}(2013){Koutoulidis}, {Plionis},
  {Georgantopoulos}, \& {Fanidakis}}]{Koutoulidis13}
{Koutoulidis}, L., {Plionis}, M., {Georgantopoulos}, I., \& {Fanidakis}, N.
  2013, \mnras, 428, 1382

\bibitem[{{Kravtsov} {et~al.}(2004){Kravtsov}, {Berlind}, {Wechsler}, {Klypin},
  {Gottl{\"o}ber}, {Allgood}, \& {Primack}}]{Kravtsov2004}
{Kravtsov}, A.~V., {Berlind}, A.~A., {Wechsler}, R.~H., {et~al.} 2004, \apj,
  609, 35

\bibitem[{{Krumpe} {et~al.}(2010){Krumpe}, {Miyaji}, \& {Coil}}]{Krumpe10}
{Krumpe}, M., {Miyaji}, T., \& {Coil}, A.~L. 2010, \apj, 713, 558

\bibitem[{{Krumpe} {et~al.}(2012){Krumpe}, {Miyaji}, {Coil}, \&
  {Aceves}}]{Krumpe12}
{Krumpe}, M., {Miyaji}, T., {Coil}, A.~L., \& {Aceves}, H. 2012, \apj, 746, 1

\bibitem[{{Krumpe} {et~al.}(2018){Krumpe}, {Miyaji}, {Coil}, \&
  {Aceves}}]{Krumpe18}
{Krumpe}, M., {Miyaji}, T., {Coil}, A.~L., \& {Aceves}, H. 2018, \mnras, 474,
  1773

\bibitem[{{Krumpe} {et~al.}(2023){Krumpe}, {Miyaji}, {Georgakakis}, {Schulze},
  {Coil}, {Dwelly}, {Coffey}, {Comparat}, {Aceves}, {Salvato}, \&
  et~al.}]{Krumpe23}
{Krumpe}, M., {Miyaji}, T., {Georgakakis}, A., {et~al.} 2023, arXiv e-prints,
  arXiv:2304.02036

\bibitem[{{Krumpe} {et~al.}(2015){Krumpe}, {Miyaji}, {Husemann}, {Fanidakis},
  {Coil}, \& {Aceves}}]{Krumpe15}
{Krumpe}, M., {Miyaji}, T., {Husemann}, B., {et~al.} 2015, \apj, 815, 21

\bibitem[{{Landy} \& {Szalay}(1993)}]{Landy1993ApJ...412...64L}
{Landy}, S.~D. \& {Szalay}, A.~S. 1993, \apj, 412, 64

\bibitem[{{Laureijs} {et~al.}(2011){Laureijs}, {Amiaux}, {Arduini},
  {Augu{\`e}res}, {Brinchmann}, {Cole}, {Cropper}, {Dabin}, {Duvet}, {Ealet},
  {Garilli}, {Gondoin}, {Guzzo}, {Hoar}, {Hoekstra}, {Holmes}, {Kitching},
  {Maciaszek}, {Mellier}, {Pasian}, {Percival}, {Rhodes}, {Saavedra Criado},
  {Sauvage}, {Scaramella}, {Valenziano}, {Warren}, {Bender}, {Castander},
  {Cimatti}, {Le F{\`e}vre}, {Kurki-Suonio}, {Levi}, {Lilje}, {Meylan},
  {Nichol}, {Pedersen}, {Popa}, {Rebolo Lopez}, {Rix}, {Rottgering},
  {Zeilinger}, {Grupp}, {Hudelot}, {Massey}, {Meneghetti}, {Miller}, {Paltani},
  {Paulin-Henriksson}, {Pires}, {Saxton}, {Schrabback}, {Seidel}, {Walsh},
  {Aghanim}, {Amendola}, {Bartlett}, {Baccigalupi}, {Beaulieu}, {Benabed},
  {Cuby}, {Elbaz}, {Fosalba}, {Gavazzi}, {Helmi}, {Hook}, {Irwin}, {Kneib},
  {Kunz}, {Mannucci}, {Moscardini}, {Tao}, {Teyssier}, {Weller}, {Zamorani},
  {Zapatero Osorio}, {Boulade}, {Foumond}, {Di Giorgio}, {Guttridge}, {James},
  {Kemp}, {Martignac}, {Spencer}, {Walton}, {Bl{\"u}mchen}, {Bonoli},
  {Bortoletto}, {Cerna}, {Corcione}, {Fabron}, {Jahnke}, {Ligori}, {Madrid},
  {Martin}, {Morgante}, {Pamplona}, {Prieto}, {Riva}, {Toledo}, {Trifoglio},
  {Zerbi}, {Abdalla}, {Douspis}, {Grenet}, {Borgani}, {Bouwens}, {Courbin},
  {Delouis}, {Dubath}, {Fontana}, {Frailis}, {Grazian}, {Koppenh{\"o}fer},
  {Mansutti}, {Melchior}, {Mignoli}, {Mohr}, {Neissner}, {Noddle}, {Poncet},
  {Scodeggio}, {Serrano}, {Shane}, {Starck}, {Surace}, {Taylor},
  {Verdoes-Kleijn}, {Vuerli}, {Williams}, {Zacchei}, {Altieri}, {Escudero
  Sanz}, {Kohley}, {Oosterbroek}, {Astier}, {Bacon}, {Bardelli}, {Baugh},
  {Bellagamba}, {Benoist}, {Bianchi}, {Biviano}, {Branchini}, {Carbone},
  {Cardone}, {Clements}, {Colombi}, {Conselice}, {Cresci}, {Deacon}, {Dunlop},
  {Fedeli}, {Fontanot}, {Franzetti}, {Giocoli}, {Garcia-Bellido}, {Gow},
  {Heavens}, {Hewett}, {Heymans}, {Holland}, {Huang}, {Ilbert}, {Joachimi},
  {Jennins}, {Kerins}, {Kiessling}, {Kirk}, {Kotak}, {Krause}, {Lahav}, {van
  Leeuwen}, {Lesgourgues}, {Lombardi}, {Magliocchetti}, {Maguire}, {Majerotto},
  {Maoli}, {Marulli}, {Maurogordato}, {McCracken}, {McLure}, {Melchiorri},
  {Merson}, {Moresco}, {Nonino}, {Norberg}, {Peacock}, {Pello}, {Penny},
  {Pettorino}, {Di Porto}, {Pozzetti}, {Quercellini}, {Radovich}, {Rassat},
  {Roche}, {Ronayette}, {Rossetti}, {Sartoris}, {Schneider}, {Semboloni},
  {Serjeant}, {Simpson}, {Skordis}, {Smadja}, {Smartt}, {Spano}, {Spiro},
  {Sullivan}, {Tilquin}, {Trotta}, {Verde}, {Wang}, {Williger}, {Zhao},
  {Zoubian}, \& {Zucca}}]{Laureijs_EUCLID_2011arXiv1110.3193L}
{Laureijs}, R., {Amiaux}, J., {Arduini}, S., {et~al.} 2011, arXiv e-prints,
  arXiv:1110.3193

\bibitem[{{Leauthaud} {et~al.}(2015){Leauthaud}, {J. Benson}, {Civano}, {L.
  Coil}, {Bundy}, {Massey}, {Schramm}, {Schulze}, {Capak}, {Elvis}, {Kulier},
  \& {Rhodes}}]{Leauthaud15}
{Leauthaud}, A., {J. Benson}, A., {Civano}, F., {et~al.} 2015, \mnras, 446,
  1874

\bibitem[{{Leauthaud} {et~al.}(2011){Leauthaud}, {Tinker}, {Behroozi}, {Busha},
  \& {Wechsler}}]{Leauthaud2011ApJ...738...45L}
{Leauthaud}, A., {Tinker}, J., {Behroozi}, P.~S., {Busha}, M.~T., \&
  {Wechsler}, R.~H. 2011, \apj, 738, 45

\bibitem[{{Li} {et~al.}(2022){Li}, {Miyatake}, {Luo}, {More}, {Oguri},
  {Hamana}, {Mandelbaum}, {Shirasaki}, {Takada}, {Armstrong}, {Kannawadi},
  {Takita}, {Miyazaki}, {Nishizawa}, {Plazas Malagon}, {Strauss}, {Tanaka}, \&
  {Yoshida}}]{Li2022PASJ...74..421L}
{Li}, X., {Miyatake}, H., {Luo}, W., {et~al.} 2022, \pasj, 74, 421

\bibitem[{{Lilly} {et~al.}(2009){Lilly}, {Le Brun}, {Maier}, {Mainieri},
  {Mignoli}, {Scodeggio}, {Zamorani}, {Carollo}, {Contini}, {Kneib}, {Le
  F{\`e}vre}, {Renzini}, {Bardelli}, {Bolzonella}, {Bongiorno}, {Caputi},
  {Coppa}, {Cucciati}, {de la Torre}, {de Ravel}, {Franzetti}, {Garilli},
  {Iovino}, {Kampczyk}, {Kovac}, {Knobel}, {Lamareille}, {Le Borgne}, {Pello},
  {Peng}, {P{\'e}rez-Montero}, {Ricciardelli}, {Silverman}, {Tanaka}, {Tasca},
  {Tresse}, {Vergani}, {Zucca}, {Ilbert}, {Salvato}, {Oesch}, {Abbas},
  {Bottini}, {Capak}, {Cappi}, {Cassata}, {Cimatti}, {Elvis}, {Fumana},
  {Guzzo}, {Hasinger}, {Koekemoer}, {Leauthaud}, {Maccagni}, {Marinoni},
  {McCracken}, {Memeo}, {Meneux}, {Porciani}, {Pozzetti}, {Sanders},
  {Scaramella}, {Scarlata}, {Scoville}, {Shopbell}, \&
  {Taniguchi}}]{Lilly2009ApJS..184..218L}
{Lilly}, S.~J., {Le Brun}, V., {Maier}, C., {et~al.} 2009, \apjs, 184, 218

\bibitem[{{Liske} {et~al.}(2015){Liske}, {Baldry}, {Driver}, {Tuffs},
  {Alpaslan}, {Andrae}, {Brough}, {Cluver}, {Grootes}, {Gunawardhana},
  {Kelvin}, {Loveday}, {Robotham}, {Taylor}, {Bamford}, {Bland-Hawthorn},
  {Brown}, {Drinkwater}, {Hopkins}, {Meyer}, {Norberg}, {Peacock}, {Agius},
  {Andrews}, {Bauer}, {Ching}, {Colless}, {Conselice}, {Croom}, {Davies}, {De
  Propris}, {Dunne}, {Eardley}, {Ellis}, {Foster}, {Frenk}, {H{\"a}u{\ss}ler},
  {Holwerda}, {Howlett}, {Ibarra}, {Jarvis}, {Jones}, {Kafle}, {Lacey},
  {Lange}, {Lara-L{\'o}pez}, {L{\'o}pez-S{\'a}nchez}, {Maddox}, {Madore},
  {McNaught-Roberts}, {Moffett}, {Nichol}, {Owers}, {Palamara}, {Penny},
  {Phillipps}, {Pimbblet}, {Popescu}, {Prescott}, {Proctor}, {Sadler},
  {Sansom}, {Seibert}, {Sharp}, {Sutherland}, {V{\'a}zquez-Mata}, {van Kampen},
  {Wilkins}, {Williams}, \& {Wright}}]{Liske2015MNRAS.452.2087L}
{Liske}, J., {Baldry}, I.~K., {Driver}, S.~P., {et~al.} 2015, \mnras, 452, 2087

\bibitem[{{Liu} {et~al.}(2022{\natexlab{a}}){Liu}, {Bulbul}, {Ghirardini},
  {Liu}, {Klein}, {Clerc}, {{\"O}zsoy}, {Ramos-Ceja}, {Pacaud}, {Comparat},
  {Okabe}, {Bahar}, {Biffi}, {Brunner}, {Br{\"u}ggen}, {Buchner}, {Ider
  Chitham}, {Chiu}, {Dolag}, {Gatuzz}, {Gonzalez}, {Hoang}, {Lamer}, {Merloni},
  {Nandra}, {Oguri}, {Ota}, {Predehl}, {Reiprich}, {Salvato}, {Schrabback},
  {Sanders}, {Seppi}, \& {Thibaud}}]{LiuAng2022AA...661A...2L}
{Liu}, A., {Bulbul}, E., {Ghirardini}, V., {et~al.} 2022{\natexlab{a}}, \aap,
  661, A2

\bibitem[{{Liu} {et~al.}(2022{\natexlab{b}}){Liu}, {Buchner}, {Nandra},
  {Merloni}, {Dwelly}, {Sanders}, {Salvato}, {Arcodia}, {Brusa}, {Wolf},
  {Georgakakis}, {Boller}, {Krumpe}, {Lamer}, {Waddell}, {Urrutia}, {Schwope},
  {Robrade}, {Wilms}, {Dauser}, {Comparat}, {Toba}, {Ichikawa}, {Iwasawa},
  {Shen}, \& {Medel}}]{Liu_SPEC2022AA...661A...5L}
{Liu}, T., {Buchner}, J., {Nandra}, K., {et~al.} 2022{\natexlab{b}}, \aap, 661,
  A5

\bibitem[{{Liu} {et~al.}(2022{\natexlab{c}}){Liu}, {Merloni}, {Comparat},
  {Nandra}, {Sanders}, {Lamer}, {Buchner}, {Dwelly}, {Freyberg}, {Malyali},
  {Georgakakis}, {Salvato}, {Brunner}, {Brusa}, {Klein}, {Ghirardini}, {Clerc},
  {Pacaud}, {Bulbul}, {Liu}, {Schwope}, {Robrade}, {Wilms}, {Dauser},
  {Ramos-Ceja}, {Reiprich}, {Boller}, \& {Wolf}}]{Liu_SIM_2022AA...661A..27L}
{Liu}, T., {Merloni}, A., {Comparat}, J., {et~al.} 2022{\natexlab{c}}, \aap,
  661, A27

\bibitem[{{Luo} {et~al.}(2015){Luo}, {Zhao}, {Zhao}, {Deng}, {Liu}, {Jing},
  {Wang}, {Zhang}, {Shi}, {Cui}, {Chu}, {Li}, {Bai}, {Wu}, {Cai}, {Cao}, {Cao},
  {Carlin}, {Chen}, {Chen}, {Chen}, {Chen}, {Chen}, {Chen}, {Chen},
  {Christlieb}, {Chu}, {Cui}, {Dong}, {Du}, {Fan}, {Feng}, {Fu}, {Gao}, {Gong},
  {Gu}, {Guo}, {Han}, {He}, {Hou}, {Hou}, {Hou}, {Hu}, {Hu}, {Hu}, {Huo},
  {Jia}, {Jiang}, {Jiang}, {Jiang}, {Jin}, {Kong}, {Kong}, {Lei}, {Li}, {Li},
  {Li}, {Li}, {Li}, {Li}, {Li}, {Li}, {Li}, {Li}, {Li}, {Li}, {Liang}, {Lin},
  {Liu}, {Liu}, {Liu}, {Liu}, {Lu}, {Luo}, {Mao}, {Newberg}, {Ni}, {Qi}, {Qi},
  {Shen}, {Shi}, {Song}, {Song}, {Su}, {Su}, {Tang}, {Tao}, {Tian}, {Wang},
  {Wang}, {Wang}, {Wang}, {Wang}, {Wang}, {Wang}, {Wang}, {Wang}, {Wang},
  {Wang}, {Wang}, {Wang}, {Wang}, {Wang}, {Wang}, {Wang}, {Wang}, {Wang},
  {Wang}, {Wei}, {Wei}, {Wu}, {Wu}, {Wu}, {Wu}, {Xing}, {Xu}, {Xu}, {Xu},
  {Yan}, {Yang}, {Yang}, {Yang}, {Yang}, {Yao}, {Yu}, {Yuan}, {Yuan}, {Yuan},
  {Yuan}, {Zhai}, {Zhang}, {Zhang}, {Zhang}, {Zhang}, {Zhang}, {Zhang},
  {Zhang}, {Zhang}, {Zhao}, {Zhou}, {Zhou}, {Zhu}, {Zhu}, {Zou}, \&
  {Zuo}}]{Luo2015RAA....15.1095L}
{Luo}, A.~L., {Zhao}, Y.-H., {Zhao}, G., {et~al.} 2015, Research in Astronomy
  and Astrophysics, 15, 1095

\bibitem[{{Luo} {et~al.}(2022){Luo}, {Silverman}, {More}, {Goulding},
  {Miyatake}, {Nishimichi}, {Hikage}, {Kawinwanichakij}, {Li}, {Li},
  {Medezinski}, {Oguri}, {Oogi}, \& {Sifon}}]{Luo2022arXiv220403817L}
{Luo}, W., {Silverman}, J.~D., {More}, S., {et~al.} 2022, arXiv e-prints,
  arXiv:2204.03817

\bibitem[{{Mandelbaum} {et~al.}(2018{\natexlab{a}}){Mandelbaum}, {Lanusse},
  {Leauthaud}, {Armstrong}, {Simet}, {Miyatake}, {Meyers}, {Bosch}, {Murata},
  {Miyazaki}, \& {Tanaka}}]{Mandelbaum2018MNRAS.481.3170M}
{Mandelbaum}, R., {Lanusse}, F., {Leauthaud}, A., {et~al.} 2018{\natexlab{a}},
  \mnras, 481, 3170

\bibitem[{{Mandelbaum} {et~al.}(2018{\natexlab{b}}){Mandelbaum}, {Miyatake},
  {Hamana}, {Oguri}, {Simet}, {Armstrong}, {Bosch}, {Murata}, {Lanusse},
  {Leauthaud}, {Coupon}, {More}, {Takada}, {Miyazaki}, {Speagle}, {Shirasaki},
  {Sif{\'o}n}, {Huang}, {Nishizawa}, {Medezinski}, {Okura}, {Okabe}, {Czakon},
  {Takahashi}, {Coulton}, {Hikage}, {Komiyama}, {Lupton}, {Strauss}, {Tanaka},
  \& {Utsumi}}]{Mandelbaum2018PASJ...70S..25M}
{Mandelbaum}, R., {Miyatake}, H., {Hamana}, T., {et~al.} 2018{\natexlab{b}},
  \pasj, 70, S25

\bibitem[{{Mandelbaum} {et~al.}(2005){Mandelbaum}, {Tasitsiomi}, {Seljak},
  {Kravtsov}, \& {Wechsler}}]{Mandelbaum2005MNRAS.362.1451M}
{Mandelbaum}, R., {Tasitsiomi}, A., {Seljak}, U., {Kravtsov}, A.~V., \&
  {Wechsler}, R.~H. 2005, \mnras, 362, 1451

\bibitem[{{Martini} {et~al.}(2013){Martini}, {Miller}, {Brodwin}, {Stanford},
  {Gonzalez}, {Bautz}, {Hickox}, {Stern}, {Eisenhardt}, {Galametz}, \&
  et~al.}]{MartiniMillerBrodwin_2013ApJ...768....1M}
{Martini}, P., {Miller}, E.~D., {Brodwin}, M., {et~al.} 2013, \apj, 768, 1

\bibitem[{{Marulli} {et~al.}(2013){Marulli}, {Bolzonella}, {Branchini},
  {Davidzon}, {de la Torre}, {Granett}, {Guzzo}, {Iovino}, {Moscardini},
  {Pollo}, {Abbas}, {Adami}, {Arnouts}, {Bel}, {Bottini}, {Cappi}, {Coupon},
  {Cucciati}, {De Lucia}, {Fritz}, {Franzetti}, {Fumana}, {Garilli}, {Ilbert},
  {Krywult}, {Le Brun}, {Le F{\`e}vre}, {Maccagni}, {Ma{\l}ek}, {McCracken},
  {Paioro}, {Polletta}, {Schlagenhaufer}, {Scodeggio}, {Tasca}, {Tojeiro},
  {Vergani}, {Zanichelli}, {Burden}, {Di Porto}, {Marchetti}, {Marinoni},
  {Mellier}, {Nichol}, {Peacock}, {Percival}, {Phleps}, {Wolk}, \&
  {Zamorani}}]{Marulli2013AA...557A..17M}
{Marulli}, F., {Bolzonella}, M., {Branchini}, E., {et~al.} 2013, \aap, 557, A17

\bibitem[{{McLure} {et~al.}(2013){McLure}, {Cirasuolo}, {Dunlop}, {Almaini}, \&
  {Foucaud}}]{Mclure2013ASSP...37..323M}
{McLure}, R.~J., {Cirasuolo}, M., {Dunlop}, J.~S., {Almaini}, O., \& {Foucaud},
  S. 2013, in Astrophysics and Space Science Proceedings, Vol.~37, Thirty Years
  of Astronomical Discovery with UKIRT, 323

\bibitem[{{Medezinski} {et~al.}(2018){Medezinski}, {Oguri}, {Nishizawa},
  {Speagle}, {Miyatake}, {Umetsu}, {Leauthaud}, {Murata}, {Mandelbaum},
  {Sif{\'o}n}, {Strauss}, {Huang}, {Simet}, {Okabe}, {Tanaka}, \&
  {Komiyama}}]{Medezinski2018PASJ...70...30M}
{Medezinski}, E., {Oguri}, M., {Nishizawa}, A.~J., {et~al.} 2018, \pasj, 70, 30

\bibitem[{{Mendez} {et~al.}(2016){Mendez}, {Coil}, {Aird}, {Skibba},
  {Diamond-Stanic}, {Moustakas}, {Blanton}, {Cool}, {Eisenstein}, {Wong}, \&
  {Zhu}}]{Mendez16}
{Mendez}, A.~J., {Coil}, A.~L., {Aird}, J., {et~al.} 2016, \apj, 821, 55

\bibitem[{{Merloni} {et~al.}(2019){Merloni}, {Alexander}, {Banerji}, {Boller},
  {Comparat}, {Dwelly}, {Fotopoulou}, {McMahon}, {Nandra}, {Salvato}, {Croom},
  {Finoguenov}, {Krumpe}, {Lamer}, {Rosario}, {Schwope}, {Shanks}, {Steinmetz},
  {Wisotzki}, \& {Worseck}}]{Merloni2019}
{Merloni}, A., {Alexander}, D.~A., {Banerji}, M., {et~al.} 2019, The Messenger,
  175, 42

\bibitem[{{Mishra} \& {Dai}(2020)}]{MishraDai_2020AJ....159...69M}
{Mishra}, H.~D. \& {Dai}, X. 2020, \aj, 159, 69

\bibitem[{{Miyaji} {et~al.}(2011){Miyaji}, {Krumpe}, {Coil}, \&
  {Aceves}}]{Miyaji2011ApJ...726...83M}
{Miyaji}, T., {Krumpe}, M., {Coil}, A.~L., \& {Aceves}, H. 2011, \apj, 726, 83

\bibitem[{{Miyatake} {et~al.}(2019){Miyatake}, {Battaglia}, {Hilton},
  {Medezinski}, {Nishizawa}, {More}, {Aiola}, {Bahcall}, {Bond}, {Calabrese},
  {Choi}, {Devlin}, {Dunkley}, {Dunner}, {Fuzia}, {Gallardo}, {Gralla},
  {Hasselfield}, {Halpern}, {Hikage}, {Hill}, {Hincks}, {Hlo{\v{z}}ek},
  {Huffenberger}, {Hughes}, {Koopman}, {Kosowsky}, {Louis}, {Madhavacheril},
  {McMahon}, {Mandelbaum}, {Marriage}, {Maurin}, {Miyazaki}, {Moodley},
  {Murata}, {Naess}, {Newburgh}, {Niemack}, {Nishimichi}, {Okabe}, {Oguri},
  {Osato}, {Page}, {Partridge}, {Robertson}, {Sehgal}, {Sherwin}, {Shirasaki},
  {Sievers}, {Sif{\'o}n}, {Simon}, {Spergel}, {Staggs}, {Stein}, {Takada},
  {Trac}, {Umetsu}, {van Engelen}, \& {Wollack}}]{Miyatake2019ApJ...875...63M}
{Miyatake}, H., {Battaglia}, N., {Hilton}, M., {et~al.} 2019, \apj, 875, 63

\bibitem[{{Momcheva} {et~al.}(2016){Momcheva}, {Brammer}, {van Dokkum},
  {Skelton}, {Whitaker}, {Nelson}, {Fumagalli}, {Maseda}, {Leja}, {Franx},
  {Rix}, {Bezanson}, {Da Cunha}, {Dickey}, {F{\"o}rster Schreiber},
  {Illingworth}, {Kriek}, {Labb{\'e}}, {Ulf Lange}, {Lundgren}, {Magee},
  {Marchesini}, {Oesch}, {Pacifici}, {Patel}, {Price}, {Tal}, {Wake}, {van der
  Wel}, \& {Wuyts}}]{Momcheva2016ApJS..225...27M}
{Momcheva}, I.~G., {Brammer}, G.~B., {van Dokkum}, P.~G., {et~al.} 2016, \apjs,
  225, 27

\bibitem[{{More} {et~al.}(2015){More}, {Miyatake}, {Mandelbaum}, {Takada},
  {Spergel}, {Brownstein}, \& {Schneider}}]{More2015ApJ...806....2M}
{More}, S., {Miyatake}, H., {Mandelbaum}, R., {et~al.} 2015, \apj, 806, 2

\bibitem[{{Moster} {et~al.}(2013){Moster}, {Naab}, \& {White}}]{Moster13}
{Moster}, B.~P., {Naab}, T., \& {White}, S.~D.~M. 2013, \mnras, 428, 3121

\bibitem[{{Mountrichas} {et~al.}(2021){Mountrichas}, {Buat}, {Yang}, {Boquien},
  {Burgarella}, \& {Ciesla}}]{Mountrichas2021AA...646A..29M}
{Mountrichas}, G., {Buat}, V., {Yang}, G., {et~al.} 2021, \aap, 646, A29

\bibitem[{{Mountrichas} {et~al.}(2019){Mountrichas}, {Georgakakis}, \&
  {Georgantopoulos}}]{Mountrichas19}
{Mountrichas}, G., {Georgakakis}, A., \& {Georgantopoulos}, I. 2019, \mnras,
  483, 1374

\bibitem[{{Myers} {et~al.}(2007){Myers}, {Brunner}, {Nichol}, {Richards},
  {Schneider}, \& {Bahcall}}]{Myers07}
{Myers}, A.~D., {Brunner}, R.~J., {Nichol}, R.~C., {et~al.} 2007, \apj, 658, 85

\bibitem[{{Ni} {et~al.}(2019){Ni}, {Yang}, {Brandt}, {Alexander}, {Chen},
  {Luo}, {Vito}, \& {Xue}}]{Ni2019MNRAS.490.1135N}
{Ni}, Q., {Yang}, G., {Brandt}, W.~N., {et~al.} 2019, \mnras, 490, 1135

\bibitem[{Nishimichi {et~al.}(2019)Nishimichi, Takada, Takahashi, Osato,
  Shirasaki, Oogi, Miyatake, Oguri, Murata, Kobayashi, \&
  Yoshida}]{Nishimichi_2019}
Nishimichi, T., Takada, M., Takahashi, R., {et~al.} 2019, The Astrophysical
  Journal, 884, 29

\bibitem[{{Nishizawa} {et~al.}(2020){Nishizawa}, {Hsieh}, {Tanaka}, \&
  {Takata}}]{Nishizawa2020arXiv200301511N}
{Nishizawa}, A.~J., {Hsieh}, B.-C., {Tanaka}, M., \& {Takata}, T. 2020, arXiv
  e-prints, arXiv:2003.01511

\bibitem[{{Oguri} {et~al.}(2018){Oguri}, {Lin}, {Lin}, {Nishizawa}, {More},
  {More}, {Hsieh}, {Medezinski}, {Miyatake}, {Jian}, {Lin}, {Takada}, {Okabe},
  {Speagle}, {Coupon}, {Leauthaud}, {Lupton}, {Miyazaki}, {Price}, {Tanaka},
  {Chiu}, {Komiyama}, {Okura}, {Tanaka}, \& {Usuda}}]{Oguri2018PASJ...70S..20O}
{Oguri}, M., {Lin}, Y.-T., {Lin}, S.-C., {et~al.} 2018, \pasj, 70, S20

\bibitem[{{Ohta} {et~al.}(2007){Ohta}, {Aoki}, {Kawaguchi}, \&
  {Kiuchi}}]{Ohta2007ApJS..169....1O}
{Ohta}, K., {Aoki}, K., {Kawaguchi}, T., \& {Kiuchi}, G. 2007, \apjs, 169, 1

\bibitem[{{Oke} \& {Gunn}(1983)}]{Oke1983}
{Oke}, J.~B. \& {Gunn}, J.~E. 1983, \apj, 266, 713

\bibitem[{{Oogi} {et~al.}(2020){Oogi}, {Shirakata}, {Nagashima}, {Nishimichi},
  {Kawaguchi}, {Okamoto}, {Ishiyama}, \& {Enoki}}]{oogi20}
{Oogi}, T., {Shirakata}, H., {Nagashima}, M., {et~al.} 2020, \mnras, 497, 1

\bibitem[{{Padovani} {et~al.}(2017){Padovani}, {Alexander}, {Assef}, {De
  Marco}, {Giommi}, {Hickox}, {Richards}, {Smol{\v{c}}i{\'c}},
  {Hatziminaoglou}, {Mainieri}, \& {Salvato}}]{Padovani2017AARv..25....2P}
{Padovani}, P., {Alexander}, D.~M., {Assef}, R.~J., {et~al.} 2017, \aapr, 25, 2

\bibitem[{{Paturel} {et~al.}(2003){Paturel}, {Petit}, {Prugniel}, {Theureau},
  {Rousseau}, {Brouty}, {Dubois}, \&
  {Cambr{\'e}sy}}]{Paturel2003AA...412...45P}
{Paturel}, G., {Petit}, C., {Prugniel}, P., {et~al.} 2003, \aap, 412, 45

\bibitem[{{Peluso} {et~al.}(2022){Peluso}, {Vulcani}, {Poggianti}, {Moretti},
  {Radovich}, {Smith}, {Jaff{\'e}}, {Crossett}, {Gullieuszik}, {Fritz}, \&
  et~al.}]{PelusoVulcaniPoggianti_2022ApJ...927..130P}
{Peluso}, G., {Vulcani}, B., {Poggianti}, B.~M., {et~al.} 2022, \apj, 927, 130

\bibitem[{{Peng} {et~al.}(2010){Peng}, {Lilly}, {Kova{\v{c}}}, {Bolzonella},
  {Pozzetti}, {Renzini}, {Zamorani}, {Ilbert}, {Knobel}, {Iovino}, {Maier},
  {Cucciati}, {Tasca}, {Carollo}, {Silverman}, {Kampczyk}, {de Ravel},
  {Sanders}, {Scoville}, {Contini}, {Mainieri}, {Scodeggio}, {Kneib}, {Le
  F{\`e}vre}, {Bardelli}, {Bongiorno}, {Caputi}, {Coppa}, {de la Torre},
  {Franzetti}, {Garilli}, {Lamareille}, {Le Borgne}, {Le Brun}, {Mignoli},
  {Perez Montero}, {Pello}, {Ricciardelli}, {Tanaka}, {Tresse}, {Vergani},
  {Welikala}, {Zucca}, {Oesch}, {Abbas}, {Barnes}, {Bordoloi}, {Bottini},
  {Cappi}, {Cassata}, {Cimatti}, {Fumana}, {Hasinger}, {Koekemoer},
  {Leauthaud}, {Maccagni}, {Marinoni}, {McCracken}, {Memeo}, {Meneux}, {Nair},
  {Porciani}, {Presotto}, \& {Scaramella}}]{Peng2010ApJ...721..193P}
{Peng}, Y.-j., {Lilly}, S.~J., {Kova{\v{c}}}, K., {et~al.} 2010, \apj, 721, 193

\bibitem[{{Peng} {et~al.}(2012){Peng}, {Lilly}, {Renzini}, \&
  {Carollo}}]{Peng2012ApJ...757....4P}
{Peng}, Y.-j., {Lilly}, S.~J., {Renzini}, A., \& {Carollo}, M. 2012, \apj, 757,
  4

\bibitem[{{Petter} {et~al.}(2023){Petter}, {Hickox}, {Alexander}, {Myers},
  {Geach}, {Whalen}, \& {Andonie}}]{PetterHickoxAlexander_2023arXiv230200690P}
{Petter}, G.~C., {Hickox}, R.~C., {Alexander}, D.~M., {et~al.} 2023, arXiv
  e-prints, arXiv:2302.00690

\bibitem[{{Pimbblet} {et~al.}(2013){Pimbblet}, {Shabala}, {Haines},
  {Fraser-McKelvie}, \& {Floyd}}]{PimbbletShabalaHaines_2013MNRAS.429.1827P}
{Pimbblet}, K.~A., {Shabala}, S.~S., {Haines}, C.~P., {Fraser-McKelvie}, A., \&
  {Floyd}, D.~J.~E. 2013, \mnras, 429, 1827

\bibitem[{{Planck Collaboration} {et~al.}(2020){Planck Collaboration},
  {Aghanim}, {Akrami}, {Ashdown}, {Aumont}, {Baccigalupi}, {Ballardini},
  {Banday}, {Barreiro}, {Bartolo}, {Basak}, {Battye}, {Benabed}, {Bernard},
  {Bersanelli}, {Bielewicz}, {Bock}, {Bond}, {Borrill}, {Bouchet}, {Boulanger},
  {Bucher}, {Burigana}, {Butler}, {Calabrese}, {Cardoso}, {Carron},
  {Challinor}, {Chiang}, {Chluba}, {Colombo}, {Combet}, {Contreras}, {Crill},
  {Cuttaia}, {de Bernardis}, {de Zotti}, {Delabrouille}, {Delouis}, {Di
  Valentino}, {Diego}, {Dor{\'e}}, {Douspis}, {Ducout}, {Dupac}, {Dusini},
  {Efstathiou}, {Elsner}, {En{\ss}lin}, {Eriksen}, {Fantaye}, {Farhang},
  {Fergusson}, {Fernandez-Cobos}, {Finelli}, {Forastieri}, {Frailis},
  {Fraisse}, {Franceschi}, {Frolov}, {Galeotta}, {Galli}, {Ganga},
  {G{\'e}nova-Santos}, {Gerbino}, {Ghosh}, {Gonz{\'a}lez-Nuevo}, {G{\'o}rski},
  {Gratton}, {Gruppuso}, {Gudmundsson}, {Hamann}, {Handley}, {Hansen},
  {Herranz}, {Hildebrandt}, {Hivon}, {Huang}, {Jaffe}, {Jones}, {Karakci},
  {Keih{\"a}nen}, {Keskitalo}, {Kiiveri}, {Kim}, {Kisner}, {Knox},
  {Krachmalnicoff}, {Kunz}, {Kurki-Suonio}, {Lagache}, {Lamarre}, {Lasenby},
  {Lattanzi}, {Lawrence}, {Le Jeune}, {Lemos}, {Lesgourgues}, {Levrier},
  {Lewis}, {Liguori}, {Lilje}, {Lilley}, {Lindholm}, {L{\'o}pez-Caniego},
  {Lubin}, {Ma}, {Mac{\'\i}as-P{\'e}rez}, {Maggio}, {Maino}, {Mandolesi},
  {Mangilli}, {Marcos-Caballero}, {Maris}, {Martin}, {Martinelli},
  {Mart{\'\i}nez-Gonz{\'a}lez}, {Matarrese}, {Mauri}, {McEwen}, {Meinhold},
  {Melchiorri}, {Mennella}, {Migliaccio}, {Millea}, {Mitra},
  {Miville-Desch{\^e}nes}, {Molinari}, {Montier}, {Morgante}, {Moss}, {Natoli},
  {N{\o}rgaard-Nielsen}, {Pagano}, {Paoletti}, {Partridge}, {Patanchon},
  {Peiris}, {Perrotta}, {Pettorino}, {Piacentini}, {Polastri}, {Polenta},
  {Puget}, {Rachen}, {Reinecke}, {Remazeilles}, {Renzi}, {Rocha}, {Rosset},
  {Roudier}, {Rubi{\~n}o-Mart{\'\i}n}, {Ruiz-Granados}, {Salvati}, {Sandri},
  {Savelainen}, {Scott}, {Shellard}, {Sirignano}, {Sirri}, {Spencer},
  {Sunyaev}, {Suur-Uski}, {Tauber}, {Tavagnacco}, {Tenti}, {Toffolatti},
  {Tomasi}, {Trombetti}, {Valenziano}, {Valiviita}, {Van Tent}, {Vibert},
  {Vielva}, {Villa}, {Vittorio}, {Wandelt}, {Wehus}, {White}, {White},
  {Zacchei}, \& {Zonca}}]{PlanckCosmo2020AA...641A...6P}
{Planck Collaboration}, {Aghanim}, N., {Akrami}, Y., {et~al.} 2020, \aap, 641,
  A6

\bibitem[{{Plionis} {et~al.}(2018){Plionis}, {Koutoulidis}, {Koulouridis},
  {Moscardini}, {Lidman}, {Pierre}, {Adami}, {Chiappetti}, {Faccioli},
  {Fotopoulou}, \& et~al.}]{PlionisKoutoulidisKoulouridis_2018A&A...620A..17P}
{Plionis}, M., {Koutoulidis}, L., {Koulouridis}, E., {et~al.} 2018, \aap, 620,
  A17

\bibitem[{{Poggianti} {et~al.}(2017{\natexlab{a}}){Poggianti}, {Jaff{\'e}},
  {Moretti}, {Gullieuszik}, {Radovich}, {Tonnesen}, {Fritz}, {Bettoni},
  {Vulcani}, {Fasano}, \& et~al.}]{PoggiantiJaffeMoretti_2017Natur.548..304P}
{Poggianti}, B.~M., {Jaff{\'e}}, Y.~L., {Moretti}, A., {et~al.}
  2017{\natexlab{a}}, \nat, 548, 304

\bibitem[{{Poggianti} {et~al.}(2017{\natexlab{b}}){Poggianti}, {Moretti},
  {Gullieuszik}, {Fritz}, {Jaff{\'e}}, {Bettoni}, {Fasano}, {Bellhouse}, {Hau},
  {Vulcani}, \& et~al.}]{PoggiantiMorettiGullieuszik_2017ApJ...844...48P}
{Poggianti}, B.~M., {Moretti}, A., {Gullieuszik}, M., {et~al.}
  2017{\natexlab{b}}, \apj, 844, 48

\bibitem[{{Powell} {et~al.}(2018){Powell}, {Cappelluti}, {Urry}, {Koss},
  {Finoguenov}, {Ricci}, {Trakhtenbrot}, {Allevato}, {Ajello}, {Oh},
  {Schawinski}, \& {Secrest}}]{powell18}
{Powell}, M.~C., {Cappelluti}, N., {Urry}, C.~M., {et~al.} 2018, \apj, 858, 110

\bibitem[{{Predehl} {et~al.}(2021){Predehl}, {Andritschke}, {Arefiev},
  {Babyshkin}, {Batanov}, {Becker}, {B{\"o}hringer}, {Bogomolov}, {Boller},
  {Borm}, {Bornemann}, {Br{\"a}uninger}, {Br{\"u}ggen}, {Brunner}, {Brusa},
  {Bulbul}, {Buntov}, {Burwitz}, {Burkert}, {Clerc}, {Churazov}, {Coutinho},
  {Dauser}, {Dennerl}, {Doroshenko}, {Eder}, {Emberger}, {Eraerds},
  {Finoguenov}, {Freyberg}, {Friedrich}, {Friedrich}, {F{\"u}rmetz},
  {Georgakakis}, {Gilfanov}, {Granato}, {Grossberger}, {Gueguen}, {Gureev},
  {Haberl}, {H{\"a}lker}, {Hartner}, {Hasinger}, {Huber}, {Ji}, {Kienlin},
  {Kink}, {Korotkov}, {Kreykenbohm}, {Lamer}, {Lomakin}, {Lapshov}, {Liu},
  {Maitra}, {Meidinger}, {Menz}, {Merloni}, {Mernik}, {Mican}, {Mohr},
  {M{\"u}ller}, {Nandra}, {Nazarov}, {Pacaud}, {Pavlinsky}, {Perinati},
  {Pfeffermann}, {Pietschner}, {Ramos-Ceja}, {Rau}, {Reiffers}, {Reiprich},
  {Robrade}, {Salvato}, {Sanders}, {Santangelo}, {Sasaki}, {Scheuerle},
  {Schmid}, {Schmitt}, {Schwope}, {Shirshakov}, {Steinmetz}, {Stewart},
  {Str{\"u}der}, {Sunyaev}, {Tenzer}, {Tiedemann}, {Tr{\"u}mper}, {Voron},
  {Weber}, {Wilms}, \& {Yaroshenko}}]{Predehl2021}
{Predehl}, P., {Andritschke}, R., {Arefiev}, V., {et~al.} 2021, \aap, 647, A1

\bibitem[{{Refregier}(2003)}]{Refregier2003ARAA..41..645R}
{Refregier}, A. 2003, \araa, 41, 645

\bibitem[{{Robotham} {et~al.}(2011){Robotham}, {Norberg}, {Driver}, {Baldry},
  {Bamford}, {Hopkins}, {Liske}, {Loveday}, {Merson}, {Peacock}, \&
  et~al.}]{RobothamNorbergDriver_2011MNRAS.416.2640R}
{Robotham}, A.~S.~G., {Norberg}, P., {Driver}, S.~P., {et~al.} 2011, \mnras,
  416, 2640

\bibitem[{{Rodr{\'\i}guez-Torres} {et~al.}(2017){Rodr{\'\i}guez-Torres},
  {Comparat}, {Prada}, {Yepes}, {Burtin}, {Zarrouk}, {Laurent}, {Hahn},
  {Behroozi}, {Klypin}, {Ross}, {Tojeiro}, \& {Zhao}}]{Rodrigo-Torres17}
{Rodr{\'\i}guez-Torres}, S.~A., {Comparat}, J., {Prada}, F., {et~al.} 2017,
  \mnras, 468, 728

\bibitem[{{Rowe} {et~al.}(2015){Rowe}, {Jarvis}, {Mandelbaum}, {Bernstein},
  {Bosch}, {Simet}, {Meyers}, {Kacprzak}, {Nakajima}, {Zuntz}, {Miyatake},
  {Dietrich}, {Armstrong}, {Melchior}, \& {Gill}}]{Rowe2015A&C....10..121R}
{Rowe}, B.~T.~P., {Jarvis}, M., {Mandelbaum}, R., {et~al.} 2015, Astronomy and
  Computing, 10, 121

\bibitem[{{Salvato} {et~al.}(2018){Salvato}, {Ilbert}, \&
  {Hoyle}}]{Salvato18_photoZ_review}
{Salvato}, M., {Ilbert}, O., \& {Hoyle}, B. 2018, Nature Astronomy, 68

\bibitem[{{Salvato} {et~al.}(2022){Salvato}, {Wolf}, {Dwelly}, {Georgakakis},
  {Brusa}, {Merloni}, {Liu}, {Toba}, {Nandra}, {Lamer}, {Buchner}, {Schneider},
  {Freund}, {Rau}, {Schwope}, {Nishizawa}, {Klein}, {Arcodia}, {Comparat},
  {Musiimenta}, {Nagao}, {Brunner}, {Malyali}, {Finoguenov}, {Anderson},
  {Shen}, {Ibarra-Medel}, {Trump}, {Brandt}, {Urry}, {Rivera}, {Krumpe},
  {Urrutia}, {Miyaji}, {Ichikawa}, {Schneider}, {Fresco}, {Boller}, {Haase},
  {Brownstein}, {Lane}, {Bizyaev}, \& {Nitschelm}}]{Salvato2022AA...661A...3S}
{Salvato}, M., {Wolf}, J., {Dwelly}, T., {et~al.} 2022, \aap, 661, A3

\bibitem[{{Schneider} {et~al.}(2022){Schneider}, {Freund}, {Czesla}, {Robrade},
  {Salvato}, \& {Schmitt}}]{Schneider2022AA...661A...6S}
{Schneider}, P.~C., {Freund}, S., {Czesla}, S., {et~al.} 2022, \aap, 661, A6

\bibitem[{{Scoville} {et~al.}(2007){Scoville}, {Aussel}, {Brusa}, {Capak},
  {Carollo}, {Elvis}, {Giavalisco}, {Guzzo}, {Hasinger}, {Impey}, {Kneib},
  {LeFevre}, {Lilly}, {Mobasher}, {Renzini}, {Rich}, {Sanders}, {Schinnerer},
  {Schminovich}, {Shopbell}, {Taniguchi}, \&
  {Tyson}}]{Scoville2007ApJS..172....1S}
{Scoville}, N., {Aussel}, H., {Brusa}, M., {et~al.} 2007, \apjs, 172, 1

\bibitem[{{Seljak} {et~al.}(2005){Seljak}, {Makarov}, {Mandelbaum}, {Hirata},
  {Padmanabhan}, {McDonald}, {Blanton}, {Tegmark}, {Bahcall}, \&
  {Brinkmann}}]{Seljak2005PhRvD..71d3511S}
{Seljak}, U., {Makarov}, A., {Mandelbaum}, R., {et~al.} 2005, \prd, 71, 043511

\bibitem[{{Seppi} {et~al.}(2022){Seppi}, {Comparat}, {Bulbul}, {Nandra},
  {Merloni}, {Clerc}, {Liu}, {Ghirardini}, {Liu}, {Salvato}, {Sanders},
  {Wilms}, {Dwelly}, {Dauser}, {K{\"o}nig}, {Ramos-Ceja}, {Garrel}, \&
  {Reiprich}}]{Seppi2022AA...665A..78S}
{Seppi}, R., {Comparat}, J., {Bulbul}, E., {et~al.} 2022, \aap, 665, A78

\bibitem[{{Shen} {et~al.}(2013){Shen}, {McBride}, {White}, {Zheng}, {Myers},
  {Guo}, {Kirkpatrick}, {Padmanabhan}, {Parejko}, {Ross}, {Schlegel},
  {Schneider}, {Streblyanska}, {Swanson}, {Zehavi}, {Pan}, {Bizyaev},
  {Brewington}, {Ebelke}, {Malanushenko}, {Malanushenko}, {Oravetz}, {Simmons},
  \& {Snedden}}]{Shen13}
{Shen}, Y., {McBride}, C.~K., {White}, M., {et~al.} 2013, \apj, 778, 98

\bibitem[{{Sheth} \& {Tormen}(1999)}]{Sheth1999MNRAS.308..119S}
{Sheth}, R.~K. \& {Tormen}, G. 1999, \mnras, 308, 119

\bibitem[{{Silverman} {et~al.}(2015){Silverman}, {Kashino}, {Sanders},
  {Kartaltepe}, {Arimoto}, {Renzini}, {Rodighiero}, {Daddi}, {Zahid}, {Nagao},
  {Kewley}, {Lilly}, {Sugiyama}, {Baronchelli}, {Capak}, {Carollo}, {Chu},
  {Hasinger}, {Ilbert}, {Juneau}, {Kajisawa}, {Koekemoer}, {Kovac}, {Le
  F{\`e}vre}, {Masters}, {McCracken}, {Onodera}, {Schulze}, {Scoville},
  {Strazzullo}, \& {Taniguchi}}]{Silverman2015ApJS..220...12S}
{Silverman}, J.~D., {Kashino}, D., {Sanders}, D., {et~al.} 2015, \apjs, 220, 12

\bibitem[{{Sinha} \& {Garrison}(2020)}]{Sinha2020MNRAS.491.3022S}
{Sinha}, M. \& {Garrison}, L.~H. 2020, \mnras, 491, 3022

\bibitem[{{Siudek} {et~al.}(2023){Siudek}, {Mezcua}, \&
  {Krywult}}]{Siudek2023MNRAS.518..724S}
{Siudek}, M., {Mezcua}, M., \& {Krywult}, J. 2023, \mnras, 518, 724

\bibitem[{{Skelton} {et~al.}(2014){Skelton}, {Whitaker}, {Momcheva}, {Brammer},
  {van Dokkum}, {Labb{\'e}}, {Franx}, {van der Wel}, {Bezanson}, {Da Cunha},
  {Fumagalli}, {F{\"o}rster Schreiber}, {Kriek}, {Leja}, {Lundgren}, {Magee},
  {Marchesini}, {Maseda}, {Nelson}, {Oesch}, {Pacifici}, {Patel}, {Price},
  {Rix}, {Tal}, {Wake}, \& {Wuyts}}]{Skelton2014ApJS..214...24S}
{Skelton}, R.~E., {Whitaker}, K.~E., {Momcheva}, I.~G., {et~al.} 2014, \apjs,
  214, 24

\bibitem[{{Skrutskie} {et~al.}(2006){Skrutskie}, {Cutri}, {Stiening},
  {Weinberg}, {Schneider}, {Carpenter}, {Beichman}, {Capps}, {Chester},
  {Elias}, {Huchra}, {Liebert}, {Lonsdale}, {Monet}, {Price}, {Seitzer},
  {Jarrett}, {Kirkpatrick}, {Gizis}, {Howard}, {Evans}, {Fowler}, {Fullmer},
  {Hurt}, {Light}, {Kopan}, {Marsh}, {McCallon}, {Tam}, {Van Dyk}, \&
  {Wheelock}}]{Skrutskie2006AJ....131.1163S}
{Skrutskie}, M.~F., {Cutri}, R.~M., {Stiening}, R., {et~al.} 2006, \aj, 131,
  1163

\bibitem[{{Smee} {et~al.}(2013){Smee}, {Gunn}, {Uomoto}, {Roe}, {Schlegel},
  {Rockosi}, {Carr}, {Leger}, {Dawson}, {Olmstead}, {Brinkmann}, {Owen},
  {Barkhouser}, {Honscheid}, {Harding}, {Long}, {Lupton}, {Loomis}, {Anderson},
  {Annis}, {Bernardi}, {Bhardwaj}, {Bizyaev}, {Bolton}, {Brewington}, {Briggs},
  {Burles}, {Burns}, {Castander}, {Connolly}, {Davenport}, {Ebelke}, {Epps},
  {Feldman}, {Friedman}, {Frieman}, {Heckman}, {Hull}, {Knapp}, {Lawrence},
  {Loveday}, {Mannery}, {Malanushenko}, {Malanushenko}, {Merrelli}, {Muna},
  {Newman}, {Nichol}, {Oravetz}, {Pan}, {Pope}, {Ricketts}, {Shelden},
  {Sandford}, {Siegmund}, {Simmons}, {Smith}, {Snedden}, {Schneider},
  {SubbaRao}, {Tremonti}, {Waddell}, \& {York}}]{Smee13}
{Smee}, S.~A., {Gunn}, J.~E., {Uomoto}, A., {et~al.} 2013, \aj, 146, 32

\bibitem[{{Starikova} {et~al.}(2011){Starikova}, {Cool}, {Eisenstein},
  {Forman}, {Jones}, {Hickox}, {Kenter}, {Kochanek}, {Kravtsov}, {Murray}, \&
  {Vikhlinin}}]{Starikova11}
{Starikova}, S., {Cool}, R., {Eisenstein}, D., {et~al.} 2011, \apj, 741, 15

\bibitem[{{Tegmark} \& {Bromley}(1999)}]{Tegmark1999ApJ...518L..69T}
{Tegmark}, M. \& {Bromley}, B.~C. 1999, \apjl, 518, L69

\bibitem[{{Tegmark} \& {Peebles}(1998)}]{Tegmark1998ApJ...500L..79T}
{Tegmark}, M. \& {Peebles}, P.~J.~E. 1998, \apjl, 500, L79

\bibitem[{{Tinker} {et~al.}(2010){Tinker}, {Robertson}, {Kravtsov}, {Klypin},
  {Warren}, {Yepes}, \&
  {Gottl{\"o}ber}}]{TinkerRobertsonKravtsov_2010ApJ...724..878T}
{Tinker}, J.~L., {Robertson}, B.~E., {Kravtsov}, A.~V., {et~al.} 2010, \apj,
  724, 878

\bibitem[{{Trujillo-Gomez} {et~al.}(2011){Trujillo-Gomez}, {Klypin}, {Primack},
  \& {Romanowsky}}]{Trujillo2011}
{Trujillo-Gomez}, S., {Klypin}, A., {Primack}, J., \& {Romanowsky}, A.~J. 2011,
  \apj, 742, 16

\bibitem[{{V{\'e}ron-Cetty} \& {V{\'e}ron}(2010)}]{VeronVeron2010}
{V{\'e}ron-Cetty}, M.-P. \& {V{\'e}ron}, P. 2010, \aap, 518, A10

\bibitem[{{Viitanen} {et~al.}(2019){Viitanen}, {Allevato}, {Finoguenov},
  {Bongiorno}, {Cappelluti}, {Gilli}, {Miyaji}, \& {Salvato}}]{Viitanen19}
{Viitanen}, A., {Allevato}, V., {Finoguenov}, A., {et~al.} 2019, \aap, 629, A14

\bibitem[{{Wechsler} \& {Tinker}(2018)}]{Wechsler2018ARAA..56..435W}
{Wechsler}, R.~H. \& {Tinker}, J.~L. 2018, \araa, 56, 435

\bibitem[{{Xu} {et~al.}(2021){Xu}, {Zehavi}, \&
  {Contreras}}]{Xu2021MNRAS.502.3242X}
{Xu}, X., {Zehavi}, I., \& {Contreras}, S. 2021, \mnras, 502, 3242

\bibitem[{{Yang} {et~al.}(2022){Yang}, {Boquien}, {Brandt}, {Buat},
  {Burgarella}, {Ciesla}, {Lehmer}, {Ma{\l}ek}, {Mountrichas}, {Papovich},
  {Pons}, {Stalevski}, {Theul{\'e}}, \& {Zhu}}]{Yang2022ApJ...927..192Y}
{Yang}, G., {Boquien}, M., {Brandt}, W.~N., {et~al.} 2022, \apj, 927, 192

\bibitem[{{Yang} {et~al.}(2019){Yang}, {Brandt}, {Alexander}, {Chen}, {Ni},
  {Vito}, \& {Zhu}}]{Yang2019MNRAS.485.3721Y}
{Yang}, G., {Brandt}, W.~N., {Alexander}, D.~M., {et~al.} 2019, \mnras, 485,
  3721

\bibitem[{{Yang} {et~al.}(2018{\natexlab{a}}){Yang}, {Brandt}, {Darvish},
  {Chen}, {Vito}, {Alexander}, {Bauer}, \& {Trump}}]{Yang2018MNRAS.480.1022Y}
{Yang}, G., {Brandt}, W.~N., {Darvish}, B., {et~al.} 2018{\natexlab{a}},
  \mnras, 480, 1022

\bibitem[{{Yang} {et~al.}(2018{\natexlab{b}}){Yang}, {Brandt}, {Vito}, {Chen},
  {Trump}, {Luo}, {Sun}, {Xue}, {Koekemoer}, {Schneider}, {Vignali}, \&
  {Wang}}]{Yang2018MNRAS.475.1887Y}
{Yang}, G., {Brandt}, W.~N., {Vito}, F., {et~al.} 2018{\natexlab{b}}, \mnras,
  475, 1887

\bibitem[{{York} {et~al.}(2000){York}, {Adelman}, {Anderson}, {Anderson},
  {Annis}, {Bahcall}, {Bakken}, {Barkhouser}, {Bastian}, {Berman}, {Boroski},
  {Bracker}, {Briegel}, {Briggs}, {Brinkmann}, {Brunner}, {Burles}, {Carey},
  {Carr}, {Castander}, {Chen}, {Colestock}, {Connolly}, {Crocker}, {Csabai},
  {Czarapata}, {Davis}, {Doi}, {Dombeck}, {Eisenstein}, {Ellman}, {Elms},
  {Evans}, {Fan}, {Federwitz}, {Fiscelli}, {Friedman}, {Frieman}, {Fukugita},
  {Gillespie}, {Gunn}, {Gurbani}, {de Haas}, {Haldeman}, {Harris}, {Hayes},
  {Heckman}, {Hennessy}, {Hindsley}, {Holm}, {Holmgren}, {Huang}, {Hull},
  {Husby}, {Ichikawa}, {Ichikawa}, {Ivezi{\'c}}, {Kent}, {Kim}, {Kinney},
  {Klaene}, {Kleinman}, {Kleinman}, {Knapp}, {Korienek}, {Kron}, {Kunszt},
  {Lamb}, {Lee}, {Leger}, {Limmongkol}, {Lindenmeyer}, {Long}, {Loomis},
  {Loveday}, {Lucinio}, {Lupton}, {MacKinnon}, {Mannery}, {Mantsch}, {Margon},
  {McGehee}, {McKay}, {Meiksin}, {Merelli}, {Monet}, {Munn}, {Narayanan},
  {Nash}, {Neilsen}, {Neswold}, {Newberg}, {Nichol}, {Nicinski}, {Nonino},
  {Okada}, {Okamura}, {Ostriker}, {Owen}, {Pauls}, {Peoples}, {Peterson},
  {Petravick}, {Pier}, {Pope}, {Pordes}, {Prosapio}, {Rechenmacher}, {Quinn},
  {Richards}, {Richmond}, {Rivetta}, {Rockosi}, {Ruthmansdorfer}, {Sandford},
  {Schlegel}, {Schneider}, {Sekiguchi}, {Sergey}, {Shimasaku}, {Siegmund},
  {Smee}, {Smith}, {Snedden}, {Stone}, {Stoughton}, {Strauss}, {Stubbs},
  {SubbaRao}, {Szalay}, {Szapudi}, {Szokoly}, {Thakar}, {Tremonti}, {Tucker},
  {Uomoto}, {Vanden Berk}, {Vogeley}, {Waddell}, {Wang}, {Watanabe},
  {Weinberg}, {Yanny}, {Yasuda}, \& {SDSS Collaboration}}]{York2000}
{York}, D.~G., {Adelman}, J., {Anderson}, Jr., J.~E., {et~al.} 2000, \aj, 120,
  1579

\bibitem[{{Zacharegkas} {et~al.}(2022){Zacharegkas}, {Chang}, {Prat}, {Pandey},
  {Ferrero}, {Blazek}, {Jain}, {Crocce}, {DeRose}, {Palmese}, {Seitz},
  {Sheldon}, {Hartley}, {Wechsler}, {Dodelson}, {Fosalba}, {Krause}, {Park},
  {S{\'a}nchez}, {Alarcon}, {Amon}, {Bechtol}, {Becker}, {Bernstein}, {Campos},
  {Carnero Rosell}, {Carrasco Kind}, {Cawthon}, {Chen}, {Choi}, {Cordero},
  {Davis}, {Diehl}, {Doux}, {Drlica-Wagner}, {Eckert}, {Elvin-Poole},
  {Everett}, {Fert{\'e}}, {Gatti}, {Giannini}, {Gruen}, {Gruendl}, {Harrison},
  {Herner}, {Huff}, {Jarvis}, {Kuropatkin}, {Leget}, {MacCrann}, {McCullough},
  {Myles}, {Navarro-Alsina}, {Porredon}, {Raveri}, {Rollins}, {Roodman},
  {Ross}, {Rykoff}, {Secco}, {Sevilla-Noarbe}, {Shin}, {Troxel}, {Tutusaus},
  {Varga}, {Yanny}, {Yin}, {Zhang}, {Zuntz}, {Abbott}, {Aguena}, {Allam},
  {Andrade-Oliveira}, {Annis}, {Bacon}, {Bertin}, {Brooks}, {Burke},
  {Carretero}, {Castander}, {Costanzi}, {da Costa}, {Pereira}, {Desai},
  {Dietrich}, {Doel}, {Evrard}, {Flaugher}, {Frieman}, {Garc{\'\i}a-Bellido},
  {Gaztanaga}, {Gschwend}, {Gutierrez}, {Hinton}, {Hollowood}, {Honscheid},
  {Hoyle}, {James}, {Kuehn}, {Lima}, {Maia}, {Marshall}, {Melchior},
  {Menanteau}, {Miquel}, {Muir}, {Ogando}, {Paz-Chinch{\'o}n}, {Pieres},
  {Sanchez}, {Serrano}, {Smith}, {Suchyta}, {Tarle}, {Thomas}, {To},
  {Wilkinson}, \& {DES Collaboration}}]{Zacharegkas2022MNRAS.509.3119Z}
{Zacharegkas}, G., {Chang}, C., {Prat}, J., {et~al.} 2022, \mnras, 509, 3119

\bibitem[{{Zehavi} {et~al.}(2011){Zehavi}, {Zheng}, {Weinberg}, {Blanton},
  {Bahcall}, {Berlind}, {Brinkmann}, {Frieman}, {Gunn}, {Lupton}, {Nichol},
  {Percival}, {Schneider}, {Skibba}, {Strauss}, {Tegmark}, \&
  {York}}]{Zehavi2011ApJ...736...59Z}
{Zehavi}, I., {Zheng}, Z., {Weinberg}, D.~H., {et~al.} 2011, \apj, 736, 59

\bibitem[{{Zhang} {et~al.}(2021){Zhang}, {Wang}, {Luo}, {Mo}, {Liang}, {Li},
  {Yang}, {Wang}, {Zhang}, {Hong}, {Wang}, {Wang}, {Li}, \&
  {Shi}}]{Zhang_Luo2021AA...650A.155Z}
{Zhang}, Z., {Wang}, H., {Luo}, W., {et~al.} 2021, \aap, 650, A155

\bibitem[{{Zheng} {et~al.}(2005){Zheng}, {Berlind}, {Weinberg}, {Benson},
  {Baugh}, {Cole}, {Dav{\'e}}, {Frenk}, {Katz}, \& {Lacey}}]{Zheng2005}
{Zheng}, Z., {Berlind}, A.~A., {Weinberg}, D.~H., {et~al.} 2005, \apj, 633, 791

\bibitem[{{Zheng} {et~al.}(2007){Zheng}, {Coil}, \& {Zehavi}}]{Zheng2007}
{Zheng}, Z., {Coil}, A.~L., \& {Zehavi}, I. 2007, \apj, 667, 760

\bibitem[{{Zou} {et~al.}(2022){Zou}, {Brandt}, {Chen}, {Leja}, {Ni}, {Yan},
  {Yang}, {Zhu}, {Luo}, {Nyland}, {Vito}, \& {Xue}}]{Zou2022ApJS..262...15Z}
{Zou}, F., {Brandt}, W.~N., {Chen}, C.-T., {et~al.} 2022, \apjs, 262, 15

\bibitem[{{Zu} \& {Mandelbaum}(2015)}]{Zu2015MNRAS.454.1161Z}
{Zu}, Y. \& {Mandelbaum}, R. 2015, \mnras, 454, 1161

\end{thebibliography}

\appendix

\section{X-ray data analysis}

\subsection{X-ray mask}
\label{appendix:mask:xray}

We used the region files created by \texttt{eSASS/srctool} to create X-ray masks for point sources and extended sources \citep[][]{Liu_SPEC2022AA...661A...5L, LiuAng2022AA...661A...2L}. 
Each source has its signal-to-noise ratio measured as a function of radius (circular apertures). 
An optimal radius for source extraction was found by maximizing the signal-to-noise ratio given the local background surface brightness. 
We clipped it to a minimum radius of 10\arcsec (\texttt{MINIMUM\_SOURCE\_RADIUS} parameter) and a maximum radius of the 99\% energy enclosed fraction radius of the PSF. 
We used this maximum signal-to-noise radius as a starting point to determine the area to be masked around sources. 

\begin{figure*}
    \centering
\includegraphics[width=0.95\columnwidth]{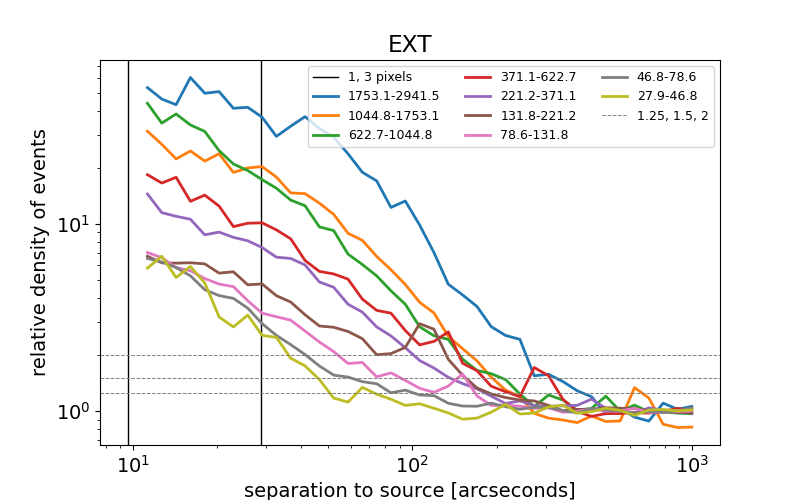}
\includegraphics[width=0.95\columnwidth]{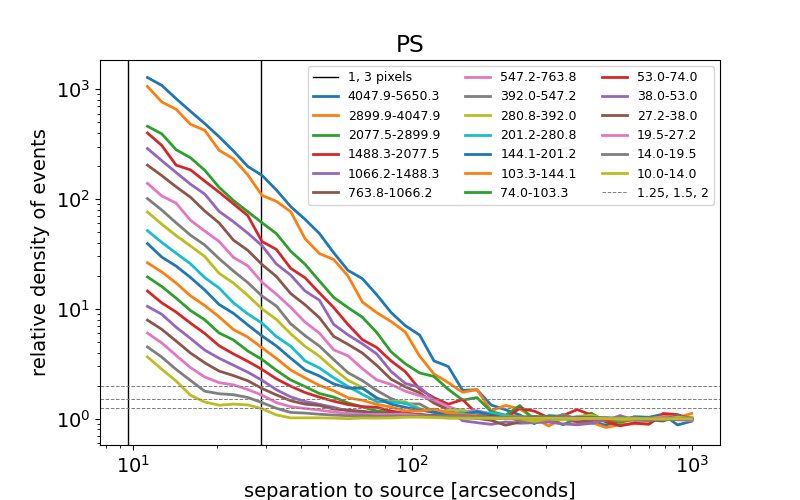}
\includegraphics[width=0.95\columnwidth]{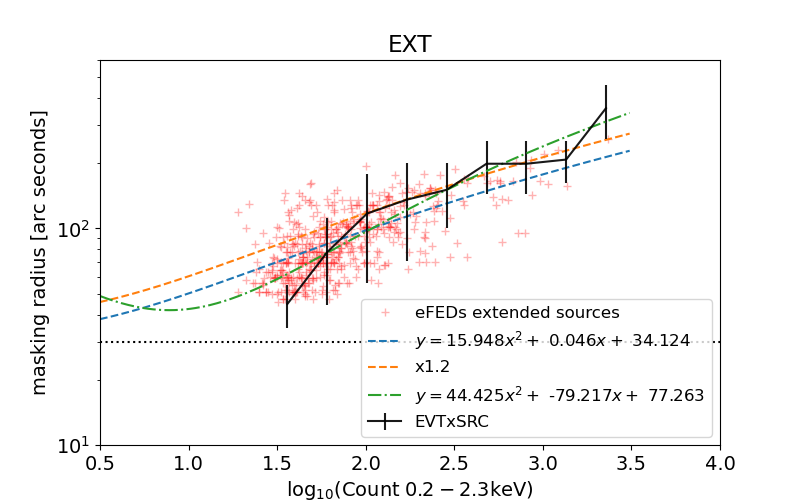}
\includegraphics[width=0.95\columnwidth]{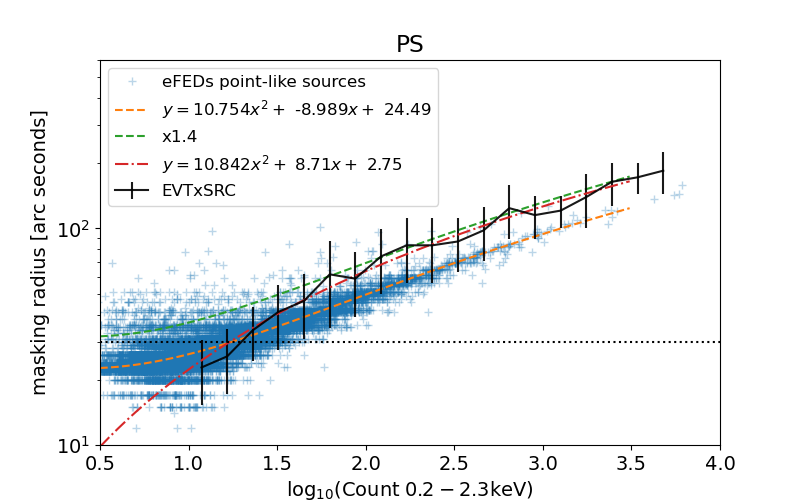}
        \caption{Cross-correlation between sources and events as a function of angular separation for extended (EXT, top-left panel) and point-like (PS, top-right panel) X-ray sources. 
        The lower panels show the masking radius vs. log10 of the counts measured. The average masking radius obtained with the cross-correlation is shown with the black line. The masking radius obtained with \texttt{eSASS/srctool} for individual sources is systematically lower than the black line. Its best-fit polynomial (dashed blue line) is multiplied by 1.2 (extended sources) or 1.4 (point sources) to align with the black line.}
    \label{fig:mask}
\end{figure*}

We measured the cross-correlation as a function of scale between events (0.2-2.3 keV) and sources in the catalog. 
We measured it for bins of the number of counts measured per source in the detection band. 
The cross-correlation becomes constant above a particular angular scale, which corresponds to a conservative masking radius of a source (with a given number of counts), that is, its average imprint on the sky (see Fig. \ref{fig:mask}, top panels). 
For each cross-correlation curve, we measured the radius at which its value is between 1.25 and two times that of the constant values measured at large separations. 
This brackets the masking radius within the black vertical error bars shown in Fig. \ref{fig:mask} (bottom panels). 
We find that this cross-correlation masking radius for point sources is, on average, 40 percent (20 for extended sources) larger than the \texttt{eSASS/srctool} radius of maximum signal-to-noise (see Fig. \ref{fig:mask}, bottom panels). 
The srctool mask is likely not conservative enough for our purpose. 
For instance, the detection of a point source just beyond the \texttt{eSASS/srctool} masking radius of another point source will be subject to biases due to the residual events measured via the cross-correlation.  
Though, if we followed the average masking radius suggested by the cross-correlation (Fig. \ref{fig:mask}, bottom panels), the large scatter in the relation between the maximum signal-to-noise radius and the total number of counts would be missed. 
So, to have a conservative mask that closely follows the data, we multiplied the masking radii from \texttt{eSASS/srctool} by a factor of 1.4 (this is more conservative than required for the extended sources, but it simplifies the procedure). That way, the masking radius will reach, on average, the line obtained from the cross-correlation. 
Doing so ensures no remaining correlation between the set of events outside the mask and the source catalog.
We conservatively masked both point sources and extended sources individually.

After applying the mask, we are left with 17,523 AGN candidates. 
Using the fraction of random points (see Sect. \ref{subsubsec:random:cat}) that fall in the masks, we estimated the area of the observed X-ray sky effectively occupied by sources. 
In all, sources occupy 9.805 deg$^2$ out of 141.97 deg$^2$. Stars occupy 0.988 deg$^2$, AGNs 6.914 deg$^2$, and extended sources 2.057 deg$^2$.

\subsection{Random catalog}
\label{appendix:random:cat} 

We used the sensitivity map produced by the eROSITA pipeline \citet[eSASS, apetool][]{Brunner2021arXiv210614517B} with a parameter $P_{thres}= e^{-8}=0.00033$ (the Poisson probability threshold, below which an excess of counts is considered a source), corresponding to a detection likelihood of 8. 
It is a pixelated fits image of size $[0,9000)\times[0,18000)$ that contains the sensitivity limit (in counts). 
Each random point falls in a pixel of this map, and we attached the corresponding count limit, $C^{lim}_X$, to the random number. 
We drew a large set of redshift and X-ray fluxes ($f_x$, $z$) from the AGN X-ray luminosity function projection to assign to each random point (\citealt{Aird15,Comparat19}). 
It is sampled down to $2\times10^{-15}$erg cm$^{-2}$ s$^{-1}$, a flux value at which the area curve is smaller than 0.5 deg$^2$.
We converted the flux into an expected number of counts, 
\begin{equation}
        CT^{expected} = f_x \times ECF \times EEF \times t_{exp} + CT^{background},
\end{equation} 
where the energy conversion factor is $ECF= 1.164\times 10^{12}$. 
The encircled energy fraction was set to $EEF=0.65$. 
The exposure time, $t_{exp}$, was obtained with the exposure map. The 
$CT^{background}$ was obtained from the background map. 
We drew a random Poisson variable, $R_v$, for each $CT^{expected}$. 
If this value exceeds the count limit, $R_v>C^{lim}_X$, the point is accepted in the random sample. 
We removed the shallower areas at the edge of the field through a minimum exposure time threshold to minimize the maximum offset between the normalized cumulative distribution of the data sample and the random sample. 
We find that an 830-second threshold minimizes the KS test values at 0.19\% for right ascension and 0.81\% for declination. 
It removes $\sim10$ deg$^2$. 
It is sufficiently accurate to estimate clustering on the photometric sample. 
After masking extended sources and stars and trimming the low exposure time region, the total number of random points remaining is 3,713,726.

\end{document}